\documentclass[%
reprint,
superscriptaddress,
preprintnumbers,
nofootinbib,
nobibnotes,
amsmath,amssymb,
aps,
prd,
floatfix,
]{revtex4-2}

\usepackage{lipsum} 
\usepackage{graphicx}
\usepackage{dcolumn}
\usepackage{bm}
\usepackage{amsmath,amsfonts,amssymb,bm,mathrsfs,longtable}
\graphicspath{ {./} }
\usepackage{hyperref}
\usepackage{array}
\usepackage{color}
\usepackage{enumitem}
\usepackage[utf8]{inputenc}
\usepackage[normalem]{ulem}
\usepackage{comment}
\usepackage{mhchem}
\usepackage{outlines}
\usepackage{mathtools}
\usepackage{cleveref}
\crefname{figure}{Fig.}{Figs.}
\crefname{section}{Sec.}{Secs.}
\Crefname{section}{Section}{Sections}
\crefname{equation}{Eq.}{Eqs.}
\crefname{table}{Table}{Tables}
\hypersetup{
	colorlinks   = true,
    linkcolor={blue!70!black},
    citecolor={blue!70!black}
}

\usepackage{xcolor}
\usepackage{booktabs}
\usepackage{multirow}

\def\be{\begin{equation}}
\def\ee{\end{equation}}

\newcommand{\Hrondo}{\mathcal{H}}

\newcommand{\f}{\frac}

\newcommand{\dd}{\mathrm{d}}
\newcommand{\sigmapsq}{\sigma'^{\,2}}

\begin{document}

\title{Cosmological effects of the Galileon term in scalar-tensor Theories}

\author{Angelo G. Ferrari}
 \email{angelo.ferrari3@unibo.it}
 \affiliation{Dipartimento di Fisica e Astronomia, Universit\`a di Bologna, viale Berti Pichat 6/2, 40127 Bologna, Italy.}
 \affiliation{INFN, Sezione di Bologna, viale C. Berti Pichat 6/2, 40127 Bologna, Italy}
 
\author{Mario Ballardini}\email{mario.ballardini@unife.it}
 \affiliation{Dipartimento di Fisica e Scienze della Terra, Universit\`a degli Studi di Ferrara, via Giuseppe Saragat 1, 44122 Ferrara, Italy}
 \affiliation{INFN, Sezione di Ferrara, via Giuseppe Saragat 1, 44122 Ferrara, Italy}
 \affiliation{INAF/OAS Bologna, via Piero Gobetti 101, 40129 Bologna, Italy}
 
\author{Fabio Finelli}\email{fabio.finelli@inaf.it}
 \affiliation{INAF/OAS Bologna, via Piero Gobetti 101, 40129 Bologna, Italy}
 \affiliation{INFN, Sezione di Bologna, viale C. Berti Pichat 6/2, 40127 Bologna, Italy}
 
\author{Daniela Paoletti}\email{daniela.paoletti@inaf.it}
 \affiliation{INAF/OAS Bologna, via Piero Gobetti 101, 40129 Bologna, Italy}
 \affiliation{INFN, Sezione di Bologna, viale C. Berti Pichat 6/2, 40127 Bologna, Italy}

\author{Nicoletta Mauri}\email{nicoletta.mauri@bo.infn.it}
 \affiliation{Dipartimento di Fisica e Astronomia, Universit\`a di Bologna, viale Berti Pichat 6/2, 40127 Bologna, Italy.}
 \affiliation{INFN, Sezione di Bologna, viale C. Berti Pichat 6/2, 40127 Bologna, Italy}

\begin{abstract}
We study the cosmological effects of a Galileon term in scalar-tensor theories of gravity.
The subset of scalar-tensor theories considered are characterized by a nonminimal coupling $F(\sigma) R$, a kinetic term with arbitrary sign $Z (\partial \sigma)^2$ with $Z = \pm 1$, a potential $V(\sigma)$, and a Galileon term $G_3(\sigma, (\partial \sigma)^2) \square \sigma$.
In addition to the modified dynamics, the Galileon term provides a screening mechanism to potentially reconcile the models with general relativity predictions inside a Vainshtein radius. 
Thanks to the Galileon term, the stability conditions, namely ghost and Laplacian instabilities, in the branch with a negative kinetic term ($Z = -1$) are fulfilled for a large volume of the parameter space.
Solving numerically the background evolution and linear perturbations, we derive the constraints on the cosmological parameters in the presence of a Galileon term for different combination of the cosmic microwave background data from {\em Planck}, baryon acoustic oscillations measurements from BOSS, and supernovae from the Pantheon compilation.
We find that the Galileon term alters the dynamics of all the studied cases.
For a standard kinetic term ($Z = 1$), we find that {\em Planck} data and a compilation of baryon acoustic oscillations data constrain the Galileon term to small values that allow screening very inefficiently.
For a negative kinetic term ($Z = -1$), a Galileon term and a non-zero potential lead to an efficient screening in a physically viable regime of the theory, with a value for the Hubble constant today which alleviates the tension between its cosmic microwave background and local determinations.
For a vanishing potential, the case with $Z=-1$ and the Galileon term driving the late acceleration of the Universe is ruled out by {\em Planck} data.
\end{abstract}

\maketitle
\section{Introduction}
The Lambda cold dark matter $\Lambda$CDM model is the standard cosmological model, it provides a remarkably good fit to most cosmological observations such as the measurements of cosmic microwave background (CMB) anisotropies in temperature and polarization \cite{Planck:2018nkj}, baryon acoustic oscillations (BAO) in the galaxy and cluster distributions \cite{eBOSS:2020yzd,Veropalumbo:2013cua}, cosmic shear measurements of the galaxy distribution \cite{KiDS:2020suj,DES:2021wwk} and of the CMB \cite{Planck:2018lbu,Sherwin:2016tyf,Wu:2019hek,Darwish:2020fwf}, and measurements of the luminosity distances of type Ia supernovae (SN Ia) \cite{SupernovaSearchTeam:1998fmf,SupernovaCosmologyProject:1998vns,Carr:2021lcj}.
Nevertheless, in recent years its enduring success has been threatened by the growing tension between the inference of $H_0$ within $\Lambda$CDM from the {\em Planck} measurement of CMB anisotropy \cite{Planck:2018vyg} and its determination with Supenovae at low redshift, which has grown to a maximum level of $5\sigma$ with the latest SH0ES release with SN Ia calibrated using Cepheids \cite{Riess:2021jrx}.

Whereas inferences of $H_0$ from different CMB experiments are very consistent with one another \cite{Planck:2018vyg,ACT:2020gnv,SPT-3G:2021wgf}, different approaches in calibrating SN provide different estimates and different uncertainties in the determination of $H_0$, as the latest SH0ES result based on Cepheids is $H_0 = 73.04 \pm 1.04\, \rm km\,s^{-1}\,Mpc^{-1}$ at 68\% confidence level (CL) \cite{Riess:2021jrx}, while the latest Carnegie-Chicago Hubble program estimate, using the tip of the Red Giant Branch calibration technique is $H_0 = 69.8 \pm 0.8\, \text{(stat)} \pm 1.7\, \text{(sys)}\,\rm km\,s^{-1}\,Mpc^{-1}$ at 68\% CL \cite{Freedman:2019jwv}.
These differences in the local determinations of $H_0$ and their error obviously affects the tension with the CMB inference in a statistically significant way and might point to unknown systematics or uncertainties not properly taken into account.
Nevertheless, this tension that emerged with the first {\em Planck} cosmological release \cite{Planck:2013pxb} has fueled interest in models explaining a larger value of $H_0$ compared to $\Lambda$CDM \cite{Knox:2019rjx,DiValentino:2020zio,DiValentino:2021izs,Perivolaropoulos:2021jda,Schoneberg:2021qvd,Shah:2021onj,Abdalla:2022yfr}.

Modified gravity (MG) is one of the most promising fundamental physics route to reconcile the inference of $H_0$ from the {\em Planck} measurement of CMB anisotropy and its inferred value determined with the calibration of supernovae at low redshift within $\Lambda$CDM model.
A simple mechanism to increase $H_0$ by an evolving gravitational constant is at play even in the simplest scalar-tensor models, i.e. induced gravity (IG) (equivalent to the Brans-Dicke model by a field redefinition).
Without the fine-tuning of parameters, the scalar field regulating the gravitational strength moves around recombination because of its coupling to pressureless matter.
This mechanism, which can be also seen as a dynamical and self-consistent variation in the gravitational constant, induces a degeneracy between $H_0$ and the nonminimal coupling of the scalar field to the curvature leading to a larger $H_0$ compared to $\Lambda$CDM \cite{Umilta:2015cta,Ballardini:2016cvy,Rossi:2019lgt,Ballesteros:2020sik,Zumalacarregui:2020cjh,Braglia:2020iik,Ballardini:2020iws,Braglia:2020auw,Lee:2022cyh}.

Although this mechanism is completely general, the increase of $H_0$ with respect to $\Lambda$CDM depends on the model and also on the number of additional parameters with respect to $\Lambda$CDM.
With {\em Planck} DR3 likelihoods, we have $\Delta H_0 = H_0 - H_0^{\Lambda {\mathrm CDM} } \sim 2 \rm\, km\,s^{-1}\,Mpc^{-1}$ for IG \cite{Umilta:2015cta,Ballardini:2016cvy,Ballardini:2020iws} and $\Delta H_0 \sim 4\rm\, km\,s^{-1}\,Mpc^{-1}$ for a nonminimally coupled scalar field with a quartic potential \cite{Braglia:2020auw}, dubbed early modified gravity (EMG), which is among the best solutions to the $H_0$ tension according to \cite{Schoneberg:2021qvd}.

All the scalar-tensor models of gravity mentioned above do not have Vainshtein screening and a nearly negligible chameleon screening due to the coupling to matter.
For this reason, one has to compare the volume of the parameter space of the nonminimal coupling required to increase $H_0$ to the Solar System bounds on the corresponding post-Newtonian parameters \cite{Bertotti:2003rm}.
As for the $H_0$ degeneracy, also in this case, the comparison is nontrivial and model dependent.
Whereas for the simplest IG models, the Solar System bounds constrain the nonminimal coupling to values which do not lead to an appreciable increase of $H_0$ compared to $\Lambda$CDM \cite{Umilta:2015cta,Ballardini:2016cvy,Ballardini:2020iws}, it is sufficent to consider a nonminimally coupled coupled scalar field to have a large volume of parameter space where viable models on Solar System scales can lead to a value of $H_0$ larger than in $\Lambda$CDM \cite{Rossi:2019lgt,Braglia:2020iik}.
Early modified gravity is a further step in the construction of a scalar-tensor viable model on Solar System scales with large $H_0$ by adding a self-interaction term which implements an attractor mechanism towards general relativity (GR) with a cosmological constant.

In this paper, we instead consider the cosmology of IG, with the addition of a Galileon term \cite{Nicolis:2008in,Deffayet:2009wt,Kobayashi:2009wr}, whose purpose is to reconcile the theory with GR on scales below the so-called Vainshtein radius.
We therefore would like to investigate the implications of the addition of an explicit term implementing a screening mechanism on the cosmology of the simplest scalar-tensor theories of gravity (see also \cite{Barreira:2014jha,Renk:2016olm,Frusciante:2019puu} for previous studies with cosmological bounds on Galileon gravity).

The paper is organized as follows. In \cref{sec:model} we describe the models considered and their background dynamics together with the stability conditions and the screening mechanism.
In \cref{sec:lin_pert} we present evolution of the linear fluctuations, CMB anisotropies and matter power spectrum.
\Cref{sec:results} is devoted to the presentation of the datasets and the details of our Markov chain Monte-Carlo (MCMC) analysis, joint with the presentation of the results. We draw our conclusions in \cref{sec:conclusions}.
\section{The model} \label{sec:model}
We study the following subclass of Horndeski theories \cite{Horndeski:1974wa,Deffayet:2009wt,Deffayet:2011gz,Kobayashi:2011nu,Kase:2018aps}:
\be \label{eq:action}
	{\cal S} = \int {\dd}^4x\sqrt{|g|} \left[G_4(\sigma)R + G_2(\sigma,X) +G_3(\sigma,X)\square\sigma + \mathcal{L_{\rm M}}\right]
\ee

where $|g|$ is the absolute value of the determinant of the metric $g_{\mu\nu}$ for which we use a $(-+++)$ signature, $\sigma$ is a scalar field, $X \equiv -\nabla_{\mu}\sigma\nabla^{\mu}\sigma/2 = -\partial_{\mu}\sigma\partial^{\mu} \sigma/2$, and $\Box~\equiv~\nabla_\mu \nabla^\mu$ is the covariant d'Alambert operator.
$\mathcal{L_{\rm M}}$ is the matter Lagrangian which does not depend on the scalar field $\sigma$ and it is minimally coupled with the metric $g$.
The action \eqref{eq:action} predicts that gravitational waves travel at the speed of light \cite{Kase:2018aps}, consistently with the current measurements \cite{LIGOScientific:2017zic}.
We consider the following form for the $G$ functions:
\begin{align}\label{eq:BDG_Gis}
\begin{split}
G_4 &= \gamma\sigma^2/2 \\
G_3 &= -2 g(\sigma)X \\ 
G_2 &= Z X-V(\sigma) + 4\zeta(\sigma)X^2. \\
\end{split}
\end{align}
where $Z=\pm 1$ is the sign of the kinetic term.
This Lagrangian is equivalent to the extension of the Brans-Dicke (BD) model \cite{jordan55,PhysRev.124.925} with a Galileon term.
In fact with the field redefinition $\phi=\gamma\sigma^2/2$, with $\gamma=Z/(4\omega_{\rm BD}) > 0 \text{ and } Z=\pm 1$, the $G$ functions become  
\begin{align}\label{eq:BDG_Gis_phi}
\begin{split}
G_4 &= \phi \\
G_3 &= -2 f(\phi)\chi \\ 
G_2 &= 2 \frac{\omega_{\rm BD}}{\phi}\chi-V(\phi), \\
\end{split}
\end{align}
where $\chi\equiv -\nabla_\mu \phi\, \nabla^\mu \phi/2$ and the relationship between $g, \zeta \text{ and} f$ is $g(\sigma) = \gamma\sigma^3 f(\sigma)$; $\zeta(\sigma)=\sigma^{-1}g(\sigma)$.
Therefore, we refer to the model defined by \cref{eq:BDG_Gis} as {\it Brans-Dicke Galileon} (BDG), while the model with $G_2 = Z X - V(\sigma)$ is the {\it induced gravity Galileon} (IGG), as it is the extension of induced gravity (IG) \cite{PhysRevLett.42.417}  with a Galileon term $G_3 = -2 g(\sigma)X$.
Therefore even if BD and IG are equivalent, their extensions with Galileon terms are not. In particular, IGG corresponds to BDG given by \cref{eq:BDG_Gis} with, formally, $\zeta=0$, but it is not simply a special case of BDG when $\zeta = 0$.
In fact, as described above, in the BDG model the functions $g$ and $\zeta$ are not independent and setting $\zeta = 0$ would mean setting $g = 0$ as well.

We study two different versions of BDG and IGG, relating to the sign of the kinetic term $Z$:

\begin{outline}
\1 Standard branch: $Z=+1$ in which the kinetic term has a standard sign.
In this branch we consider both IGG and BDG (IGGst and BDGst, respectively) and we take $V(\sigma) \text{ and } g(\sigma)$ to be power laws of the scalar field: $V(\sigma) = \lambda_n \sigma^n$, $g(\sigma) = \alpha_m \sigma^m$, for several combinations of $n \text{ and } m$. In this scenario the potential dominates the dynamics and provides the acceleration of the expansion of the Universe, while the $G_3$ enters as a small correction to the standard BD theory.
\1 Phantom branch: $Z=-1$ and $\gamma<1/6$, equivalent to $\omega_{\rm BD} <-3/2$, in which the kinetic term has a nonstandard sign.
In this branch we consider only BDG with $g(\sigma) = \alpha\sigma^{-1}$, either with or without a potential (cosmological constant).
If there is no potential the burden to provide comic acceleration, a healthy theory and effective screening on small scales, is on the $G_3$ term \cite{Silva:2009km}.
Reinserting the potential into the theory, both the Galileon term and the cosmological constant contribute to the dynamics and give rise to the late-time acceleration of the Universe.
Even if we have a nonzero potential, the $G_3$ is still able to provide a healthy theory and effective screening on small scales, where we recover (GR).
\end{outline}

%
%

\subsection{Covariant equations}
By varying the action with respect to the metric we obtain the modified Einstein field equation
\begin{align}\label{eq:EE}
    \begin{split}
        G_{\mu\nu} = \frac{1}{F(\sigma)} \bigg[ &T_{\mu\nu}^{\rm (M)} + T_{\mu\nu}^{\rm (G)} + Z \big( \partial_{\mu}\sigma\partial_{\nu}\sigma - \frac{1}{2}g_{\mu\nu}\partial^{\rho}\sigma\partial_{\rho}\sigma \big) \\ &- g_{\mu\nu} V(\sigma) + (\nabla_\mu \nabla_\nu - g_{\mu\nu} \Box) F(\sigma) \bigg]\,,
    \end{split}
\end{align}
where $F(\sigma)=\gamma\sigma^2$, $G_{\mu\nu}=R_{\mu\nu}-\frac{1}{2}R$ is the Einstein tensor and $T_{\mu\nu}^{\rm (M)}~=~-\frac{2}{\sqrt{|g}}\frac{\delta(\sqrt{|g|}\mathcal{L}_{\rm M})}{\delta g^{\mu\nu}}$ is the energy-momentum tensor of matter. $T_{\mu\nu}^{(G)}$ is defined as 
\begin{align}
\begin{split}
T_{\mu\nu}^{\rm (G)} = -2 \Big\{ \, &g(\sigma) \, \nabla_\mu \sigma \nabla _\nu \sigma \, \Box \sigma - \nabla_{(\mu} \, \sigma \, \nabla_{\nu)} \big[\, g(\sigma) (\partial \sigma)^2\,  \big] \\ 
&+ \frac{1}{2} g_{\mu\nu} \nabla_\alpha\, \sigma \nabla^\alpha \big[\, g(\sigma) (\partial \sigma)^2 \, \big] - \frac{\zeta(\sigma)}{2} g_{\mu\nu} (\partial \sigma)^4 \\ 
&+2\zeta(\sigma) \nabla_\mu \sigma \nabla _\nu \sigma \ (\partial \sigma)^2 \, \Big\},
\end{split}
\end{align}
with $\nabla_{(\mu} \, \sigma \, \nabla_{\nu)} =\frac{1}{2} \big(\nabla_{\mu} \, \sigma \, \nabla_{\nu} + \nabla_{\nu} \, \sigma \, \nabla_{\mu} \big)$.
\vspace{1ex}
It is useful to write down the trace of the Einstein equation \eqref{eq:EE}, which is 
\be \label{eq:Etrace}
R = \frac{1}{F} \Big[ - T^{(\rm M)} - T^{\rm (G)} + Z (\partial \sigma)^2 + 4V + 3\,\square F(\sigma)  \Big],
\ee
with $T^{(G)}$ as 
\be
T^{\rm (G)} = - 2 \bigg\{ g(\sigma)\, (\partial \sigma)^2 \Box \sigma + \nabla_\mu \sigma \nabla^\mu \big[ g(\sigma)\, (\partial \sigma)^2 \big] \bigg\}.
\ee
Note that $T^{(G)}$ does not depend on $\zeta(\sigma)$. 

The equation for the evolution of the scalar field, obtained by varying the action \eqref{eq:action} with respect to the field, is
\begin{align}
    \begin{split}
		&\Box\sigma \big[ Z - 4\zeta\, (\partial\sigma)^2 \big] -2 g\, \big\{ (\Box \sigma)^2 - \nabla^\mu\nabla^\nu \sigma \nabla_\mu\nabla_\nu \sigma \\
		&- \nabla^\mu \sigma \nabla^\nu \sigma R_{\mu\nu} \big\} +4 g,_{\sigma} \nabla^\mu \sigma \nabla^\nu \sigma \nabla_\mu \nabla_\nu \sigma 
		+ g,_{\sigma \sigma} (\partial\sigma)^4 \\
		&- 3\zeta,_{\sigma} (\partial\sigma)^4 - 4\zeta(\sigma)\nabla_\mu [ (\partial\sigma)^2 ]\nabla^\mu \sigma + \gamma\sigma R -V,_\sigma = 0.
    \end{split}
\end{align}

%
%

\subsection{Background evolution}
Working in a spatially flat Friedmann-Lema\^itre-Robertson-Walker (FLRW) universe described by the line element
\be
\dd s^2 = -\dd t^2 + a^2(t) \, \dd x^2\,,
\ee
the Einstein field equations reduce to:
\begin{align}
\begin{split}
		3 F H^2 =  &\rho + \frac{1}{2} Z \dot\sigma^2 - 3 H \dot{F} + V(\sigma) \\ 
		&+ \dot\sigma^3\big[6 g(\sigma) H - \dot{g}(\sigma)  + 3\zeta(\sigma) \dot\sigma \big]\\
		\equiv &\rho + \rho_{\sigma}\,,
\end{split}
\label{eq:Friedmann}
\end{align}
\begin{align}
\begin{split}
    -2 F \dot{H} = &\rho + p + Z\dot\sigma^2 + \ddot{F} - H\dot{F}\\ 
    &+ \dot\sigma^2\big[(6 g H - 2 g,_\sigma\dot\sigma)\dot\sigma + 4\zeta\dot\sigma^2 -2 g \ddot\sigma  \big]\\
    \equiv &\rho + p + \rho_\sigma + p_\sigma\,,
\end{split}
\end{align}
where:
\be
\rho_\sigma = \frac{Z}{2} \dot\sigma^2 - 3 H \dot{F} + V(\sigma) 
		+ \dot\sigma^3\big[6 g(\sigma) H - \dot{g}(\sigma)  + 3\zeta(\sigma) \dot\sigma \big],
\ee
\be
p_\sigma = \frac{Z}{2} \dot\sigma^2 - V(\sigma) + \ddot{F} + 2 H \dot{F} - \dot\sigma^4 ( g,_{\sigma} - \zeta ) - 2 g \dot\sigma^2 \ddot\sigma.
\ee

The scalar field equation in the FLRW metric takes the form:
\begin{align}
\begin{split}
&\ddot\sigma \big(Z+12 g H \dot\sigma - 4 (g,_{\sigma} -3\zeta)\dot\sigma^2 \big)- 6\gamma\sigma(2H^2+\dot{H}) +V,_\sigma \\
&+ 3 Z H \dot\sigma+ 6 g (3H^2+\dot{H})\dot\sigma^2 +12 H \zeta \dot\sigma^3\\
&- (g,_{\sigma\sigma} - 3\zeta,_{\sigma}) \dot\sigma^4 = 0.
\label{eq:scalar_field_flrw}
\end{split}    
\end{align}

Because of the nonminimal coupling between the scalar field and the Ricci scalar in the Lagrangian, the Newton constant in the Friedmann equations is replaced by $G_{\rm cosm} = (8\pi F)^{-1}$ that now varies with time.

Following the notation of \cite{Finelli:2007wb} we define the density parameters for radiation (r), pressureless matter (m) and the scalar field ($\sigma$) as
\be
\widetilde\Omega_i = \frac{\rho_i}{3FH^2} \equiv \frac{\rho_i}{\rho_{\rm crit}}\quad (i={\rm r, m}, \sigma).
\label{eq:tilde_Omegas}
\ee
It is also convenient to define dark energy density and pressure parameters in a framework which mimics Einstein gravity at the present time, by rewriting the Friedmann equations as \cite{Torres:2002pe,Gannouji:2006jm,Finelli:2007wb}
\begin{align}
\begin{split}
		3 F_0 H^2 &= \rho + \rho_{\rm DE}\,, \\
		-2 F_0 \dot{H} &= \rho  + p + \rho_{\rm DE} + p_{\rm DE},
\end{split}
\end{align}
which leads to
\be
\rho_{\rm DE} = \frac{F_0}{F}\rho_{\sigma} + \rho\bigg(\frac{F_0}{F}-1\bigg)\, ,
\ee
\be
p_{\rm DE} = \frac{F_0}{F} p_{\sigma} + p\bigg(\frac{F_0}{F}-1\bigg)\,.
\ee
In this way, the effective parameter of state for dark energy can be defined as $w_{\rm DE} \equiv p_{\rm DE}/\rho_{\rm DE}$. In this context the density parameters mimicking radiation, matter and dark energy (DE) in Einstein gravity are
\be
\Omega_i = \frac{\rho_i}{3 F_0 H^2} \quad (i={\rm r, m, DE}).
\label{eq:Omegas}
\ee
The two definitions in \cref{eq:tilde_Omegas,eq:Omegas} coincide at $z=0$: $\widetilde\Omega_{0, i}=\Omega_{0, i}$.

%
%

\subsubsection{Standard branch $(Z=1)$}
In the standard branch we consider IGG: the theory without $\zeta(\sigma)$: $G_4=\gamma\sigma^2/2,\, G_2= X-V(\sigma), \, G_3= -2 g(\sigma)X$.
In the region in parameter space that produces a reasonable cosmological background evolution BDG and IGG are nearly indistinguishable.
For this reason we restrict ourselves to IGG for $Z=1$.

We consider a monomial potential $V(\sigma)=\lambda_n \sigma^n$ and $g(\sigma)=\alpha_m \sigma^m$.
For this choice of $V(\sigma) \text{ and } g(\sigma)$ the IGG model presents exact solutions with accelerated expansion in the absence of matter where the scale factor is $a(t)\propto t^p$, with $n$ and $m$ related by $m=1-n$, and 
\begin{align}
\label{eq:p_ana_sol}
&p = \frac{2 (-2 + n + 4 \beta - 4 \gamma + n^2 \gamma) } { ( 24 \beta - 16 \gamma + 20 n \gamma - 8 n^2 \gamma + n^3 \gamma) }\, , \\ 
&\sigma(t) = c_0 t^{ -2/(n-2) },
\end{align}
where $\beta$, defined by $\alpha\equiv\beta c_0^{n-2}$, is a reparametrization of $\alpha$, useful to show that in the limit of $\alpha \rightarrow 0$ ($\beta \rightarrow 0$) we recover the analogous solution found in IG \cite{Ballardini:2016cvy}.

For $n=4$ and $n=2$, there are de Sitter solutions $a(t)\propto{e^{Ht}}$ with a constant $H >0$.
For $n=4$, the solution found for IG \cite{Ballardini:2016cvy}: $\sigma =\pm H\sqrt{3\gamma/\lambda_4}$ is still valid in IGG for every possible form of $g(\sigma)$, as this term does not contribute under the ansatz $\sigma = \text{constant}$.
For $n=2$, when $g(\sigma)\propto\sigma^{-1}$, there are solutions of the form $\sigma\propto e^{H t \delta}$, with $\delta\equiv\delta(\alpha,\gamma,H)$. Its explicit form is
\be
\delta = -1,\, \text{ or } \delta = \frac{-1-4\gamma\pm \sqrt{(1+4\gamma)^2 + 48 H^2 \gamma \alpha }}{12 H^2 \alpha}.
\ee
For $\alpha\rightarrow 0$ these solutions reduce to the ones discussed in \cite{CerioniPhd}:
\be
\delta=-1\, \text{ or } \delta=\frac{2\gamma}{1+4\gamma}\, .
\ee

\begin{figure}
\includegraphics[width=0.47\textwidth]{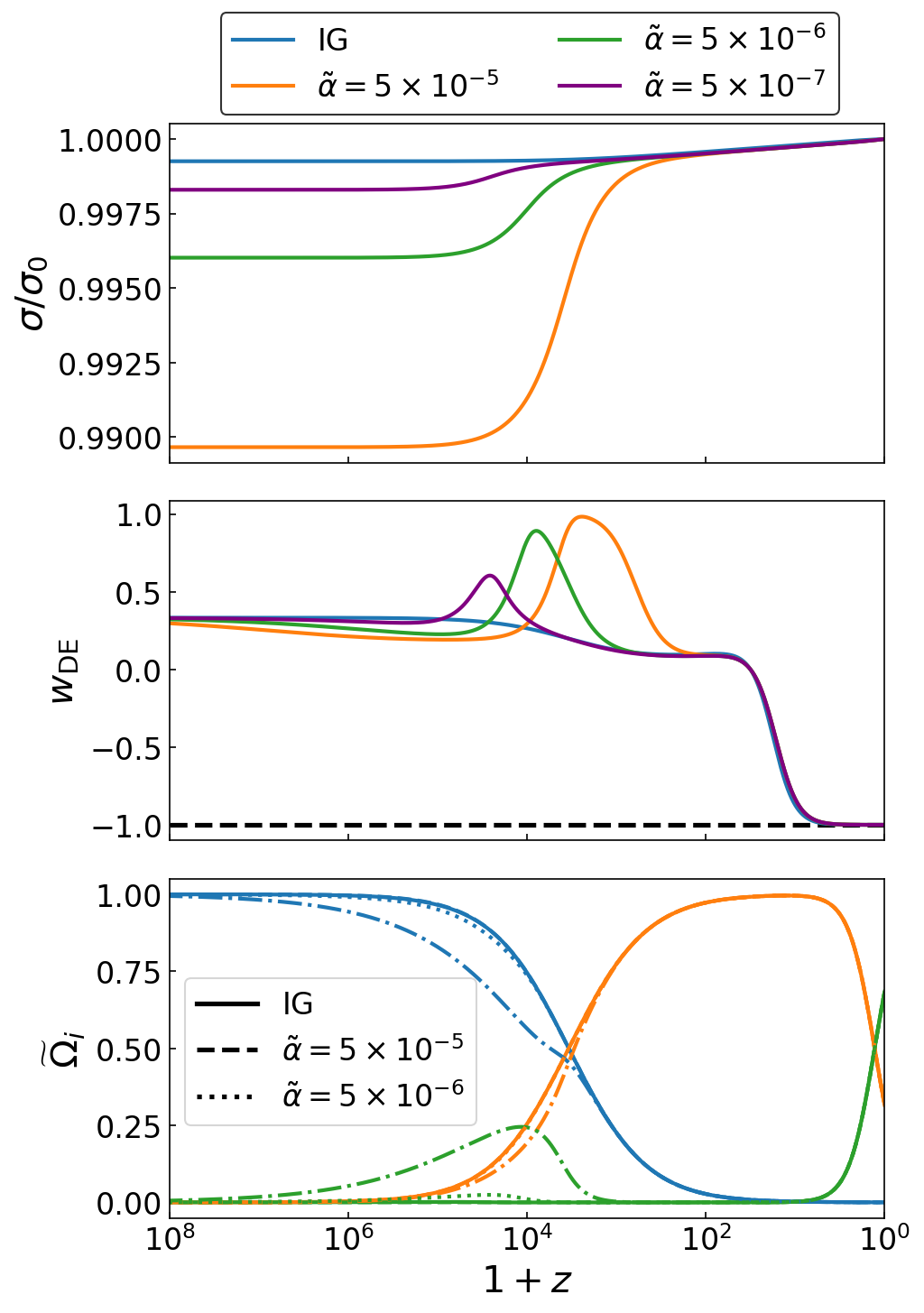}
\caption{Time evolution of the scalar field (top panel), the equation of state parameter for dark energy (middle panel) and the density parameters (bottom panel) in IGG standard ($Z=1$)  with $\gamma = 5\times 10^{-5}$, constant $V(\sigma) = \Lambda$ and $g(\sigma)=\alpha$, for different values of $\tilde\alpha$.}
\label{fig:iggV0_sigma_wde_omegas}
\end{figure}

The evolution of the scalar field is shown in the upper panel of \cref{fig:iggV0_sigma_wde_omegas}, in the model with a constant potential $V(\sigma)=\lambda_0\equiv \Lambda$ and $g(\sigma)=\alpha$, compared with the analogous IG model with $\gamma=5\times 10^{-5}$. We consider the rescaled, adimensional quantity $\tilde\alpha = \alpha \times {\rm (Mpc\, [GeV]^{-1} )^{-2} \times (M_{\rm Pl}\, [GeV] )^{1+m} }$ and it can be seen that the departure from IG is significant for larger $\alpha$, i.e. stronger gravity at early times.
Deep in the radiation era the field is nearly at rest but then it grows steeply (the larger $\alpha$ the steeper the growth) reaching the value expected in IG in the matter era and evolving until today in the same way as it does in IG.

The value of the field at $z=0$ is fixed by requiring that the effective gravitational constant \cite{Hohmann:2015kra,Quiros:2019gbt} 
\be\label{eq:GeffG3}
G_{\rm eff} = \frac{1}{16 \pi G_4} \Bigg[ \frac{ 4 G_{4,\sigma}^2 + G_4 ( G_{2,X} -2 G_{3,\sigma} ) }{ 3 G_{4,\sigma}^2 + G_4 ( G_{2,X} -2 G_{3,\sigma})}  \Bigg]
\ee
coincide at $z=0$ with the measured value of the Newton’s constant $G$.
For $m=0$, the $G_3$ does not contribute to \cref{eq:GeffG3}, as it enters only through the first derivative with respect to the field, while, for $m\neq0$ one can check that the contribution is negligible due to the redshift evolution of the scalar field in the matter era and we can therefore use the IG approximation \cite{Boisseau:2000pr}
\be\label{eq:Geff_ig}
8\pi G_{\rm eff}(z=0) = \frac{1}{\gamma\sigma_0^2}\frac{1+8\gamma}{1+6\gamma}
\ee
to fix the value of $\sigma (z=0) = \sigma_0$.

In the bottom plot of \cref{fig:iggV0_sigma_wde_omegas} we show the evolution of density parameters in IGG with a constant potential $V(\sigma)=\Lambda$ and $g(\sigma)=\alpha$.
The values of $\alpha$ are chosen large enough to show the effect of the $G_3$ term in this scenario, which causes a different evolution of $\widetilde\Omega_{\rm r}$ and $\widetilde\Omega_{\sigma}$ in the early Universe with respect to IG and $\Lambda \rm CDM$.

The middle panel of \cref{fig:iggV0_sigma_wde_omegas} presents the evolution of the parameter of state of dark energy $w_{\rm DE}$ defined above, it tracks the IG behavior at late times but departs from it at $z\geq10^3$, in correspondence to the analogous uptick in the evolution of $\widetilde\Omega_{\sigma}$.
In fact, $w_{\rm DE}$ follows the dominant component: deep in the radiation epoch it has a value close to $1/3$ , then in the matter era it decreases towards zero; finally, at present epoch, it becomes negative, $w_{\rm DE} \simeq -1$, mimicking a cosmological constant.
The bump which approximately occurs in the radiation era corresponds to the epoch in which the energy density of the field grows and becomes of the same order of that of radiation. The growth of $\Omega_\sigma$ so early in time is due to inefficient cosmological Vainshtein screening \cite{Chow:2009fm} in this model for the range of parameters considered.

\begin{figure}
\includegraphics[width=0.47\textwidth]{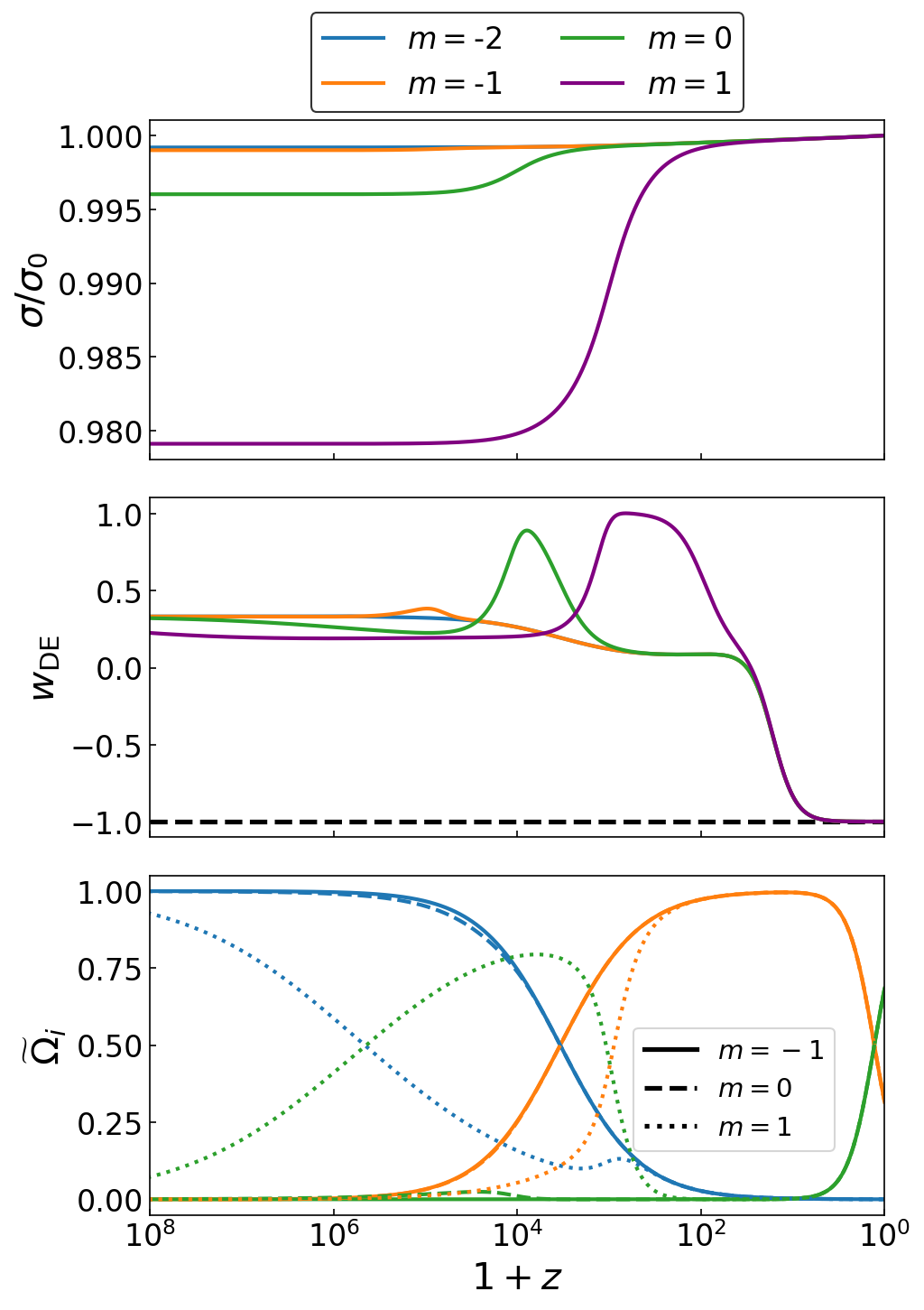}
\caption{Time evolution of the scalar field (top panel), the equation of state parameter for dark energy (middle panel) and the density parameters (bottom panel) in IGG standard ($Z=1$) with $\gamma = 5\times 10^{-5}$, constant $V(\sigma) = \Lambda$ and $g(\sigma)~=~\alpha\sigma^m$, for different values of the exponent $m$ and fixed $\tilde\alpha~=~5\times 10^{-6}$.}
\label{fig:iggV0_vary_m_sigma_wde_omegas}
\end{figure}
In \cref{fig:iggV0_vary_m_sigma_wde_omegas} we show the dependence of the background quantities ($\sigma,\, w_{\rm DE},\, \widetilde\Omega_i$) on the exponent $m$ in the function $g(\sigma) = \alpha \sigma^m$, while keeping $\gamma = 5\times 10^{-5} \text{ and } \tilde\alpha = 5\times 10^{-6}$ fixed, together with the exponent $n=0$ in $V(\sigma)\propto\sigma^n$.
We point out that changing $m$ while keeping $\tilde\alpha$ fixed modifies the actual value of $\alpha$ and its dimensions since $\tilde\alpha = \alpha \times {\rm (Mpc\, [GeV]^{-1} )^{-2} \times (M_{\rm Pl} \, [GeV] )^{1+m}}$. 
We can see from the figures that for small values of $m$ the time evolution of the scalar field is suppressed and with that the equation of state parameter and the density parameters resemble more those of $\Lambda\rm CDM$.
For larger values of the exponent $m$ the scalar field tends to dominate the dynamics both in the early Universe, reaching equality with radiation at $z=10^{6}$ for $m=1$, and in the late Universe where it mimics a cosmological constant.
This early epoch of scalar field domination also delays the matter radiation equality that happens at $z<10^4$, with $\widetilde\Omega_{\rm r}=\widetilde\Omega_{\rm m} \simeq 0.1$, and leads to two different epochs in which $\widetilde\Omega_{\rm m}$ is equal to the density parameter of the scalar field.
One of these epochs is the beginning of the matter era, around $z\simeq 10^{-3}$ and one is the onset of the late-time acceleration of the expansion of the Universe in which the field dominates the dynamics once again.

%
%

\subsubsection{BDG phantom}\label{sec:bdgphL}
In the phantom branch we study BDG with $g(\sigma)=\alpha\sigma^{-1}$ and $\zeta(\sigma) = \sigma^{-1} g(\sigma)$, because, for this choice, it was shown in \cite{Silva:2009km} that, for a null potential and in absence of matter and radiation, $\rho=p=0$, there is a self-accelerating solution with $\dot{H}=\dot{Q}=0$, with $Q=\dot\sigma/\sigma$.
This solution satisfies 
\be
y\equiv\frac{Q}{H}=\gamma \frac{-4\pm \sqrt{-32 - 6 Z/\gamma}}{Z+8\gamma};
\ee
and it is real for $Z=-1 \text{ and } \gamma<3/16$ (equivalent to $\omega_{\rm BD}<-4/3$), for which \cref{eq:Friedmann,eq:scalar_field_flrw} give
\be\label{eq:H2selfacc}
H^2=\frac{\gamma}{\alpha}\,\frac{3+6y+y^2/(2\gamma)}{2y^3(3+2y)}.
\ee
For a full study of the theory with $\Lambda = 0$ which exactly respects \cref{eq:H2selfacc} see \cref{sec:app_bdgph_noL}.

Inspired by these analytical solutions, we study an intermediate case always considering $g(\sigma)=\alpha\sigma^{-1}$ and reinserting the potential.
In doing so we obtain a theory that has a $\Lambda\rm CDM$ limit. In order to provide the late-time cosmic acceleration the parameters $\alpha$ and $\Lambda$ should be fine-tuned, the procedure is the following: we pick a value for $\alpha$ and use a shooting algorithm on the cosmological constant to obtain, in a flat Universe, $\Omega_{0, {\rm DE }}(\Lambda, \alpha)=1-\Omega_{0, \rm m}-\Omega_{0, \rm r}$.
In this way the burden to provide the late-time acceleration of the expansion of the Universe is shared between the potential and the Galileon term.
Today's value of the scalar field is such that the Planck mass is
\be
M_{\rm Pl}^2(z=0) \equiv \gamma\sigma_0^2 = 1.
\label{eq:field_today_when_screening}
\ee
 We do not set $\sigma_0$ following Eq. \eqref{eq:Geff_ig} or \eqref{eq:GeffG3} because, due to the Vainshtein screening mechanism, the theory reduces to GR at small scales, in the sense that the post-Newtonian parameters are those of GR and the gravitational constant on small scales is $G_{\rm N}=G_{\rm cosm}$ \cite{Kimura:2011dc}.
 In this way, thanks to the screening mechanism we recover GR with the correct value of the gravitational constant on small scales today (see \cref{sec:screening} for details).
In the following, instead of working directly with the parameter $\alpha$, we use $\tilde\alpha$ defined previously, rescaled by a factor $10^8$: $\tilde\alpha_{8} \equiv 10^{-8} \tilde\alpha$.
All the plots and the MCMC constraints will be expressed as a function of $1/\tilde\alpha_8$. The reason for using the inverse of $\tilde\alpha$ is that the $\Lambda\rm CDM$ limit is obtained when $\tilde\alpha \rightarrow \infty \, (1/\tilde\alpha \rightarrow 0)$. 

\begin{figure}
\includegraphics[width=0.48\textwidth]{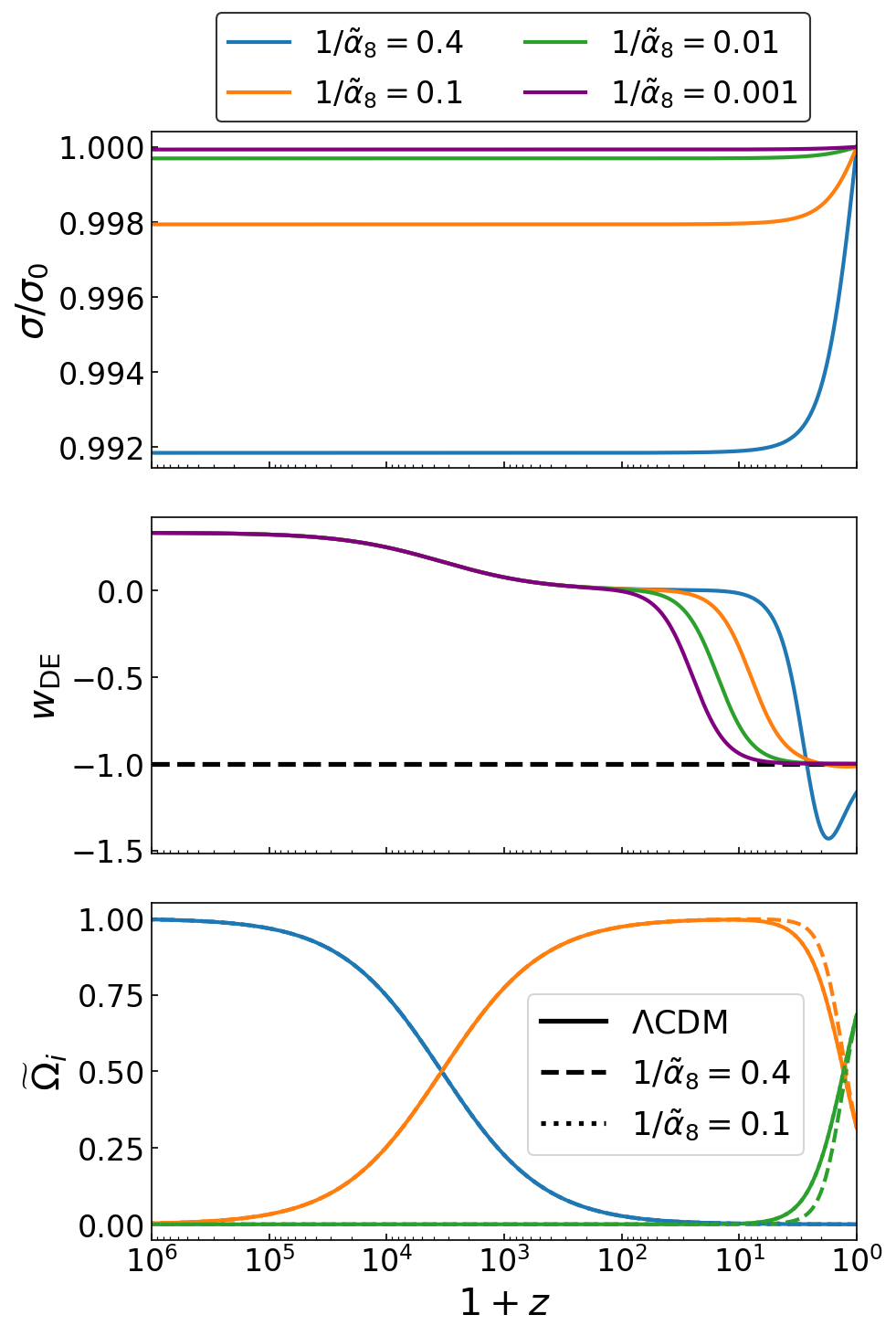}
\caption{Time evolution of the scalar field (top panel), the equation of state parameter for dark energy (middle panel) and the density parameters (bottom panel) in BDG phantom ($Z=-1$) with $\gamma = 5\times 10^{-5}$, constant $V(\sigma) = \Lambda$ and $g(\sigma)=\alpha\sigma^{-1}$, for different values of $1/\tilde\alpha_8$.}
\label{fig:bdgphL_field_wde_Omegas}
\end{figure}

In the top panel of \cref{fig:bdgphL_field_wde_Omegas} we show the evolution of the scalar field with redshift as a function of $1/\tilde\alpha_8$: for larger values of this parameter the field starts at a lower initial value and it grows more steeply in the late Universe to reach its value at $z=0$ fixed by the requirement $\gamma\sigma_0^2=1$.
Contrary to the standard branch where the field started to evolve deep in the radiation era, here the field is frozen approximately until the epoch in which the $\sigma$ energy density starts to grow. 
This is a manifestation of the cosmological Vainshtein screening mechanism \cite{Chow:2009fm}.
The fact that the field is frozen means that for a large portion of the history of the universe we have a gravitational constant which is effectively constant but different from the value measured here on Earth today.
In the bottom panel of \cref{fig:bdgphL_field_wde_Omegas} we show the evolution of the density parameters and it can be seen that already for $1/\tilde\alpha_8=0.1$ the background expansion in this model is indistinguishable from $\Lambda \rm CDM$.
Only for more extreme values we see departures from $\Lambda \rm CDM$ as the growth of the dark energy density is delayed and then steeper for $1/\tilde\alpha_8=0.4$.
This behavior is confirmed also by the time evolution of $w_{\rm DE}$ that approaches and reaches the value $-1$ earlier for smaller values of $1/\tilde\alpha_8$; while for the larger values we observe a phantom behavior with $w_{\rm DE} < -1$ and $w_{\rm DE} \neq -1$ today, reaching $-1$ eventually in the future. In this scenario, therefore the $\Lambda \rm CDM$ limit is approached when $1/\tilde\alpha_8 \rightarrow 0$.
This is the reason why we are considering the inverse of $\alpha$ as a parameter: in the Markov chain Monte-Carlo analysis we will sample on $1/\tilde\alpha_8$ in order to have the $\Lambda \rm CDM$ limit at zero and not at infinity, as we would if we had sampled on $\alpha$.

%
%

\subsection{Stability conditions in the phantom branch}\label{sec:stability}
\begin{figure}
\includegraphics[width=0.48\textwidth]{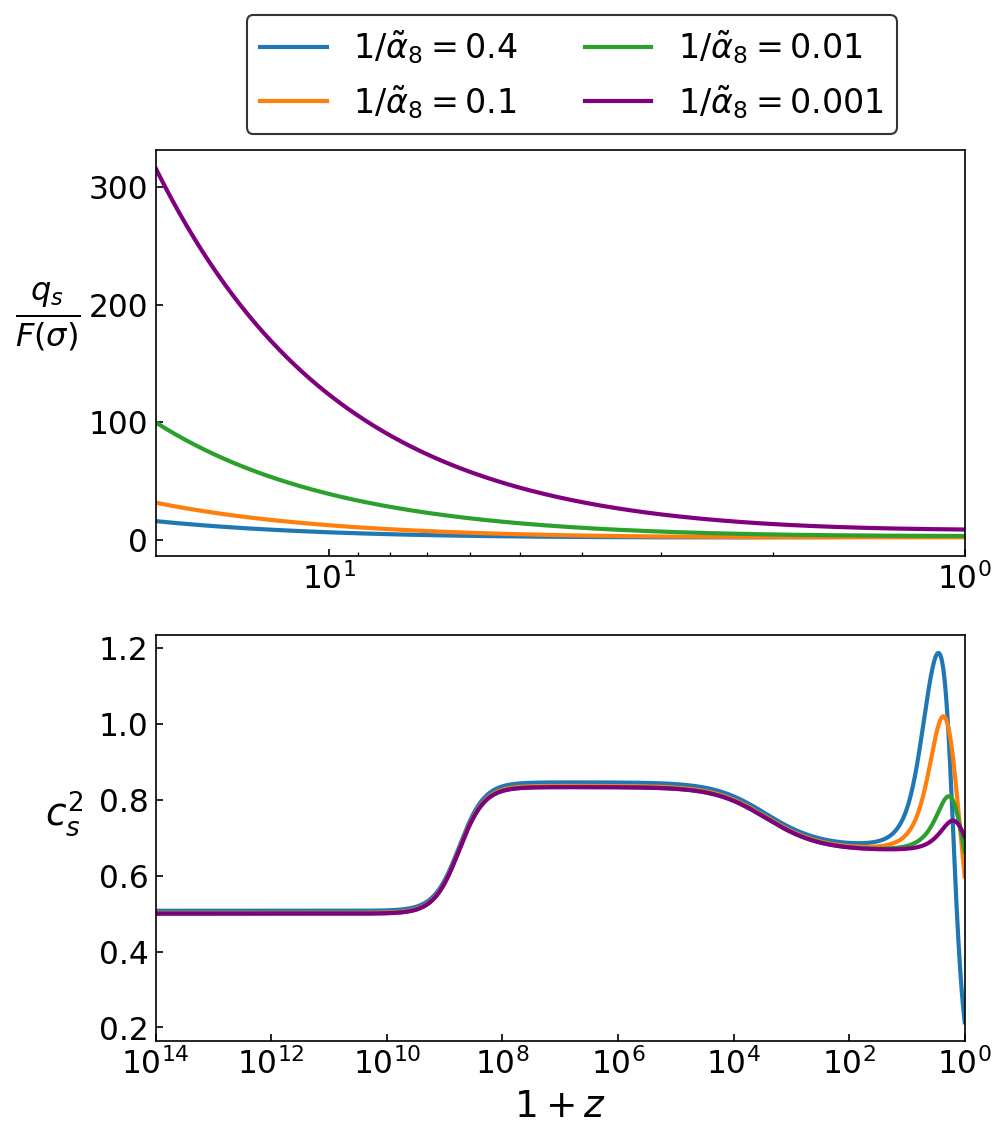}
\caption{Stability conditions: ghost instability (top panel), Laplacian instability (bottom panel) for BDG in the phantom branch ($Z=-1$) with $\gamma = 5\times 10^{-5}$, constant $V(\sigma) = \Lambda$ and $g(\sigma)=\alpha\sigma^{-1}$, for different values of $1/\tilde\alpha_8$.}
\label{fig:stability}
\end{figure}

When the sign of the kinetic term in the Lagrangian is nonstandard, i.e., $Z=-1$, and $G_3=0$, scalar-tensor theories can suffer from ghost and Laplacian instabilities. 
Ghost instabilities occur when the sign of the kinetic term is negative in the second-order action for perturbations of the Horndeski action, whereas Laplacian instabilities occur when the speed of sound squared becomes negative \cite{Bellini:2014fua}. In general, the theory in \cref{eq:action} is free from ghost and Laplacian instabilities if \cite{Bellini:2014fua,DeFelice:2011bh}
\begin{align}\label{eq:noghost}
\begin{split}
q_s\equiv \,&4 G_4 \big\{ G_{2X} + 2 G_{3\sigma} +\dot\sigma \big[(G2_{XX} + G3_{X\sigma})\dot\sigma \\
&-6 G_{3X} H \big] \big\} + 3 (2G_{4\sigma} + G_{3X}\dot\sigma^2)^2 > 0,
\end{split}
\end{align}
\begin{align}\label{eq:noLapl}
\begin{split}
c_s^2\equiv \,\big[&4 G_{2X} G_4 +8 G_{3\sigma} G_{4} \\ &+ (6 G_{4\sigma}^2 - G_{3X}\dot\sigma^2) (2 G_{4\sigma}^2 + G_{3X}\dot\sigma^2) \\
&-8 G_4 (G_{3X}\ddot\sigma + 2G_{3X}H\dot\sigma + G_{3X\sigma}\dot\sigma^2) \big]/q_s > 0,
\end{split}
\end{align}
where, $G_{i\,\sigma}, G_{i\, X}$ indicate the derivative of the $i$th $G$ function with respect to the scalar field and the kinetic term, respectively.
From \cref{eq:noghost,eq:noLapl}, it can be seen that even IG 
with $Z = -1$ and $0< \gamma < 1/6$ ($\omega_{\rm BD}<-3/2$) - a region of the parameter space which would contain otherwise a ghost - can be stable thanks to the addition of the $G_3$ term in the Lagrangian.
This is fully confirmed by \cref{fig:stability} where we show \cref{eq:noghost,eq:noLapl} as functions of redshift for the BDG in the phantom branch. Both conditions are satisfied at all times for the values of $1/\tilde\alpha_8$ we have considered.

Note, however, that in BDGph the speed of sound of scalar perturbations can become temporarily superluminal for large values of $1/\tilde\alpha_8$ at redshifts close to the transition to dark energy domination, as can be seen in the bottom panel of \cref{fig:stability}. 
This was first noticed in \cite{Silva:2009km} for the case $\Lambda=0$ and it could potentially put a constraint on the value of $1/\tilde\alpha_8$.
If a theory with superluminal propagation is viable or not is still a controversial issue: some authors claim it is not problematic \cite{Babichev:2007dw,Gorini:2007ta}, others argue the opposite \cite{Adams:2006sv,Bonvin:2006vc,Ellis:2007ic}.
We choose conservatively to not impose any theoretical prior $c_s^2 \le 1$ in our MCMC analysis.
Note that the stability conditions in the tensor sector are automatically satisfied by the choice of our Lagrangian and parameters: the regime $\gamma > 0$ we consider ensures that the cosmological Planck mass and the square of the tensor speed of sound are positive.

%
%

\subsection{Screening}\label{sec:screening}
The scalar degree of freedom in scalar-tensor theories of gravity can leave imprints also on scales which have detached from the cosmological expansion where deviations from GR are strongly constrained.
Thus, a theory of MG should have either small deviations from GR on Solar System scales or be equipped with a screening mechanism.
The Vainshtein screening is a mechanism that operates in theories with a self-interaction of the form $X\Box\sigma$, like the models considered in this paper: around local sources the nonlinear interactions lead to the decoupling of the scalar field from the remaining degrees of freedom within the so-called Vainshtein radius $r_V$.
For scales $r<r_V$ the theory of gravity reduces to GR: the post-Newtonian parameters are those of GR and the gravitational constant on small scales is $G_{\rm N}=G_{\rm cosm}$ \cite{Kimura:2011dc}.

 As it was demonstrated in \cite{Kimura:2011dc}, the Vainshtein mechanism cannot suppress the cosmological evolution of the scalar field: the mechanism is not able to determine the coupling constant of the theory (Newton gravitational constant or $\sigma_0$).
 The cosmological evolution of the scalar field sets the value of the coupling constant of GR today and on small scales. For this reason, for models with effective Vainshtein screening, we fix $\sigma_0$ according to \cref{eq:field_today_when_screening} in our background evolution. 
 
 The Vainshtein radius, which is the radius within which the theory reduces to GR, in a cosmological background, for a spherical astronomical object of mass $\delta M$, is given by \cite{Kimura:2011dc}
\be\label{eq:Vainshtein_radius}
r_V = \left(\mathcal{B C} \mu /H^2\right)^{1/3},
\ee
where $\mu = \delta M /(16\pi G_4)$, and $\mathcal{B} \text{ and } \mathcal{C}$ are functions of the Hordenski functions $G_i$s: 
\be
{\cal B}\equiv\frac{4\beta_0}{\alpha_0+2\alpha_1\alpha_2+\alpha_2^2},\quad
\ee
\be
{\cal C}\equiv\frac{\alpha_1+\alpha_2}{\alpha_0+2\alpha_1\alpha_2+\alpha_2^2},
\ee
where
\be
\alpha_i(t)\equiv\frac{A_i}{{\cal G}_T},
\quad
\beta_0(t)\equiv\frac{B_0}{{\cal G}_T},
\quad
\ee
with \\
\be
\beta_0\equiv\frac{\alpha_1}{2}+\alpha_2\;(\neq 0),\label{relation0}
\ee
and \\
\begin{eqnarray}
{\cal F}_T&\equiv&2G_4 \equiv {\cal G}_T
\\
{\cal E}&\equiv&2XG_{2X}-G_2-6X\dot\sigma HG_{3X}+2XG_{3\sigma}
\nonumber\\
&&-6H^2G_4-6H\dot\sigma G_{4\sigma }
\\
{\cal P}&\equiv&G_2+2X\left(G_{3\sigma}+\ddot\sigma G_{3X} \right)
+2\left(3H^2+2\dot H\right) G_4
\nonumber\\&&
+2\left(\ddot\sigma+2H\dot\sigma\right) G_{4\sigma}+4XG_{4\sigma\sigma}
\\
\Theta&\coloneqq& \dot\sigma XG_{3X}+
2HG_4 + \dot\sigma G_{4\sigma},
\\
A_0&\equiv&\frac{\dot\Theta}{H^2}+\frac{\Theta}{H}
-{\cal G}_T-2\frac{\dot{\cal G}_T}{H}-\frac{{\cal E}+{\cal P}}{2H^2},
\\
A_1&\equiv&\frac{1}{H}\dot{\cal G}_T + {\cal G}_T-{\cal F}_T,
\\
A_2&\equiv& {\cal G}_T-\frac{\Theta}{H},
\\
B_0&\equiv&-\frac{X}{H} \dot\sigma G_{3X}.
\end{eqnarray}
\begin{figure}
\includegraphics[width=0.48\textwidth]{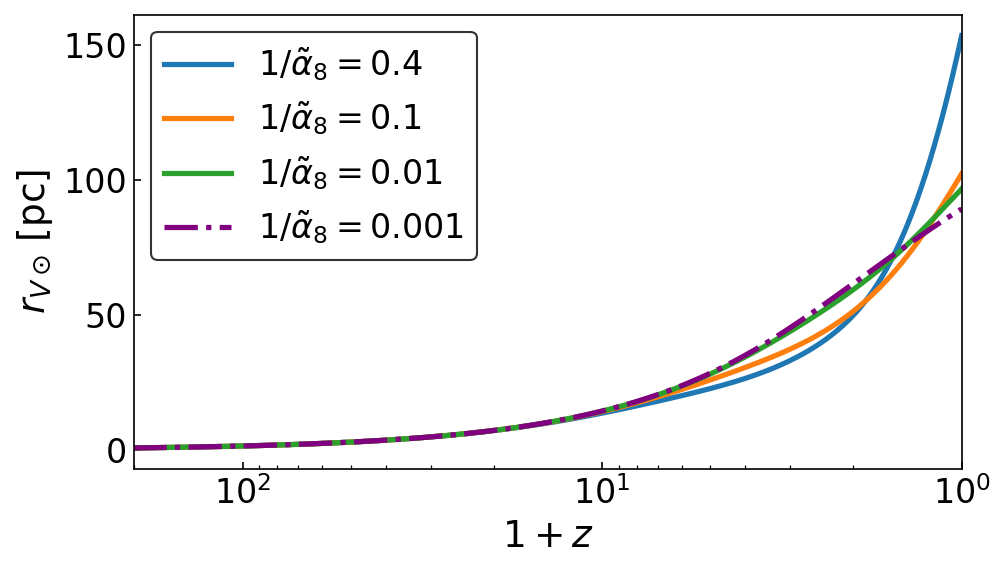}
\caption{Screening: Vainshtein radius for the sun in BDG phantom ($Z=-1$) with $\gamma = 5\times 10^{-5}$, constant $V(\sigma) = \Lambda$ and $g(\sigma)=\alpha\sigma^{-1}$, for different values of $1/\tilde\alpha_8$.}
\label{fig:Vainshtein_radius}
\end{figure}
In \cref{fig:Vainshtein_radius} we show the cosmological evolution of the Vainshtein radius for an object with $\delta M = M_\odot$, for BDG in the phantom branch.
The screening is effective and the Vainshtein radius starts growing for $z \lesssim 10$, reaching $r_{V\odot} \gtrsim 90\, {\rm pc}$ today, successfully recovering GR in the Solar System.
We do not present Vainshtein radius in the standard branch since the screening is not effective and the Vainshtein radius is very small.
This is the reason why $\sigma_0$ is fixed using \cref{eq:Geff_ig} in the background evolution in the standard branch.
\section{Linear perturbations and CMB anisotropies}\label{sec:lin_pert}
In the synchronous gauge the perturbed FLRW metric is to first order
\be \label{eqn:pert_metric}
	\dd s^2 = a^2(\tau) \left[-\dd^2 \tau + (\delta_{ij} + h_{ij}) \dd x^i \dd x^j \right]
\ee
where $\tau$ is the conformal time, ${\bf x}$ the spatial comoving coordinate, and $h_{ij}$ is the metric perturbation.

From this point on, we move in Fourier space for the calculation of the perturbed quantities.
The scalar mode of $h_{ij}$, following the conventions of Ref.~\cite{Ma_1995}, can be expressed as a Fourier integral:
\begin{align}
\begin{split}
	h_{ij}(\tau, {\bf x}) = \int \dd^3k e^{\imath {\bf k\cdot x}}
	\bigg[ &\hat{k}_i\hat{k}_j\, h(\tau, {\bf k}) 
	\\ &+ \left(\hat{k}_i\hat{k}_j-\frac{\delta_{ij}}{3} \right) 6\, \eta(\tau, {\bf k}) \bigg]
\end{split}
\end{align}
where $\hat{k}_i \equiv k_i/k$ with $k \equiv |{\bf k}|$ and $h\equiv \delta^{ij} h_{ij}$ is the Fourier transform of trace of $h_{ij}(\tau,\,{\bf x})$.
The scalar field perturbation $\delta\sigma$ is, in this notation,
\be 
\delta\sigma(\tau,\,{\bf x}) = \int \dd^3 k e^{\imath {\bf k\cdot x}}\,\delta\sigma(\tau,\,{\bf k}).
\ee \\

%
%

\subsubsection{The perturbed Einstein and scalar field equations}
Following \cite{Ma_1995} we use the definitions of the velocity potential $\theta$ and the anisotropic stress perturbation $\Theta$
\begin{align}
    &(\bar{\rho}+\bar{p})\theta \equiv \imath k^i \delta T^0_{\ i} \,,\\
    &(\bar{\rho}+\bar{p})\Theta \equiv -\left(\hat{k}_i\hat{k}_j-\frac{\delta_{ij}}{3}\right) \Sigma^i_{\ j} \,,\\
    &\Sigma^i_{\ j} \equiv T^i_{\ j} - \delta^i_j \frac{T^k_{\ k}}{3} \,,
\end{align}
to write down the perturbed Einstein equations in the following form:
\begin{align}
&k^2 \eta -\frac{1}{2} \Hrondo h' = -\frac{a^2}{2} \big(\delta \tilde{\rho} + \delta \tilde{\rho}^{(G)}\big) \label{eq:tt_pert_einstein_bdg}\, , \\[1ex]
&k^2 \eta' = \frac{a^2}{2} \Big[\big(\tilde\rho+\tilde P\big)\tilde\theta +\big(\tilde{\rho}^{(G)}+\tilde{P}^{(G)}\big) \tilde{\theta}^{(G)}	\Big] \label{eq:ts_pert_einstein_bdg}\, ,\\[1ex]
&h'' + 2\Hrondo h' -2k^2\eta = -3a^2 \big(\delta\tilde P + \delta\tilde{P}^{(G)}\big) \label{eq:trace_ss_pert_einstein_bdg}\, , \\[1ex]
&h''+6\eta'' +2\Hrondo(h'+6\eta')-2k^2\eta = -3a^2 \big(\tilde\rho +\tilde P\big)\tilde\Theta \label{eq:traceless_ss_pert_einstein_bdg} \,,
\end{align}
where a tilde denotes the effective perturbations defined in \cref{sec:app_pert}.
The equation for the evolution of the scalar field fluctuation in Fourier space is
\begin{align}
\begin{split}
\delta \sigma'' = \frac{\delta \tilde{\cal{G}}}{ Z + 6\gamma + \frac{2\sigma'}{a^2} \bigg[ 3\,g\,\Big(2\Hrondo -\frac{\sigma'}{\sigma}\Big) + \sigma' (6\zeta-2g_{,\sigma}) \bigg]}\, ,
\end{split}
\end{align}
where the explicit expression for $\delta\tilde{\cal G}$ can be found in \cref{sec:app_pert}.

In \cref{eq:tt_pert_einstein_bdg,eq:ts_pert_einstein_bdg,eq:trace_ss_pert_einstein_bdg,eq:traceless_ss_pert_einstein_bdg} we distinguished explicitly between the quantities coming from IG and the ones arising from the Galileon term to make as manifest as possible the reduction of IGG and BDG to the IG equations of \cite{Paoletti:2018xet,Ballardini:2023mzm} when $g(\sigma)=\zeta(\sigma)=0$. 

%
%

\subsubsection{Imprints on CMB anisotropies and large scale structure}
In order to obtain the predictions for the CMB anisotropy
angular power spectra and the matter power spectrum we have used an extension of the code {\tt CLASSig} \cite{Umilta:2015cta} which includes the models studied in our paper. The basis for this code is the publicly available Einstein-Boltzmann solver {\tt CLASS}\footnote{\url{https://lesgourg.github.io/class_public/class.html}} \cite{class1}.

\begin{figure*}
\includegraphics[width=0.95\textwidth]{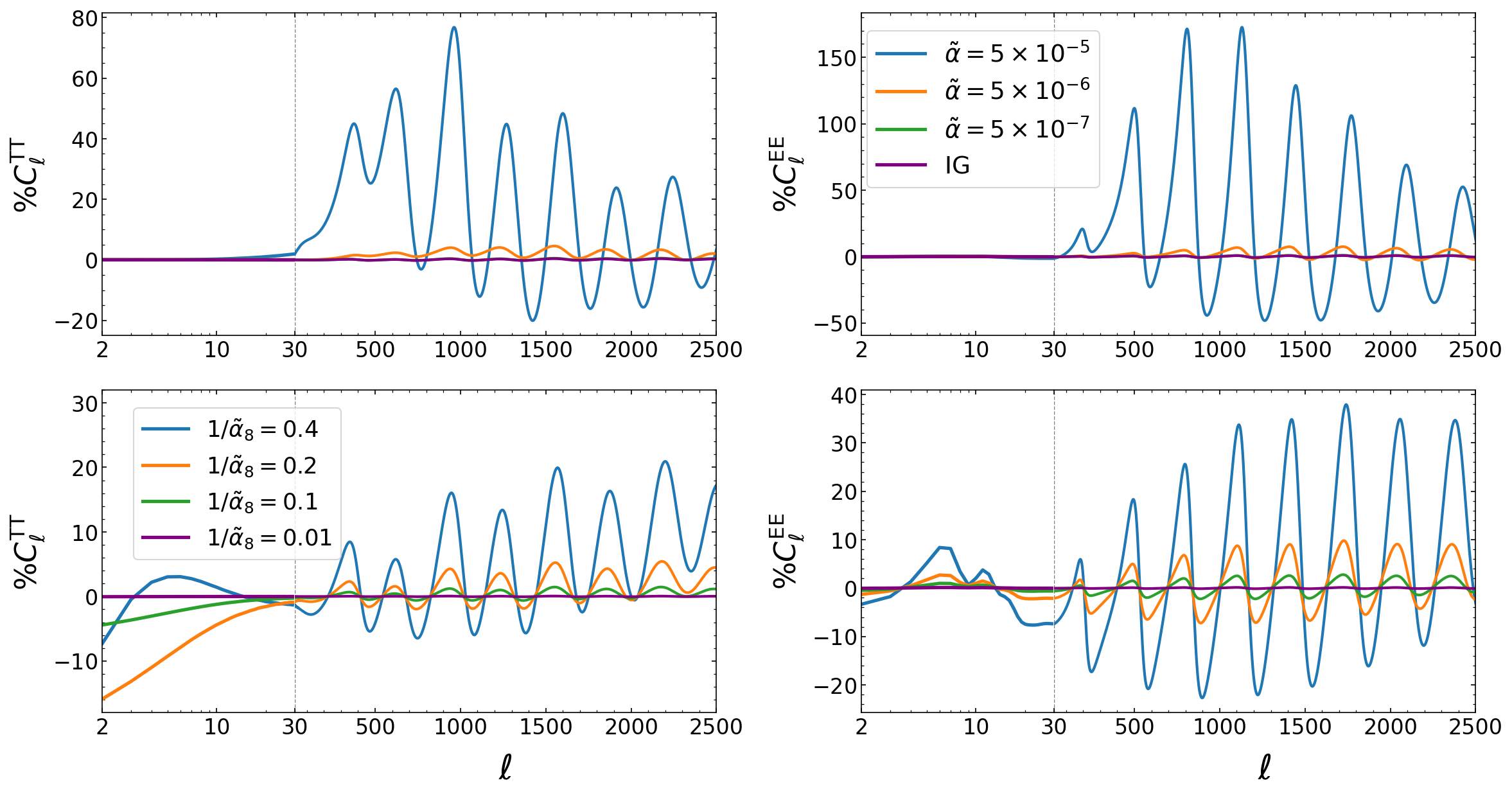}
\caption{Relative differences with respect to $\Lambda \rm CDM$ of CMB TT (left panels) and EE (right panels) power spectrum. Top panel: IGG standard with $\gamma = 5\times 10^{-5},\; V(\sigma) = \Lambda,\; g(\sigma) = \alpha$; bottom panel: BDG phantom with $\gamma = 5\times 10^{-5},\; V(\sigma) = \Lambda \text{ and } g(\sigma)=\alpha\sigma^{-1}$.} 
\label{fig:Delta_CMB_spectra}
\end{figure*}

\begin{figure*}
\includegraphics[width=0.95\textwidth]{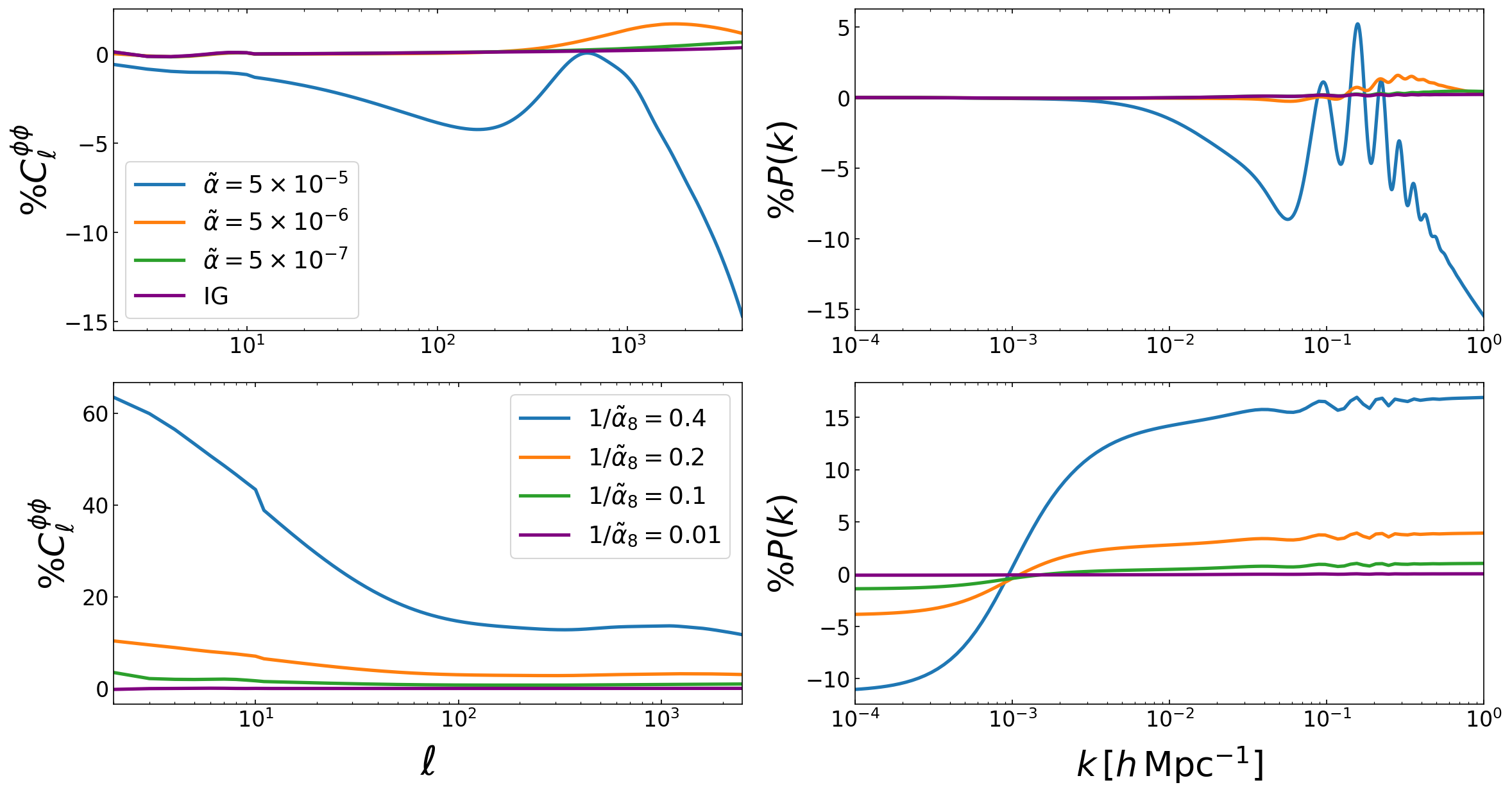}
\caption{Relative differences with respect to $\Lambda \rm CDM$ in CMB lensing potential power spectrum (left panels) and matter power spectrum (right panels). Top panel: IGG standard with $\gamma = 5\times 10^{-5},\; V(\sigma) = \Lambda,\; g(\sigma) = \alpha$; bottom panel: BDG phantom with $\gamma = 5\times 10^{-5},\; V(\sigma) = \Lambda \text{ and } g(\sigma)=\alpha\sigma^{-1}$.}
\label{fig:Delta_LSS_spectra}
\end{figure*}

\begin{figure}
\includegraphics[width=0.48\textwidth]{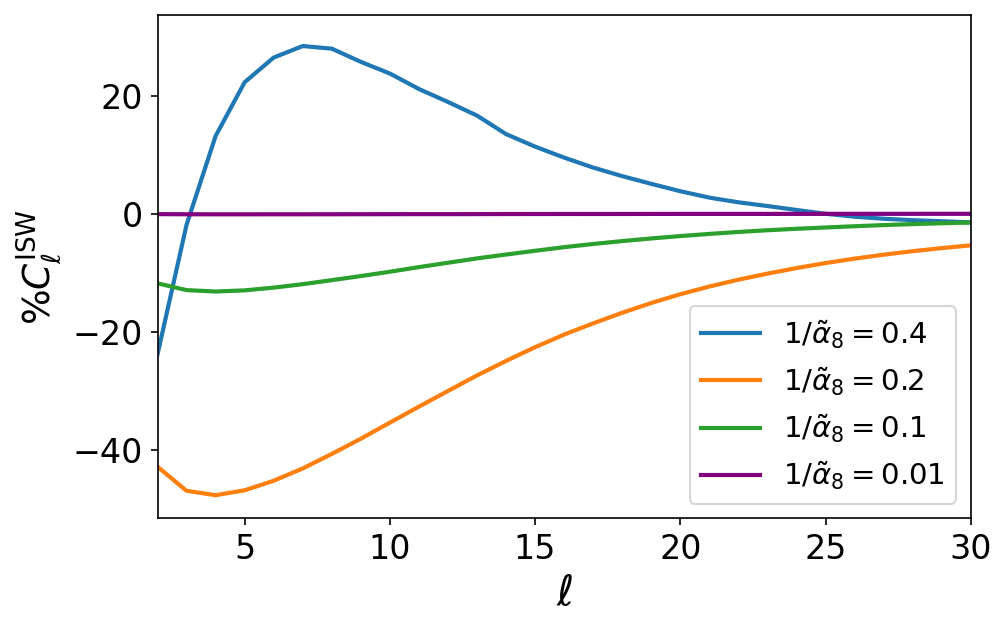}
\caption{Relative differences with respect to $\Lambda \rm CDM$ of the ISW contribution to the CMB TT angular power spectrum for BDG phantom with $\gamma = 5\times 10^{-5},\; V(\sigma) = \Lambda \text{ and } g(\sigma)=\alpha\sigma^{-1}$.} 
\label{fig:Delta_ISW_bdgphL}
\end{figure}

In \cref{fig:Delta_CMB_spectra} we show the deviations in the temperature (TT) and and E-mode polarization (EE) CMB angular power spectra with respect to $\Lambda\rm CDM$, whereas in \cref{fig:Delta_LSS_spectra} we show how the CMB lensing potential angular power spectrum and the total linear matter power spectrum at redshift $z=0$ differ in these theories with respect to the standard model.
In the top panels of \cref{fig:Delta_CMB_spectra,fig:Delta_LSS_spectra} we have IGG in the standard branch, while in the bottom panels we show BDG phantom with potential $V(\sigma)=\Lambda$. 
 
We note how in the standard branch there are small effects in the TT power spectrum at at low multipoles, resulting in an ISW effect extremely similar to the one observed in $\Lambda \rm CDM$.
This behavior is common to all the spectra in IGG at low multipoles, for which the differences with respect to GR are small.
The departure from $\Lambda \rm CDM$ is more evident on the scales of the acoustic peaks in TT and on smaller scales, where, due to the modification of gravity around the time of recombination, the contribution of the scalar field shifts and enhances the peaks of the power spectrum. 

On the contrary, in the model with $Z=-1$ we can observe large departures from $\Lambda \rm CDM$ at low multipoles in TT, due to the interplay of $G_3$ and the ISW effect \cite{Kobayashi:2009wr,Renk:2016olm}. 
We show the relative difference in the ISW contribution to the TT angular power spectrum in \cref{fig:Delta_ISW_bdgphL}. 

For the BDG in the phantom branch, while we see differences on all scales in all power spectra of \cref{fig:Delta_CMB_spectra,fig:Delta_LSS_spectra}, due to the presence of the cosmological constant, it is clear that as $1/\tilde\alpha_8$ gets smaller we recover the predictions of $\Lambda \rm CDM$ and there is therefore room to provide a good fit to the data even with $1/\tilde\alpha_8 \neq 0$ but small.

The differences in the matter power spectrum with respect to the standard model, presented in the right panels of \cref{fig:Delta_LSS_spectra}, are interesting targets for upcoming large scale structure data such as DESI\footnote{\url{https://www.desi.lbl.gov}}, Euclid\footnote{\url{https://www.esa.int/Science_Exploration/Space_Science/Euclid}}, or Vera Rubin observatory\footnote{\url{https://www.vro.org/},\, \url{https://www.lsst.org/}}, which will significantly tighten constraints on these models.
\section{Constraints from cosmological observations}\label{sec:results}
In this section, we present the constraints on the cosmological parameters of the models studied: IGG in the standard branch (IGGst) and BDG in the phantom branch with a potential (BDGph). In the following results the potential is always given by a $V(\sigma)=\Lambda$, while the function $g(\sigma)$ is a constant in the standard branch and $g(\sigma) = \alpha\sigma^{-1}$ for $Z=-1$. 
Note that in the following we consider only IGG in the standard branch since, for the same range of parameters, BDG gives exactly the same results. 

%
%

\subsection{Methodology and datasets}
We study these models first with CMB data only and we extend the analysis to other datasets when necessary in order to constrain a parameter within our prior or when the model is not ruled out by CMB alone and the results are interesting enough justify the study with additional likelihoods.
We perform a Markov chain Monte Carlo analysis using the publicly available sampling code {\tt MontePython-v3}\footnote{\url{https://github.com/brinckmann/montepython_public}} \cite{Audren_2013,Brinckmann:2018cvx} connected to the aforementioned extension of {\tt CLASSig} \cite{Umilta:2015cta}.
For the sampling, we use the Metropolis-Hastings algorithm with a Gelman-Rubin \cite{Gelman:1992zz} convergence criterion $R-1 < 0.01$.
The reported mean values and uncertainties on the parameters, together with the contour plots have been obtained using \texttt{GetDist}\footnote{\url{https://getdist.readthedocs.io/en/latest/}} \cite{Lewis:2019xzd}.
We make use of CMB data in temperature, polarization, and lensing from {\em Planck} Data Release $3$ \cite{Planck:2019nip,Planck:2018lbu}.
We consider the {\em Planck} baseline likelihood (hereafter P18) which is composed by the {\tt Plik} likelihood on high multipoles, $\ell>30$, the {\tt commander} likelihood on the lower multipoles for temperature and {\tt SimAll} for the E-mode polarization \cite{Planck:2019nip}, and the conservative multiple range, $8< \ell <400$, for the CMB lensing. 
As complementary data to {\em Planck}, we use BAO data from to the post-reconstruction measurements from BOSS DR12 \cite{BOSS:2016wmc}; low-$z$ BAO measurements from SDSS DR7, 6dF, and MGS \cite{Beutler_2011,Ross:2014qpa}; Ly$\alpha$ BAO measurements from eBOSS DR14, and a combination of those \cite{deSainteAgathe:2019voe,Blomqvist:2019rah,Cuceu:2019for}.
Hereafter we call this combination of datasets simply BAO. 
For the BDGph model, on top of BAO, we will also consider the combination of P18 with the Pantheon catalog of SN Ia in the redshift range $0.01 < z < 2.3$ \cite{Pan-STARRS1:2017jku}\footnote{\url{https://github.com/dscolnic/Pantheon}} (hereafter SN) using a prior on the supernovae peak absolute magnitude $M_{\rm B}$ \cite{Camarena:2021jlr} [hereafter p(M)] of
\be
M_{\rm B} = -19.2435 \pm 0.0373\; \rm mag.
\ee

We sample on six standard parameters: $\omega_{\rm b}$, $\omega_{\rm c}$, $H_0$, $\tau_{\rm reio}$, $\ln \left(10^{10} A_s\right)$, $n_s$, and the modified gravity parameters, in particular
\begin{itemize}
    \item For IGGst we sample on $\tilde\alpha$ at fixed $\gamma=5\times10^{-5}$.
    \item For BDGph we sample on $1/\tilde\alpha_8$ with fixed $\gamma~=~5~\times~10^{-5}$.
\end{itemize}
In addition to the cosmological parameters we sample on the nuisance and foregrounds parameters of the P18 likelihoods and, when considering the Phanteon dataset, on $M_{\rm B}$.
In our analysis we assume two massless neutrino with $N_{\rm eff} = 2.0328$ and a massive one with fixed minimum mass $m_\nu = 0.06 \rm eV$.
As in \cite{Ballardini:2016cvy}, we fix the primordial \ce{^{4}He} mass fraction $Y_{\rm p}$ taking into account the different value of the effective gravitational constant during big bang nucleosynthesis, and the baryon fraction, $\omega_{\rm b}$, tabulated in the public code
{\tt PArthENoPE} \cite{Pisanti:2007hk,Consiglio:2017pot}. 

Moreover, for each run we compute the best-fit values, obtained minimizing the $\chi^2$ following the methods of \cite{Schoneberg:2021qvd}, and quote the difference in the model $\chi^2$ with respect to the $\Lambda \rm CDM$ one, i.e. $\Delta\chi^2 \equiv \chi^2-\chi^2_{\Lambda\rm CDM}$. Thus, negative values of $\Delta\chi^2$ indicate an improvement in the fit with respect to $\Lambda\rm CDM$. We also compute the Akaike information criterion (AIC) \cite{1100705} of the extended model $\mathcal M$ relative to that of $\Lambda$CDM: 
$\Delta {\rm AIC} = \Delta \chi^2 + 2 (N_{\mathcal M} - N_{\Lambda\rm CDM})$, where $N_{\mathcal M}$ is the number of free parameters of the model.

%
%

\subsection{Results}

\subsubsection{Induced gravity Galileon standard}
\begin{figure*}
\includegraphics[width=\textwidth]{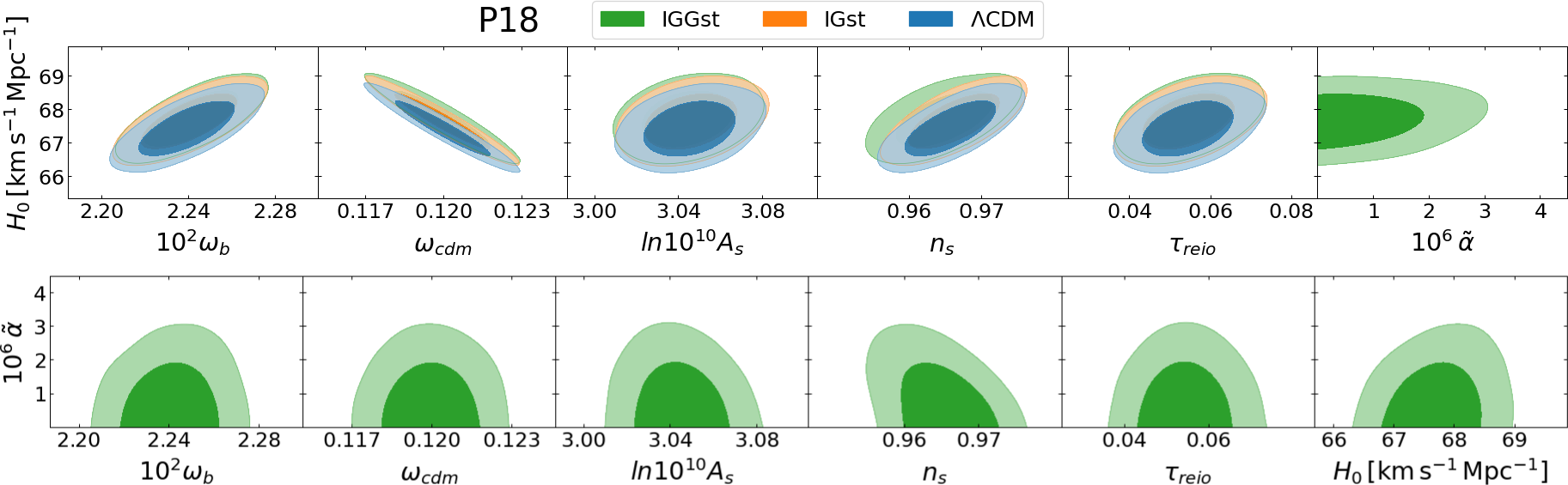}
\caption{Marginalized $68\%$ and $95\%$ CL 2D regions using CMB data alone (P18). IGG standard ($Z = 1$) with $\gamma = 5\times 10^{-5},\; V(\sigma) = \Lambda,\; g(\sigma)=\alpha$ in green; IG standard with $\gamma = 5\times 10^{-5},\; V(\sigma) = \Lambda$, in orange and $\Lambda \rm CDM$ in blue.}
\label{fig:rect_H0_alpha_IGGst_P18}
\end{figure*}

For IGGst we present the results obtained by considering $n=m=0 \,, \gamma=5\times 10^{-5}$, with $\alpha$ allowed to vary.
We show the P18 results for this model in \cref{fig:rect_H0_alpha_IGGst_P18,tab:P18_IGGst_IGst_LCDM}, with a comparison to $\Lambda$CDM and IG with $\gamma = 5\times 10^{-5}$.
We obtain a tight constrain on the Galileon term, i.e. $\tilde\alpha < 2.5 \times 10^{-6}$ at 95 \% CL with P18 data. This result shows how CMB data are sufficient to strongly constrain IGG around IG.
All the contours for the other parameters overlap with the IG ones. 

The marginalized mean and uncertainty for the Hubble constant $H_0\, [{\rm km\, s^{-1} Mpc^{-1}}]$ at $68\%$ CL is $67.72 \pm 0.54$ without any hint at a possible reduction of the Hubble tension, as there is no significant increase either in the mean or in the uncertainty with respect to the $\Lambda \rm CDM$ value of $67.36 \pm 0.54$ \cite{Planck:2018vyg}, or the IGst value $67.64 \pm 0.54$ (when $\gamma=5\times 10^{-5}$ is fixed in the MCMC).
When combining BAO with P18 we obtain tighter constraints but the same qualitative behavior (\cref{tab:P18_IGGst_IGst_LCDM,tab:P18_BAO}).

The cosmological constraints from P18 and P18+BAO on $\tilde\alpha$ allow a Vainshtein mechanism to occur only at subparsec scale for an object of a solar mass.
\begin{table*}[htb]
\setlength{\tabcolsep}{9pt}
\begin{tabular} { l  c c c}
\toprule
& IGGst & IGst & $\Lambda$CDM\\
\midrule
{$10^{2}\omega{}_{\rm b }$}    & $2.242\pm 0.014$   & $2.240\pm 0.014$   & $2.237\pm 0.015$   \\
{$\omega{}_{\rm cdm }$}        & $0.1200\pm 0.0012$ & $0.1200\pm 0.0012$ & $0.1200\pm 0.0012$ \\
{$H_0$} 
$[{\rm km\, s^{-1} Mpc^{-1}}]$ & $67.72\pm 0.54 $   & $67.64\pm 0.54$    & $67.36\pm 0.54$    \\
{$\ln (10^{10}A_{\rm s})$}     & $3.044\pm 0.014$   & $3.046\pm 0.014$   & $3.044\pm 0.014$   \\
{$n_{\rm s }$}                 & $0.9647\pm 0.0044$ & $0.9662\pm 0.0041$ & $0.9649\pm 0.0042$ \\
{$\tau{}_{\rm reio }$}         & $0.0545\pm 0.0073$ & $0.0546\pm 0.0075$ & $0.0544\pm 0.0073$ \\
{$10^6\tilde\alpha{}$}         & $< 2.5$ (95\% CL) & ---                & --- \\
{$\gamma{}$}                   & $5\times 10^{-5}$  & $5\times 10^{-5}$  & --- \\
\midrule
$\Delta\chi^2$ & 0.2 & 0.2 & --- \\
$\Delta\rm AIC$ & 2.2 & 0.2 & --- \\
\bottomrule
\end{tabular}
\caption{Constraints on the main parameters (at $68\%$ CL unless otherwise
stated) considering \emph {Planck} 2018 data alone for {\bf IGG standard} ($Z=1$) with $\gamma = 5\times 10^{-5},\; V(\sigma) = \Lambda,\; g(\sigma) = \alpha$, {\bf IGG standard} with $\gamma = 5\times 10^{-5},\; V(\sigma) = \Lambda$ and $\Lambda$CDM.}
\label{tab:P18_IGGst_IGst_LCDM}
\end{table*}

%
%

\subsubsection{Brans-Dicke Galileon phantom}
\begin{figure*}
\includegraphics[width=\textwidth]{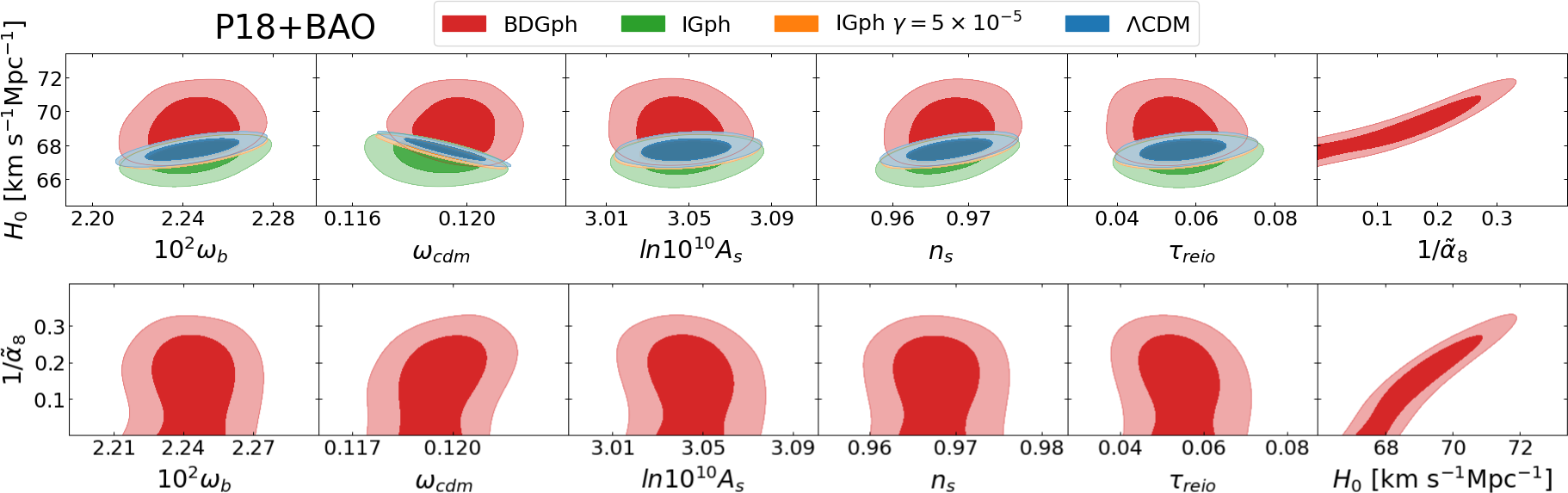}
\caption{Marginalized $68\%$ and $95\%$ CL 2D regions, using the combination P18 + BAO. BDG phantom ($Z = -1$) with $\gamma=5\times10^{-5},\; \; g(\sigma)=\alpha\sigma^{-1}$ in red, IGph and IGph with fixed $\gamma=5\times 10^{-5}$ respectively in green and orange; in all cases $V(\sigma) = \Lambda$. $\Lambda \rm CDM$ in blue.}
\label{fig:rec_H0_1oa_BDGphL_P18_BAO}
\end{figure*}

For this model in the phantom branch with a cosmological constant, we add BAO to P18 data in order to constrain the parameter $1/\tilde\alpha_8$ within our prior range, $1/\tilde\alpha_8 = 10^8 / \tilde\alpha \in [0, 0.4]$.
Current CMB data are indeed not enough to constrain this seven parameter model with $\gamma$ fixed.
The presence of the cosmological constant helps in providing posterior distributions closer to $\Lambda \rm CDM$ and similar values for most of the cosmological parameters with the exception of $H_0$: due to the shape of the posterior in the plane $1/\tilde\alpha_8$-$H_0$ (see \cref{fig:rec_H0_1oa_BDGphL_P18_BAO}), higher values of the Hubble constant compared to the $\Lambda \rm CDM$ ones are allowed (as it happens for nonminimally coupled scalar fields with standard kinetic terms due to the degeneracy $\gamma$-$H_0$).  
The marginalized mean and uncertainty for the Hubble constant $H_0\, [{\rm km\, s^{-1} Mpc^{-1}}]$ at $68\%$ CL using the combination P18+BAO (\cref{tab:P18_BAO}) is $69.1^{+0.9}_{-1.3}$. For the same dataset, IG phantom with $\gamma=5\times10^{-5}$ fixed at the same value considered here for BDGph, gives $67.62\pm 0.42$, while it gives $67.20^{+0.68}_{-0.55}$ when $\gamma$ is a free parameter. {\em Thus, BDGph can raise the value of the Hubble constant relatively to IGph and it reduces the Hubble tension to a significance of $2.5\sigma$.} 

\begin{figure*}
\includegraphics[width=\textwidth]{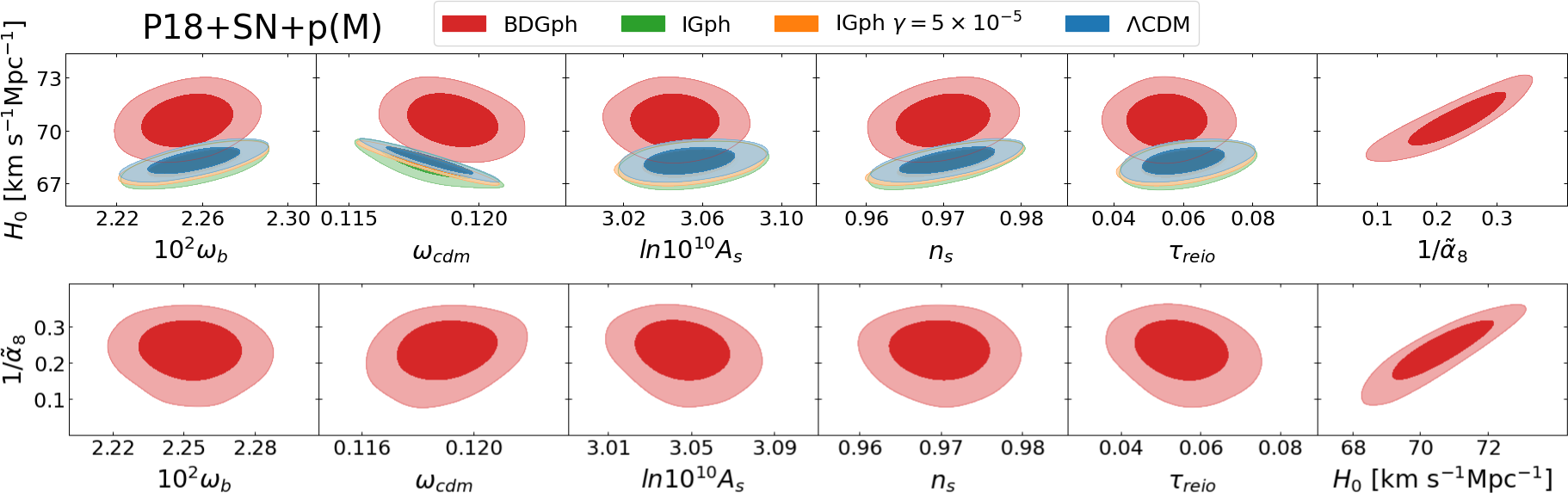}
\caption{Marginalized $68\%$ and $95\%$ CL 2D regions, using the combination P18 + SN + p(M). BDG phantom ($Z = -1$) with $\gamma=5\times10^{-5},\; \; g(\sigma)=\alpha\sigma^{-1}$ in red, IGph and IGph with fixed $\gamma=5\times 10^{-5}$ respectively in green and orange; in all cases $V(\sigma) = \Lambda$. $\Lambda \rm CDM$ in blue.}
\label{fig:rec_H0_1oa_BDGphL_P18_SN_pM}
\end{figure*}

This result motivates our further analysis using the Pantheon catalog with the addition of a prior on the peak absolute magnitude of SN Ia as in \cite{Ballardini:2021evv}. We present the outcomes of this analysis in \cref{tab:P18_SN_pM,fig:rec_H0_1oa_BDGphL_P18_SN_pM}. When considering this dataset the Hubble tension has a significance of only $1.7\sigma$ in BDGph, as the marginalized mean and uncertainty at $68\%$ CL is $H_0 = 70.58\pm 0.97\, {\rm km\, s^{-1}\, Mpc^{-1}}$. The corresponding values in IGph, with $\gamma=5\times 10^{-5}$ and $\gamma$ free to vary in the MCMC are, respectively, $68.15\pm 0.52$ and $68.07\pm 0.56$. {\em This means that the Galileon term is not only necessary to avoid instabilities present in IGph (see \cref{sec:stability}) but it also plays an important role in alleviating the Hubble tension.} Using this dataset we obtain $\Delta \chi^2 = -8.5$ and $\Delta\rm AIC = -6.5$, showing a weak preference for the model with respect to $\Lambda$CDM while in all other cases there is no significant improvement, if any, with respect to $\Lambda$CDM.

We would also like to emphasize another aspect of our findings in terms of the Vainshtein radius.
We obtain $1/\tilde\alpha_8=0.23^{+0.06}_{-0.05}$ at $68\%$ CL with P18 + SN + p(M); this means that we see high statistical significance for $1/\tilde\alpha_8\neq 0$, and consequently a Vainshtein radius of $\mathcal{O}(100)\, \rm pc$ for a solar mass. 
This is in contrast with the case P18+BAO where $1/\tilde\alpha_8$ is consistent with zero at $1\sigma$ and we only have an upper limit $1/\tilde\alpha_8 < 0.28$ at $95\%$ CL, perfectly consistent with the $\Lambda \rm CDM$ limit of the theory.
\begin{table*}[htb]
\setlength{\tabcolsep}{2pt}
\footnotesize
\begin{tabular} {| l |c|c|c | c|c|c|}
\cline{2-7}
\multicolumn{1}{c|}{} & IGGst & BDGph & IGst & IGph $\gamma = 5\times 10^{-5}$ & IGph & $\Lambda$CDM \\
\hline
{$10^{2}\omega{}_{\rm b }$} & $2.245\pm 0.014$ & $2.244\pm 0.014$ & $2.244\pm 0.013$ & $2.245\pm 0.013$ & $2.245\pm 0.014$ & $2.244\pm 0.013$\\

{$\omega{}_{\rm cdm }$} & $0.11940\pm 0.00096$ & $0.11966\pm 0.00097$ & $0.11937\pm 0.00093$ & $0.11910\pm 0.00093$ & $0.11884\pm 0.00099$ & $0.11925\pm 0.00094$\\

{$H_0$} 
$[{\rm km\, s^{-1} Mpc^{-1}}]$ 
& $67.97\pm 0.44$ & $69.1^{+0.9}_{-1.3}$ & $67.91\pm 0.42$ & $67.62\pm 0.42$ & $67.20^{+0.68}_{-0.55}$ & $67.75\pm 0.43$\\

{$\ln (10^{10}A_{\rm s})$} & $3.047\pm 0.014$ & $3.045\pm 0.014$ & $3.048\pm 0.014$ & $3.048^{+0.014}_{-0.015}$ & $3.048\pm 0.015$ &  $3.049\pm 0.014$\\

{$n_{\rm s }$} & $0.9662\pm  0.0041$ & $0.9678\pm 0.0038$ & $0.9676\pm 0.0036$ & $0.9675\pm 0.0037$ & $0.9671\pm 0.0038$ & $0.9675\pm 0.0037$\\

{$\tau{}_{\rm reio }$} & $0.0562\pm 0.0073$ & $0.0543\pm 0.0073$ & $0.0563\pm 0.0071$ & $0.0568^{+0.0068}_{-0.0077}$ & $0.0575\pm 0.0075$ & $0.0568\pm 0.0072$\\

{$10^6\tilde\alpha{}$} & $< 2.2$ (95\% CL) & --- & --- & --- & --- & --- \\

{$1/\tilde\alpha_8{}$} &  ---  &  $< 0.28$ (95\% CL) &  --- & --- &  --- & --- \\

{$\gamma{}$} & $5 \times 10^{-5}$ & $5 \times 10^{-5}$ & $5 \times 10^{-5}$ & $5 \times 10^{-5}$ & $< 0.00047$ (95\% CL) & --- \\
\hline

\multicolumn{1}{|l}{$\Delta\chi^2$} & \multicolumn{1}{c}{-0.1} & \multicolumn{1}{c}{-1.1} & \multicolumn{1}{c}{-0.1} & \multicolumn{1}{c}{0.2} & \multicolumn{1}{c}{0} & \multicolumn{1}{c|}{---}\\
\multicolumn{1}{|l}{$\Delta\rm AIC$} & \multicolumn{1}{c}{1.9} & \multicolumn{1}{c}{0.9} & \multicolumn{1}{c}{-0.1} & \multicolumn{1}{c}{0.2} & \multicolumn{1}{c}{2} & \multicolumn{1}{c|}{---}\\
\hline

\end{tabular}
\caption{Constraints on the main parameters (at $68\%$ CL unless otherwise stated) considering the combination {\bf P18+BAO} for {\bf IGG standard} ($Z=1$) with $g(\sigma) = \alpha$, {\bf IG standard} and {\bf phantom} ($Z=-1$) with $\gamma = 5\times 10^{-5}$ fixed, {\bf IG phantom} with $\gamma$ free to vary, {\bf BDG phantom} with $g(\sigma) = \alpha\sigma^{-1}$ and $\Lambda$CDM. In all cases unless otherwise specified $\gamma = 5 \times 10^{-5}$ and $V(\sigma)=\Lambda$.}
\label{tab:P18_BAO}
\end{table*}
\begin{table*}[htb]
\setlength{\tabcolsep}{9pt}
\begin{tabular} { l  c c c c}
\toprule
& BDGph & IGph $\gamma = 5\times 10^{-5}$ &  IGph & $\Lambda$CDM\\
\midrule
{$10^{2}\omega{}_{\rm b }$}    & $2.253\pm 0.014$   & $2.256\pm 0.014$   & $2.257\pm 0.014$ & $2.256\pm 0.014$   \\
{$\omega{}_{\rm cdm }$}        & $0.1190\pm 0.0012$ & $0.1180\pm 0.0011$ & $0.1180\pm 0.0011$ & $0.1181\pm 0.0011$ \\
{$H_0$}  
$[{\rm km\, s^{-1} Mpc^{-1}}]$ & $70.58\pm 0.97$    & $68.15\pm 0.52$    & $68.07\pm 0.56$ & $68.31\pm 0.50$    \\
{$\ln (10^{10}A_{\rm s})$}     & $3.046\pm 0.015$   & $3.053\pm 0.015$   & $3.054\pm 0.015$ & $3.054^{+0.014}_{-0.016}$   \\
{$n_{\rm s }$}                 & $0.9699\pm 0.0040$ & $0.9703\pm 0.0041$ & $0.9702\pm 0.0041$ & $0.9705\pm 0.0040$ \\
{$\tau{}_{\rm reio }$}         & $0.0555\pm 0.0077$ & $0.0602^{+0.0076}_{-0.0085}$ & $0.0608\pm 0.0079$ & $0.0602^{+0.0072}_{-0.0083}$ \\
{$1/\tilde\alpha_8{}$}         & $0.23^{+0.06}_{-0.05}$ & --- & --- & --- \\
{$\gamma{}$}         & $5 \times 10^{-5}$ & $5 \times 10^{-5}$ &  $< 0.00022$ (95\% CL) & --- \\
\midrule
$\Delta\chi^2$ & -8.5 & 1.1 & 0 & --- \\
$\Delta\rm AIC$ & -6.5 & 1.1 & 2 & --- \\
\bottomrule

\end{tabular}
\caption{Constraints on the main parameters (at $68\%$ CL unless otherwise stated) considering the combination {\bf P18+SN+p(M)} for {\bf IG phantom} ($Z=-1$) with $\gamma = 5\times 10^{-5}$ fixed and free to vary, {\bf BDG phantom} with $\gamma = 5\times 10^{-5},\; g(\sigma) = \alpha\sigma^{-1}$ and $\Lambda$CDM. In all cases $V(\sigma)=\Lambda$.}
\label{tab:P18_SN_pM}
\end{table*}

\section{Conclusions} \label{sec:conclusions}
We have studied the cosmological effects of a model where a scalar field, $\sigma$, with a coupling to the Ricci scalar of the type $F(\sigma) = \gamma\sigma^2$, has a cubic Galileon term of the form $\alpha \, \sigma^m (\partial \sigma)^2 \square \sigma$. 
Since the extensions of IG and BD with a Galileon term are not equivalent theories up to a field redefinition, differing by a term $4 \zeta(\sigma)X^2$ in the Lagrangian, we have studied the two models separately: IGG and BDG.

This Galileon term leads to the Vaishtein mechanism which can screen gravity, potentially reconciling the theory with GR inside the so-called Vainshtein radius also for values of the coupling to the Ricci scalar - or of the Brans-Dicke parameter $\omega_\mathrm{BD}$ - which evade the Solar System constraints.
Moreover, thanks to the presence of the Galileon term, the theory is free of ghost and Laplacian instabilities even for a noncanonical sign of the kinetic term in the Lagrangian.
We have therefore considered the theory \eqref{eq:BDG_Gis} in the two branches, called respectively, the standard ($Z=1$) and phantom ($Z=-1$) branches, where $Z$ enters the kinetic term in the Lagrangian as $Z X$, with $X = - \nabla_\mu\sigma \nabla^\mu \sigma / 2$. 

We have shown that the Galileon term leads to important cosmological effects in all the cases considered. In this paper we have computed the theoretical predictions for cosmological observables such as the CMB anisotropy and matter power spectra.
We have also compared these predictions with observations by restricting the analysis to fixed $\gamma$ and $m$, if not otherwise stated.

For a standard kinetic term, we find an instability in the radiation dominated era due to the presence of the Galileon term. This instability is dissipated in the subsequent matter dominated era, during which the Galileon term becomes subleading, and then eventually the scalar field energy density is no more negligible contributing to the acceleration of the Universe. 
We find that the CMB anisotropy pattern is sensitive to the dissipation of the instability in the matter era. This effect constrains the Galileon term to be small close to the CMB last scattering surface.

In the branch $Z=-1$ we have considered only BDG for the sake of simplicity.
The phenomenology of a kinetic term with a negative sign - corresponding to a BD theory in the parameter range $\omega_\mathrm{BD}< 0$ - is rather different.
First, the presence of a Galileon term leads to an healthy theory for all the values of $\gamma$, i.e. for any negative value of $\omega_\mathrm{BD}$, therefore rescuing the range $\omega_\mathrm{BD} < -3/2$ which would contain a ghost in the BD theory with no Galileon term.
The Galileon term leads to a dynamics very different from the corresponding branch of induced gravity, by freezing the scalar field for most of the matter dominated era and releasing it at lower redshift. 

In the standard branch, {\em Planck} 2018 and BAO data constrain the amplitude of the Galileon parameter as $\tilde\alpha < 2.2 \times 10^{-6}$ at 95 \% CL (\cref{tab:P18_BAO}) for IGG with $\gamma=5\times 10^{-5}$, $m=0$, and a constant term $\Lambda$ as a potential.
The constraints for BDG in the standard branch are identical to the ones of IGG, we have therefore reported them only once for IGG.
The cosmological constraint  on $\tilde\alpha$ allows a Vainshtein mechanism to occur only at subparsec scale for an object of a solar mass. Summarizing, for a standard kinetic term, we have therefore found tight constraints on the Galileon term and resulting posterior probabilities for cosmological parameters very similar to those of IG.    

For BDG with $Z=-1$ and a constant term $\Lambda$ as a potential, we have found that {\em Planck} 2018 and BAO data constrain $H_0 = 69.1^{+0.9}_{-1.3}\,{\rm km\, s^{-1}\, Mpc^{-1}}$ at 68\% CL and $\tilde\alpha^{-1} < 0.281\times 10^{-8}$ at 95\% CL, for $\gamma = 5 \times 10^{-5}$ and $m=-1$ (\cref{tab:P18_BAO,fig:rec_H0_1oa_BDGphL_P18_BAO}).
The addition of a Galileon term in the phantom branch is therefore crucial to obtain a value of $H_0$ larger than $\Lambda$CDM.
We have therefore a modified gravity model which leads to a value of $H_0$ larger than in $\Lambda$CDM with a screening of $\mathcal{O}(100)\, \rm pc$ for a solar mass, as desired.
By adding the {\em Pantheon} dataset with a prior on the supernovae peak absolute magnitude $M_{\rm B} = -19.2435 \pm 0.0373\; \rm mag$, we find $H_0 = 70.58\pm 0.97 \, {\rm km\, s^{-1}\, Mpc^{-1}}$ and $\tilde\alpha^{-1} = 0.23^{+0.06}_{-0.05} \times 10^{-8}$ at 68\% CL, always for $\gamma = 5\times 10 ^{-5}$ and $m=-1$ (\cref{tab:P18_SN_pM,fig:rec_H0_1oa_BDGphL_P18_SN_pM}).
As it happens for the degeneracy between $H_0$ and $\gamma$ in IG \cite{Ballardini:2021eox}, the SH0ES measurement pulls $H_0$ along the degeneracy with $\alpha^{-1}$.
By adding a Galileon term, the theory inherits also characteristics of a late model in increasing $H_0$.
In addition, the value and the posterior of $S_8$ are unchanged with respect to $\Lambda$CDM, not aggravating the so-called $\sigma_8$ tension \cite{DiValentino:2020vvd}. 

We have also analyzed the theoretical predictions of the model introduced by Silva and Koyama \cite{Silva:2009km} in which there is no potential and the late-time acceleration is driven exclusively by the Galileon term.
This particular model is physically viable and provides screening on Solar System scales. We have shown that it leads to CMB predictions which are at odds with the {\em Planck} data, with a $\Delta \chi^2 = 30.6 $ with respect to $\Lambda$CDM. Therefore, although theoretically interesting because the acceleration is not driven by an effective cosmological constant, the model is ruled out by observations.

In conclusion, we have shown how a Galileon term leads to rather nontrivial cosmological effects to the simplest scalar-tensor theories of gravity.
It can lead simultaneously to a value of $H_0$ larger than in $\Lambda$CDM and to effective screening for a large volume of parameter space.
It is therefore interesting to find that another term in the Horndeski Lagrangian can lead to a value larger then the $\Lambda$CDM one, as previously found for the basic evolving Newton's constant in \cite{Umilta:2015cta,Ballardini:2016cvy,Rossi:2019lgt,Paoletti:2018xet,Ballardini:2020iws,Braglia:2020auw,Braglia:2020iik, Ballardini:2021eox,Ballardini:2021evv,Ballardini:2023mzm}.
A more general exploration of the parameter space of this theory in which the speed of gravitational waves is $c$ without any fine tuning, by allowing to vary simultaneously the coupling to the Ricci scalar, $\gamma$, and the Galileon term is in progress. 
\section*{Acknowledgments}
M.B., F.F., D.P. acknowledge financial support from the ASI/INAF Contract for the Euclid mission n.2018-23-HH.0, from the INFN InDark initiative and from the COSMOS network (\href{www.cosmosnet.it}{www.cosmosnet.it}) through the ASI (Italian Space Agency) Grants No. 2016-24-H.0 and 2016-24-H.1-2018, as well as No. 2020-9-HH.0 (participation in LiteBIRD phase A). 

This work has made use of computational resources of CNAF HPC cluster in Bologna.
\clearpage
\appendix

\section{BDG in the phantom branch with $\Lambda=0$} \label{sec:app_bdgph_noL}
In this appendix we study the model, introduced in \cite{Silva:2009km}, that exactly respects the solution \eqref{eq:H2selfacc}, considering it as a limiting case of BDGph when $\Lambda = 0$.
We discuss the background evolution, including stability conditions and the Vainshtein screening of the theory, CMB anisotropies and matter power spectrum and the constraints on the model coming from CMB observations. 

%
%

\subsection{Background evolution, stability conditions and screening}
In the absence of the cosmological constant, in order to describe today's acceleration of the expansion of the Universe, the parameter $\alpha$ should be fine-tuned, this is done by solving \cref{eq:H2selfacc} for $\alpha$ and using it as an initial guess for a shooting algorithm that fixes the value of $\alpha$ to produce the desired $\Omega_{\rm DE}=1-\Omega_{\rm m}-\Omega_{\rm r}$, given the density parameters of matter and radiation as inputs.
With $\alpha$ fixed by the requirement of cosmic acceleration, the only free parameter of the theory is $\gamma$; resulting therefore in a theory that can provide cosmic acceleration without a cosmological constant, with as many free parameters as the BD model.

\begin{figure}
\includegraphics[width=0.48\textwidth]{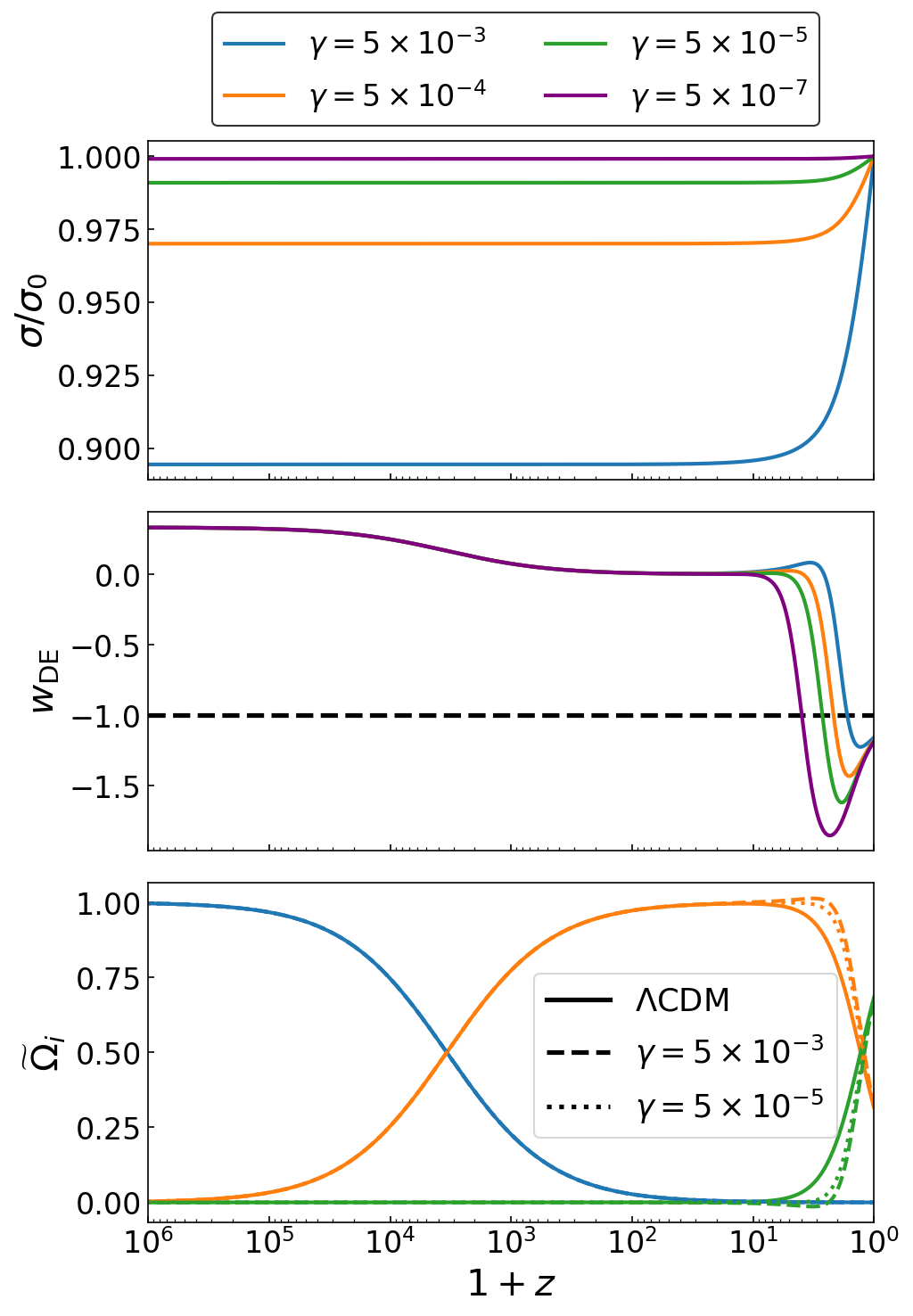}
\caption{Time evolution of the scalar field (top panel), the equation of state parameter for dark energy (middle panel) and the density parameters (bottom panel) in BDG phantom ($Z=-1$) with no potential and $g(\sigma)=\alpha\sigma^{-1}$, for different values of $\gamma$.}
\label{fig:sk_sigma_wde_Omegas_sk}
\end{figure}

Today's value of the scalar field is such that it satisfies \cref{eq:field_today_when_screening}. See \cref{sec:bdgphL,sec:screening} for details on why we do not need to set $\sigma_0$ following \cref{eq:Geff_ig} or \eqref{eq:GeffG3} thanks the Vainshtein screening mechanism, but we still need to respect the condition $M_{\rm Pl}=1$ today.
The evolution of the scalar field with redshift is plotted in the top panel of \cref{fig:sk_sigma_wde_Omegas_sk}.
The field is frozen deep in the radiation era and it starts to grow at very late times reaching the value required by consistency with GR on small scales today.
In the figure we normalized the field to its value at $z=0$ which is different depending on the chosen value of $\gamma$, since $\gamma\sigma_0^2=1$.

Departures from $\Lambda \rm CDM$ are evident already at the background level, especially in the matter era and in the field dominated era.
In fact, during the matter era, before the equality with DE, we observe, from the bottom panel of \cref{fig:sk_sigma_wde_Omegas_sk}, $\widetilde\Omega_\sigma$ becoming slightly negative.
This effect is more prominent for larger values of $\gamma$.
At the time when $\widetilde\Omega_\sigma$ is decreasing and becoming negative, the corresponding $\Omega_{\rm DE}$ in the $\Lambda \rm CDM$ model is already growing, therefore, a steeper growth of $\widetilde\Omega_\sigma$ is needed in order to reach the $\Lambda \rm CDM$ value today and for the matter-dark energy equality to happen more or less at the same redshift.
We emphasize that $\widetilde\Omega_\sigma<0$ is not a problem from the physical point of view: this parameter simply describes the contribution of the scalar field to the total expansion rate when the Friedmann equations are recast in a form that resembles Einstein gravity \cite{Torres:2002pe,Gannouji:2006jm}.

The evolution of the parameter of state of DE $w_{\rm DE}$ from the middle panel of \cref{fig:sk_sigma_wde_Omegas_sk} is also interesting, as it presents a phantom behavior $w_{\rm DE}<-1$ in correspondence to $\widetilde\Omega_\sigma$ becoming negative and growing back to dominate the energy content of the Universe.
This behavior is more prominent for a smaller value of $\gamma$ for which the dip tends to be deeper and $w_{\rm DE}$ reaches lower values.
The parameter $w_{\rm DE}$ today is slightly different from $-1$ and it eventually reaches $-1$ in the future, as shown in \cite{Silva:2009km}.

\begin{figure}
\includegraphics[width=0.48\textwidth]{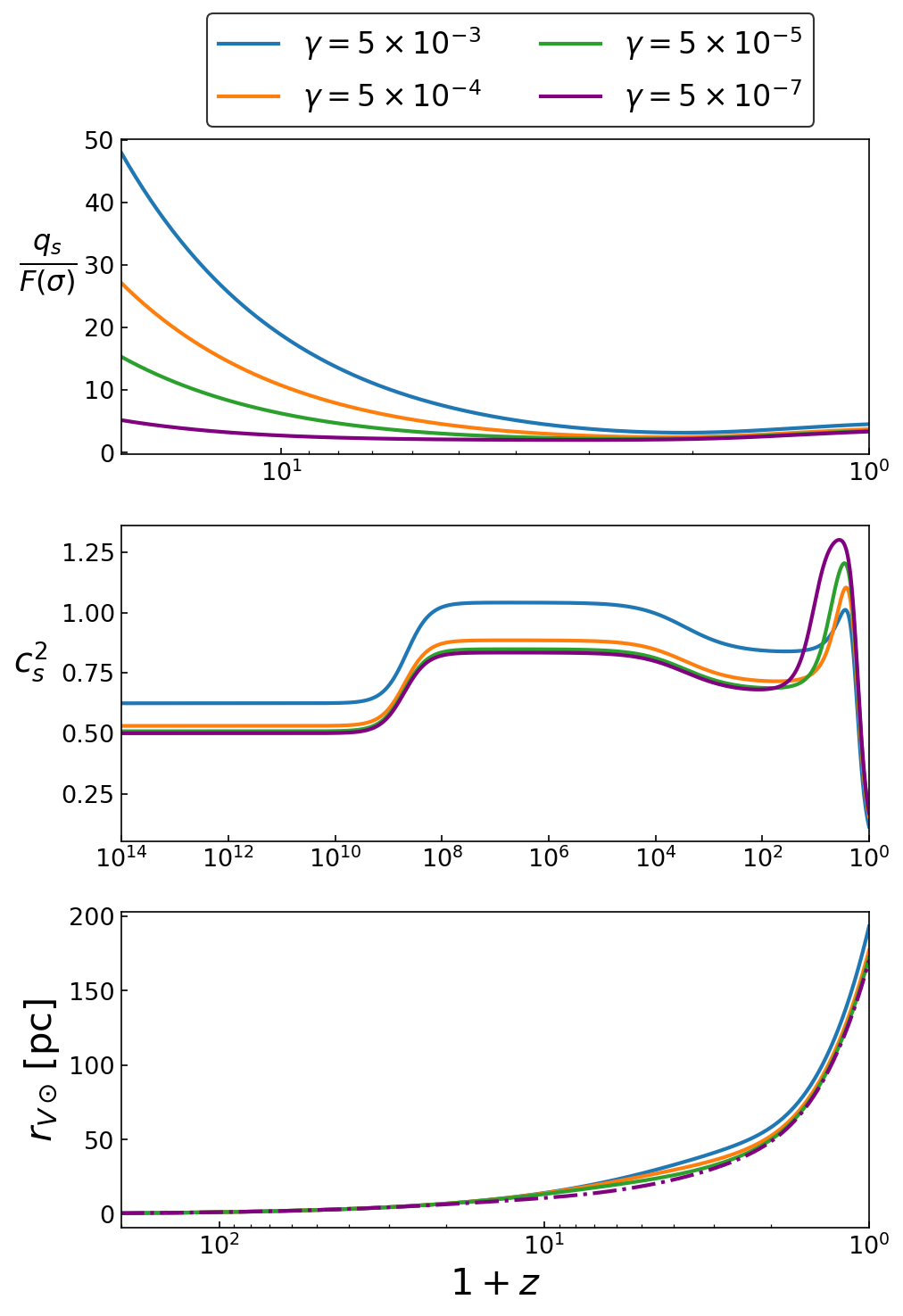}
\caption{Stability conditions: ghost instability (top panel), Laplacian instability (middle panel). Vainshtein radius for a solar mass spherical object (bottom panel) BDG in the phantom branch ($Z=-1$), without potential, $g(\sigma)=\alpha/\sigma$, for different values of $\gamma$.}
\label{fig:sk_stability_and_rV}
\end{figure}
The top two panels of \cref{fig:sk_stability_and_rV} show the evolution of the stability conditions \eqref{eq:noghost} and \eqref{eq:noLapl} with redshift. 
The conditions are satisfied for all the values of $\gamma$ considered thanks to the Galileon term.
This term is also responsible for effective screening on small scales through the Vainshtein screening mechanism that allows the theory to reduce to GR within the so-called Vainshtein radius.
Thus, within this radius the post-Newtonian parameters are those of GR. 
The time evolution of the Vainshtein radius \eqref{eq:Vainshtein_radius} for a celestial body of a solar mass is shown in the bottom panel of \cref{fig:sk_stability_and_rV}: the screening is effective and $r_{V\odot}\sim \mathcal{O}(100)\, \rm pc$ today for all the values of $\gamma$ studied.  

%
%

\subsection{CMB anisotropies and matter power spectrum}
\begin{figure*}
\includegraphics[width=0.95\textwidth]{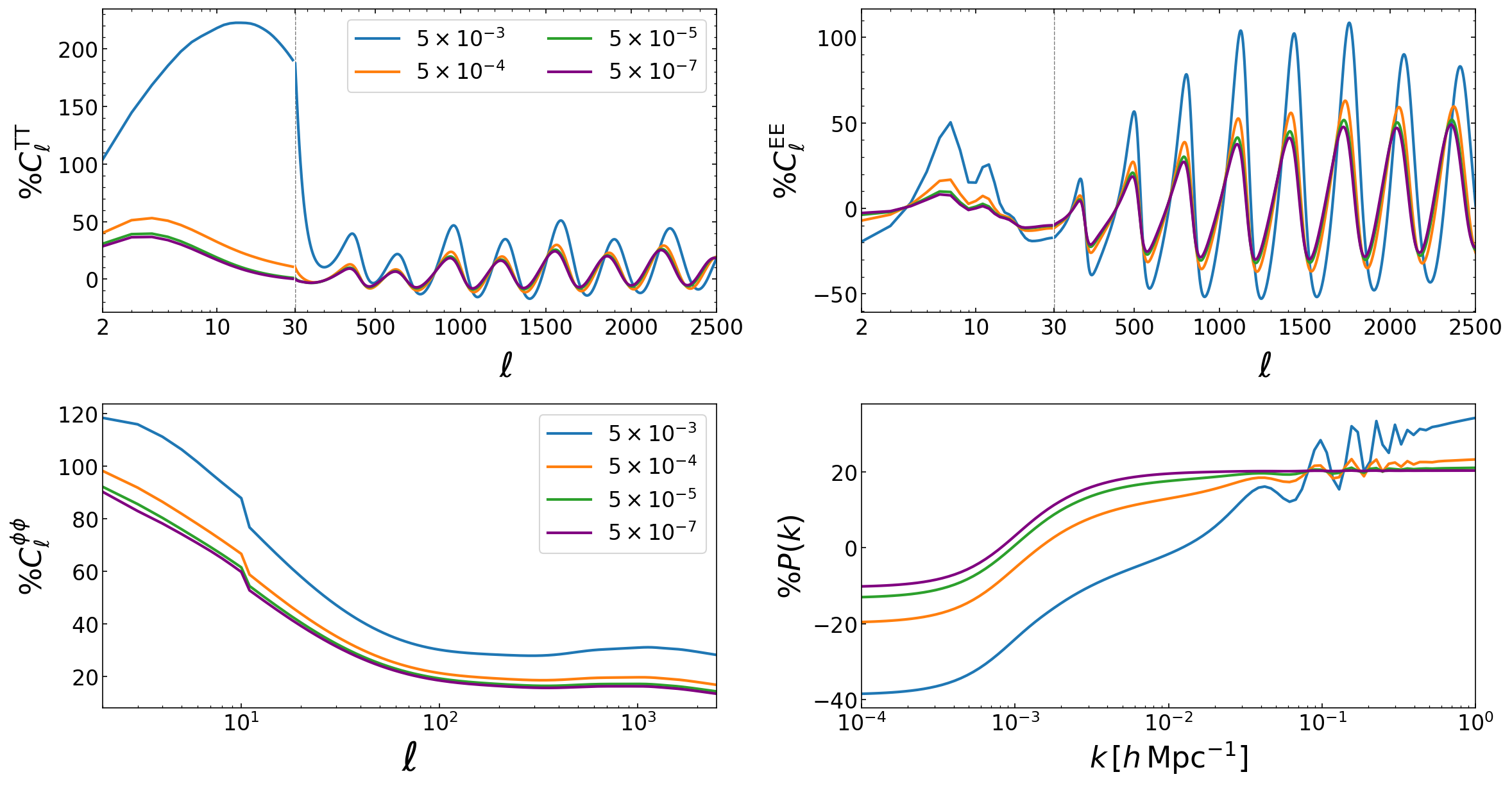}
\caption{Relative differences with respect to $\Lambda \rm CDM$ for BDG phantom ($Z=-1$) without potential and $g(\sigma)=\alpha\sigma^{-1}$, for different values of $\gamma$. CMB TT (top left), EE (top right), lensing potential (bottom left) and linear matter power spectra (bottom right).}
\label{fig:BDGph_noL_Delta_CMB_LSS_spectra}
\end{figure*}

As we saw in \cref{sec:lin_pert} the interplay of the $G_3$ and the ISW effect can result in an enhancement of the power spectrum on large scales \cite{Kobayashi:2009wr,Renk:2016olm}.
For this reason, in \cref{fig:BDGph_noL_Delta_CMB_LSS_spectra} we can observe very large departures from $\Lambda \rm CDM$ at the largest scales in TT, due to the enhanced ISW effect.
It is worth pointing out that as there is no potential, the curves do not flatten out towards zero for smaller $\gamma$, this is due to the fact that without a cosmological constant there is not a $\Lambda \rm CDM$ limit, as already discussed in \cref{sec:model}.
So, while this model provides late-time cosmic acceleration without a cosmological constant, it shows large differences with respect to $\Lambda \rm CDM$ at the level of CMB and matter power spectra; this is a disadvantage for differences $>30\%$ when comparing the predictions against the data, since $\Lambda \rm CDM$ fits {\em Planck} data quite well.

%
%

\subsection{Constraints from cosmological observations}
\begin{figure*}
\includegraphics[width=0.92\textwidth]{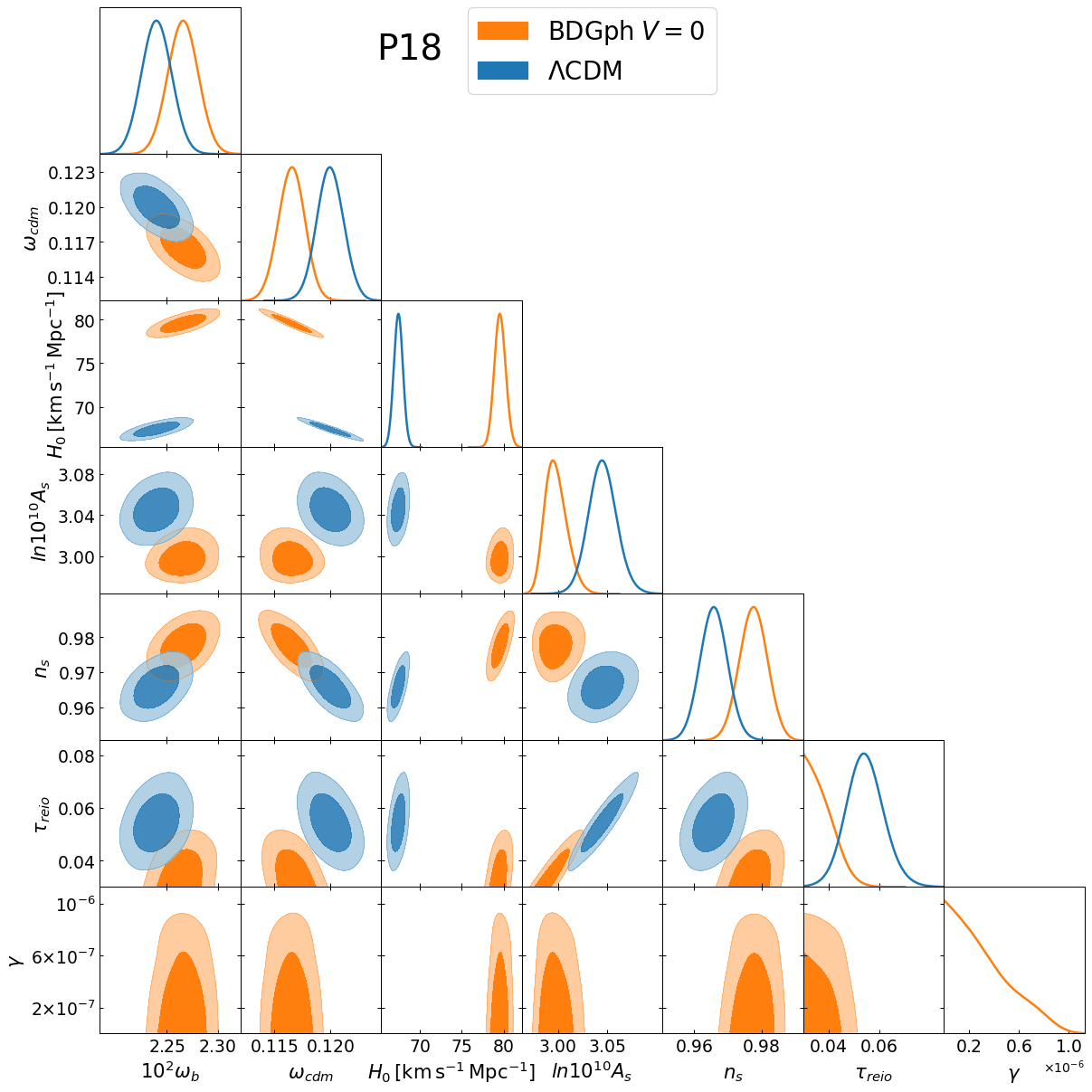}
\caption{Marginalized joint $68\%$ and $95\%$ CL regions 2D parameter space using P18 data alone. In orange BDG phantom ($Z=-1$) without potential, $g(\sigma)=\alpha\sigma^{-1}$ and $\Lambda \rm CDM$ in blue.}
\label{fig:triangle_SK_P18}
\end{figure*}

\begin{table}[htb]
\setlength{\tabcolsep}{8pt}
\begin{tabular} { l  c}
\toprule
\multicolumn{2}{c}{P18} \\
\midrule
{$10^{2}\omega{}_{\rm b }$}  & $2.266\pm 0.015$                  \\
{$\omega{}_{\rm cdm }$}      & $0.1165\pm 0.0012$                \\
{$H_0$}     
$[{\rm km s^{-1} Mpc^{-1}}]$ & $79.57\pm 0.67$                   \\
{$ln10^{10}A_{\rm s }$}      & $2.9979^{+0.0093}_{-0.013}$       \\
{$n_{\rm s }         $}      & $0.9776\pm 0.0041$                \\
{$\tau{}_{\rm reio } $}      & $< 0.0475\,$ (95\% CL)            \\
{$\gamma $}                  & $< 7.70\cdot 10^{-7}\,$ (95\% CL) \\
\midrule
$\Delta\chi^2$ & {30.6} \\
$\Delta \rm AIC $ & {32.6} \\
\bottomrule
\end{tabular}
\caption{Constraints on the main parameters (at $68\%$ CL unless otherwise
stated) considering {\em Planck} 2018 data alone for {\bf BDG phantom} ($Z=-1$) without potential and $g(\sigma) = \alpha\sigma^{-1}$.}
\label{tab:SK_table}
\end{table}
As we saw, this is the most extreme of the models considered, as it has no cosmological constant and therefore no $\Lambda \rm CDM$ limit and it cannot reproduce the CMB and LSS theoretical prediction of $\Lambda \rm CDM$, this put it at a disadvantage when it is tested against data, and a $\Delta\chi^2=30.6$ when considering P18 confirms it.
While this completely rules out the model, there is a feature that is worth highlighting in addition to the theoretical appealing characteristic of providing cosmic acceleration without a cosmological constant: it raises the Hubble constant to $H_0 = 79.57 \pm 0.67 \, {\rm km\, s^{-1}\, Mpc^{-1}}$.
While this result is still far from alleviating the tension, the ability to produce a Hubble constant larger than the SH0ES value \cite{Riess:2021jrx} is quite interesting and it might be worth to construct and investigate similar models that, while retaining this ability, might provide a better fit to the data. The results for the model are shown in \cref{fig:triangle_SK_P18} and \cref{tab:SK_table}.
\clearpage

%
%

\section{Linear perturbed equations} \label{sec:app_pert}

Here we summarize the definition of the coefficients appearing in the perturbed Einstein and scalar field equations:
\begin{widetext}

\begin{align}
\delta\tilde\rho &\equiv \frac{\delta\rho_m}{\gamma\sigma^2}  -\frac{h'\sigma'}{a^2\sigma}  -\frac{2}{a^2} \Bigg\{ \frac{\delta\sigma}{\sigma} \bigg[\frac{a^2\rho_m}{\gamma\sigma^2} +Z\frac{\sigma'^{\,2}}{2\gamma\sigma^2} +\frac{a^2}{\gamma\sigma} \bigg(\frac{V}{\sigma} -\frac{V_{,\sigma}}{2}	\bigg) - \frac{3\Hrondo\sigma'}{\sigma} +k^2	\bigg] + \frac{\delta\sigma'}{\sigma} \bigg(\, 3 \Hrondo -Z\frac{\sigma'}{2\gamma\sigma}	\bigg)
\Bigg\} \, ,
\\
\delta\tilde{\rho}^{(G)} &\equiv\, 
-\frac{2}{\gamma a^4} \frac{\sigma'^{\,2}}{\sigma^2} \Bigg\{ 
\frac{\delta\sigma}{\sigma} \bigg[
3\Hrondo\sigma' \big(2\,g - \sigma g_{,\sigma} \big) + \frac{\sigma'^{\,2}}{2}\big(\sigma g_{,\sigma,\sigma} - 2g_{,\sigma} \big) -k^2\sigma g +\frac{3\sigma'^{\,2}}{2} \big(2\zeta - \zeta_{,\sigma}\sigma \big)	
\bigg]
\nonumber\\
&+ \delta\sigma' \Big(\,\sigma' (2g_{, \sigma} - 6 \zeta ) - 9\Hrondo g  \,\Big) - \frac12 h' \sigma' g
\Bigg\}\, ,
\\
\big(\tilde\rho+\tilde P\big) \tilde\theta\, &\equiv\, \frac{(\rho_m+P_m)}{\gamma\sigma^2} \theta_m +\frac{2k^2}{a^2}\bigg\{\, \frac{\delta\sigma}{\sigma} \bigg[\, \frac{\sigma'}{2\gamma\sigma}(Z+2\gamma) -\Hrondo \bigg] +\frac{\delta\sigma'}{\sigma}	\bigg\}\, ,
\\
\big(\tilde\rho^{(G)}+\tilde P^{(G)}\big) \tilde\theta^{(G)}& \equiv \frac{2k^2}{\gamma a^4} \frac{\sigma'^{\,2}}{\sigma^2} \, \Big[\delta\sigma(\,3\Hrondo g -g_{,\sigma} \sigma' +2\zeta\sigma' \,) - \delta\sigma' g	\Big]\, ,
\\
\delta\tilde P &\equiv \frac{\delta P_m}{\gamma\sigma^2} +\frac{1}{a^2} \Bigg\{ - \frac{\delta\sigma}{\sigma}\bigg[\frac{a^2}{\gamma\sigma} \bigg(V_{,\sigma} -\frac{2V}{\sigma} \bigg) + \frac{2a^2 P_m}{\gamma\sigma^2} +\frac{\sigma'^{\,2}}{\gamma\sigma^2} (Z+4\gamma) + \frac{2\sigma''}{\sigma}+\frac{2\Hrondo\sigma'}{\sigma} - \frac{4k^2}{3}
\bigg]
\nonumber\\
&+ \delta\sigma' \bigg(\frac{\sigma'}{\gamma\sigma^2} (Z+4\gamma) +\frac{2\Hrondo}{\sigma} \bigg) +\frac{2\,\delta\sigma''}{\sigma} +\frac{2\sigma'}{3\sigma} h'
\Bigg\} \, ,
\\
\delta\tilde{P}^{(G)} &\equiv \frac{1}{\gamma a^4} \frac{\sigma'^{\,2}}{\sigma^2} \, \Bigg\{ \frac{\delta\sigma}{\sigma} \Big[ 2\, (2\,g -g_{,\sigma}\sigma) (\sigma'' - \Hrondo\sigma') +\sigma'^{\,2}\big(2\,g_{,\sigma}\, - g_{,\sigma,\sigma}\, \sigma -\, 2\zeta\, + \zeta_{,\sigma}\, \sigma \big) 
\Big]
\nonumber \\
&-\delta\sigma' \Big(\,2g_{,\sigma}\sigma' -4\Hrondo\,g +2\,g \frac{\sigma''}{\sigma'} +4\,\zeta\,\sigma'	\,\Big) -2\,g\, \delta\sigma''
\Bigg\}  \, ,
\\
(\tilde\rho+\tilde P)\tilde\Theta &\equiv \frac{(\rho_m+P_m)\Theta}{\gamma\sigma^2} + \frac{1}{3a^2}\bigg[\,\frac{4k^2\delta\sigma}{\sigma} + 2(h'+6\eta')\frac{\sigma'}{\sigma} \,\bigg] \, ,
\\
\delta\tilde{\cal G} &\equiv \f{a^2( \delta\rho_m-3\,\delta P_m)}{(1+6\gamma)\sigma} \, - \frac{h'}{2 a^2} \left[ a^2 \sigma' (Z+6\gamma) + 2\, g\sigma' \left( 3\Hrondo\sigma' -\frac{\sigma'^2}{\sigma} +2g\sigma'' \right) +4\zeta\sigma'^3 \right] - h'' g \frac{\sigma'^2}{a^2} \nonumber\\
&+\f{1}{a^2}\delta\sigma\Bigg\{ a^4 \left[ \f{3 P_m - \rho_m}{\sigma^2} -\f{4V}{\sigma^2} + \f{4 V_{,\sigma}}{\sigma} - V_{,\sigma\sigma} + (Z+6\gamma) \left(\f{\sigma'^2}{\sigma^2} -k^2\right) \right] \nonumber\\
&\;+ g \left[ 2 k^2 \left( \f{\sigma'^{\,2}}{\sigma} -2\Hrondo\sigma' -2\sigma'' \right) -6 \f{\sigma'^{\,2}}{\sigma^2} \sigma'' \right] -4k^2\zeta\,\sigma'^{\,2} + g_{,\sigma}\,\sigma' \left[ 6 \sigma'' \left( \f{\sigma'}{\sigma} -2\Hrondo \right) - 2 \sigma' \left( 3\Hrondo' +\f{\sigma'^{\,2}}{\sigma^2}  \right) \right] \nonumber\\
&\;-12 \sigma'^{\,2} \zeta_{,\sigma}\sigma'' +2g_{,\sigma\sigma}\sigma'^{\,2 }\left(\f{\sigmapsq}{\sigma} -2\Hrondo\sigma' + 2\sigma''  \right) -3\zeta_{,\sigma\sigma}\sigma'^{\,4} - g_{,\sigma\sigma\sigma} \sigma'^{\,4}  \Bigg\} \nonumber\\
&+\f{1}{a^2}\delta\sigma' \Bigg\{ a^2\left[ - 2 (Z+6\gamma)\left(\Hrondo + \f{\sigma'}{\sigma}  \right)\right]  +12\, g \left[\sigma''\left(\f{\sigma'}{\sigma} - \Hrondo \right) -\Hrondo'\sigma'  \right] -24\zeta\sigma'\sigma'' \nonumber\\
&\qquad +4\,g_{,\sigma}\sigma'\left(2\f{\sigmapsq}{\sigma} -3\Hrondo\sigma'+2\sigma'' \right) -12 \zeta_{,\sigma}\sigma'^{\,3} + 4\, g_{,\sigma\sigma}\sigma'^{\,3}
\Bigg\}\, .
\end{align}
\end{widetext}
	\clearpage
\bibliography{references}

\begin{thebibliography}{89}%
\makeatletter
\providecommand \@ifxundefined [1]{%
 \@ifx{#1\undefined}
}%
\providecommand \@ifnum [1]{%
 \ifnum #1\expandafter \@firstoftwo
 \else \expandafter \@secondoftwo
 \fi
}%
\providecommand \@ifx [1]{%
 \ifx #1\expandafter \@firstoftwo
 \else \expandafter \@secondoftwo
 \fi
}%
\providecommand \natexlab [1]{#1}%
\providecommand \enquote  [1]{``#1''}%
\providecommand \bibnamefont  [1]{#1}%
\providecommand \bibfnamefont [1]{#1}%
\providecommand \citenamefont [1]{#1}%
\providecommand \href@noop [0]{\@secondoftwo}%
\providecommand \href [0]{\begingroup \@sanitize@url \@href}%
\providecommand \@href[1]{\@@startlink{#1}\@@href}%
\providecommand \@@href[1]{\endgroup#1\@@endlink}%
\providecommand \@sanitize@url [0]{\catcode `\\12\catcode `\$12\catcode `\&12\catcode `\#12\catcode `\^12\catcode `\_12\catcode `\%12\relax}%
\providecommand \@@startlink[1]{}%
\providecommand \@@endlink[0]{}%
\providecommand \url  [0]{\begingroup\@sanitize@url \@url }%
\providecommand \@url [1]{\endgroup\@href {#1}{\urlprefix }}%
\providecommand \urlprefix  [0]{URL }%
\providecommand \Eprint [0]{\href }%
\providecommand \doibase [0]{https://doi.org/}%
\providecommand \selectlanguage [0]{\@gobble}%
\providecommand \bibinfo  [0]{\@secondoftwo}%
\providecommand \bibfield  [0]{\@secondoftwo}%
\providecommand \translation [1]{[#1]}%
\providecommand \BibitemOpen [0]{}%
\providecommand \bibitemStop [0]{}%
\providecommand \bibitemNoStop [0]{.\EOS\space}%
\providecommand \EOS [0]{\spacefactor3000\relax}%
\providecommand \BibitemShut  [1]{\csname bibitem#1\endcsname}%
\let\auto@bib@innerbib\@empty
\bibitem [{\citenamefont {Aghanim}\ \emph {et~al.}(2020{\natexlab{a}})\citenamefont {Aghanim} \emph {et~al.}}]{Planck:2018nkj}%
  \BibitemOpen
  \bibfield  {author} {\bibinfo {author} {\bibfnamefont {N.}~\bibnamefont {Aghanim}} \emph {et~al.} (\bibinfo {collaboration} {Planck}),\ }\bibfield  {title} {\bibinfo {title} {{Planck 2018 results. I. Overview and the cosmological legacy of Planck}},\ }\href {https://doi.org/10.1051/0004-6361/201833880} {\bibfield  {journal} {\bibinfo  {journal} {Astron. Astrophys.}\ }\textbf {\bibinfo {volume} {641}},\ \bibinfo {pages} {A1} (\bibinfo {year} {2020}{\natexlab{a}})},\ \Eprint {https://arxiv.org/abs/1807.06205} {arXiv:1807.06205 [astro-ph.CO]} \BibitemShut {NoStop}%
\bibitem [{\citenamefont {Alam}\ \emph {et~al.}(2021)\citenamefont {Alam} \emph {et~al.}}]{eBOSS:2020yzd}%
  \BibitemOpen
  \bibfield  {author} {\bibinfo {author} {\bibfnamefont {S.}~\bibnamefont {Alam}} \emph {et~al.} (\bibinfo {collaboration} {eBOSS}),\ }\bibfield  {title} {\bibinfo {title} {{Completed SDSS-IV extended Baryon Oscillation Spectroscopic Survey: Cosmological implications from two decades of spectroscopic surveys at the Apache Point Observatory}},\ }\href {https://doi.org/10.1103/PhysRevD.103.083533} {\bibfield  {journal} {\bibinfo  {journal} {Phys. Rev. D}\ }\textbf {\bibinfo {volume} {103}},\ \bibinfo {pages} {083533} (\bibinfo {year} {2021})},\ \Eprint {https://arxiv.org/abs/2007.08991} {arXiv:2007.08991 [astro-ph.CO]} \BibitemShut {NoStop}%
\bibitem [{\citenamefont {Veropalumbo}\ \emph {et~al.}(2014)\citenamefont {Veropalumbo}, \citenamefont {Marulli}, \citenamefont {Moscardini}, \citenamefont {Moresco},\ and\ \citenamefont {Cimatti}}]{Veropalumbo:2013cua}%
  \BibitemOpen
  \bibfield  {author} {\bibinfo {author} {\bibfnamefont {A.}~\bibnamefont {Veropalumbo}}, \bibinfo {author} {\bibfnamefont {F.}~\bibnamefont {Marulli}}, \bibinfo {author} {\bibfnamefont {L.}~\bibnamefont {Moscardini}}, \bibinfo {author} {\bibfnamefont {M.}~\bibnamefont {Moresco}},\ and\ \bibinfo {author} {\bibfnamefont {A.}~\bibnamefont {Cimatti}},\ }\bibfield  {title} {\bibinfo {title} {{An improved measurement of baryon acoustic oscillations from the correlation function of galaxy clusters at z \ensuremath{\sim} 0.3}},\ }\href {https://doi.org/10.1093/mnras/stu1050} {\bibfield  {journal} {\bibinfo  {journal} {Mon. Not. Roy. Astron. Soc.}\ }\textbf {\bibinfo {volume} {442}},\ \bibinfo {pages} {3275} (\bibinfo {year} {2014})},\ \Eprint {https://arxiv.org/abs/1311.5895} {arXiv:1311.5895 [astro-ph.CO]} \BibitemShut {NoStop}%
\bibitem [{\citenamefont {Asgari}\ \emph {et~al.}(2021)\citenamefont {Asgari} \emph {et~al.}}]{KiDS:2020suj}%
  \BibitemOpen
  \bibfield  {author} {\bibinfo {author} {\bibfnamefont {M.}~\bibnamefont {Asgari}} \emph {et~al.} (\bibinfo {collaboration} {KiDS}),\ }\bibfield  {title} {\bibinfo {title} {{KiDS-1000 Cosmology: Cosmic shear constraints and comparison between two point statistics}},\ }\href {https://doi.org/10.1051/0004-6361/202039070} {\bibfield  {journal} {\bibinfo  {journal} {Astron. Astrophys.}\ }\textbf {\bibinfo {volume} {645}},\ \bibinfo {pages} {A104} (\bibinfo {year} {2021})},\ \Eprint {https://arxiv.org/abs/2007.15633} {arXiv:2007.15633 [astro-ph.CO]} \BibitemShut {NoStop}%
\bibitem [{\citenamefont {Abbott}\ \emph {et~al.}(2022)\citenamefont {Abbott} \emph {et~al.}}]{DES:2021wwk}%
  \BibitemOpen
  \bibfield  {author} {\bibinfo {author} {\bibfnamefont {T.~M.~C.}\ \bibnamefont {Abbott}} \emph {et~al.} (\bibinfo {collaboration} {DES}),\ }\bibfield  {title} {\bibinfo {title} {{Dark Energy Survey Year 3 results: Cosmological constraints from galaxy clustering and weak lensing}},\ }\href {https://doi.org/10.1103/PhysRevD.105.023520} {\bibfield  {journal} {\bibinfo  {journal} {Phys. Rev. D}\ }\textbf {\bibinfo {volume} {105}},\ \bibinfo {pages} {023520} (\bibinfo {year} {2022})},\ \Eprint {https://arxiv.org/abs/2105.13549} {arXiv:2105.13549 [astro-ph.CO]} \BibitemShut {NoStop}%
\bibitem [{\citenamefont {Aghanim}\ \emph {et~al.}(2020{\natexlab{b}})\citenamefont {Aghanim} \emph {et~al.}}]{Planck:2018lbu}%
  \BibitemOpen
  \bibfield  {author} {\bibinfo {author} {\bibfnamefont {N.}~\bibnamefont {Aghanim}} \emph {et~al.} (\bibinfo {collaboration} {Planck}),\ }\bibfield  {title} {\bibinfo {title} {{Planck 2018 results. VIII. Gravitational lensing}},\ }\href {https://doi.org/10.1051/0004-6361/201833886} {\bibfield  {journal} {\bibinfo  {journal} {Astron. Astrophys.}\ }\textbf {\bibinfo {volume} {641}},\ \bibinfo {pages} {A8} (\bibinfo {year} {2020}{\natexlab{b}})},\ \Eprint {https://arxiv.org/abs/1807.06210} {arXiv:1807.06210 [astro-ph.CO]} \BibitemShut {NoStop}%
\bibitem [{\citenamefont {Sherwin}\ \emph {et~al.}(2017)\citenamefont {Sherwin}, \citenamefont {van Engelen}, \citenamefont {Sehgal}, \citenamefont {Madhavacheril}, \citenamefont {Addison}, \citenamefont {Aiola} \emph {et~al.}}]{Sherwin:2016tyf}%
  \BibitemOpen
  \bibfield  {author} {\bibinfo {author} {\bibfnamefont {B.~D.}\ \bibnamefont {Sherwin}}, \bibinfo {author} {\bibfnamefont {A.}~\bibnamefont {van Engelen}}, \bibinfo {author} {\bibfnamefont {N.}~\bibnamefont {Sehgal}}, \bibinfo {author} {\bibfnamefont {M.}~\bibnamefont {Madhavacheril}}, \bibinfo {author} {\bibfnamefont {G.~E.}\ \bibnamefont {Addison}}, \bibinfo {author} {\bibfnamefont {S.}~\bibnamefont {Aiola}}, \emph {et~al.},\ }\bibfield  {title} {\bibinfo {title} {Two-season atacama cosmology telescope polarimeter lensing power spectrum},\ }\href {https://doi.org/10.1103/PhysRevD.95.123529} {\bibfield  {journal} {\bibinfo  {journal} {Phys. Rev. D}\ }\textbf {\bibinfo {volume} {95}},\ \bibinfo {pages} {123529} (\bibinfo {year} {2017})},\ \Eprint {https://arxiv.org/abs/1611.09753} {arXiv:1611.09753 [astro-ph.CO]} \BibitemShut {NoStop}%
\bibitem [{\citenamefont {Wu}\ \emph {et~al.}(2019)\citenamefont {Wu} \emph {et~al.}}]{Wu:2019hek}%
  \BibitemOpen
  \bibfield  {author} {\bibinfo {author} {\bibfnamefont {W.~L.~K.}\ \bibnamefont {Wu}} \emph {et~al.},\ }\bibfield  {title} {\bibinfo {title} {{A Measurement of the Cosmic Microwave Background Lensing Potential and Power Spectrum from 500 deg$^2$ of SPTpol Temperature and Polarization Data}},\ }\href {https://doi.org/10.3847/1538-4357/ab4186} {\bibfield  {journal} {\bibinfo  {journal} {Astrophys. J.}\ }\textbf {\bibinfo {volume} {884}},\ \bibinfo {pages} {70} (\bibinfo {year} {2019})},\ \Eprint {https://arxiv.org/abs/1905.05777} {arXiv:1905.05777 [astro-ph.CO]} \BibitemShut {NoStop}%
\bibitem [{\citenamefont {Darwish}\ \emph {et~al.}(2020)\citenamefont {Darwish} \emph {et~al.}}]{Darwish:2020fwf}%
  \BibitemOpen
  \bibfield  {author} {\bibinfo {author} {\bibfnamefont {O.}~\bibnamefont {Darwish}} \emph {et~al.},\ }\bibfield  {title} {\bibinfo {title} {{The Atacama Cosmology Telescope: A CMB lensing mass map over 2100 square degrees of sky and its cross-correlation with BOSS-CMASS galaxies}},\ }\href {https://doi.org/10.1093/mnras/staa3438} {\bibfield  {journal} {\bibinfo  {journal} {Mon. Not. Roy. Astron. Soc.}\ }\textbf {\bibinfo {volume} {500}},\ \bibinfo {pages} {2250} (\bibinfo {year} {2020})},\ \Eprint {https://arxiv.org/abs/2004.01139} {arXiv:2004.01139 [astro-ph.CO]} \BibitemShut {NoStop}%
\bibitem [{\citenamefont {Riess}\ \emph {et~al.}(1998)\citenamefont {Riess} \emph {et~al.}}]{SupernovaSearchTeam:1998fmf}%
  \BibitemOpen
  \bibfield  {author} {\bibinfo {author} {\bibfnamefont {A.~G.}\ \bibnamefont {Riess}} \emph {et~al.} (\bibinfo {collaboration} {Supernova Search Team}),\ }\bibfield  {title} {\bibinfo {title} {{Observational evidence from supernovae for an accelerating universe and a cosmological constant}},\ }\href {https://doi.org/10.1086/300499} {\bibfield  {journal} {\bibinfo  {journal} {Astron. J.}\ }\textbf {\bibinfo {volume} {116}},\ \bibinfo {pages} {1009} (\bibinfo {year} {1998})},\ \Eprint {https://arxiv.org/abs/astro-ph/9805201} {arXiv:astro-ph/9805201} \BibitemShut {NoStop}%
\bibitem [{\citenamefont {Perlmutter}\ \emph {et~al.}(1999)\citenamefont {Perlmutter} \emph {et~al.}}]{SupernovaCosmologyProject:1998vns}%
  \BibitemOpen
  \bibfield  {author} {\bibinfo {author} {\bibfnamefont {S.}~\bibnamefont {Perlmutter}} \emph {et~al.} (\bibinfo {collaboration} {Supernova Cosmology Project}),\ }\bibfield  {title} {\bibinfo {title} {{Measurements of $\Omega$ and $\Lambda$ from 42 high redshift supernovae}},\ }\href {https://doi.org/10.1086/307221} {\bibfield  {journal} {\bibinfo  {journal} {Astrophys. J.}\ }\textbf {\bibinfo {volume} {517}},\ \bibinfo {pages} {565} (\bibinfo {year} {1999})},\ \Eprint {https://arxiv.org/abs/astro-ph/9812133} {arXiv:astro-ph/9812133} \BibitemShut {NoStop}%
\bibitem [{\citenamefont {Carr}\ \emph {et~al.}(2022)\citenamefont {Carr}, \citenamefont {Davis}, \citenamefont {Scolnic}, \citenamefont {Scolnic}, \citenamefont {Said}, \citenamefont {Brout}, \citenamefont {Peterson},\ and\ \citenamefont {Kessler}}]{Carr:2021lcj}%
  \BibitemOpen
  \bibfield  {author} {\bibinfo {author} {\bibfnamefont {A.}~\bibnamefont {Carr}}, \bibinfo {author} {\bibfnamefont {T.~M.}\ \bibnamefont {Davis}}, \bibinfo {author} {\bibfnamefont {D.}~\bibnamefont {Scolnic}}, \bibinfo {author} {\bibfnamefont {D.}~\bibnamefont {Scolnic}}, \bibinfo {author} {\bibfnamefont {K.}~\bibnamefont {Said}}, \bibinfo {author} {\bibfnamefont {D.}~\bibnamefont {Brout}}, \bibinfo {author} {\bibfnamefont {E.~R.}\ \bibnamefont {Peterson}},\ and\ \bibinfo {author} {\bibfnamefont {R.}~\bibnamefont {Kessler}},\ }\bibfield  {title} {\bibinfo {title} {{The Pantheon+ analysis: Improving the redshifts and peculiar velocities of Type Ia supernovae used in cosmological analyses}},\ }\href {https://doi.org/10.1017/pasa.2022.41} {\bibfield  {journal} {\bibinfo  {journal} {Publ. Astron. Soc. Austral.}\ }\textbf {\bibinfo {volume} {39}},\ \bibinfo {pages} {e046} (\bibinfo {year} {2022})},\ \Eprint {https://arxiv.org/abs/2112.01471} {arXiv:2112.01471 [astro-ph.CO]} \BibitemShut {NoStop}%
\bibitem [{\citenamefont {Aghanim}\ \emph {et~al.}(2020{\natexlab{c}})\citenamefont {Aghanim} \emph {et~al.}}]{Planck:2018vyg}%
  \BibitemOpen
  \bibfield  {author} {\bibinfo {author} {\bibfnamefont {N.}~\bibnamefont {Aghanim}} \emph {et~al.} (\bibinfo {collaboration} {Planck}),\ }\bibfield  {title} {\bibinfo {title} {{Planck 2018 results. VI. Cosmological parameters}},\ }\href {https://doi.org/10.1051/0004-6361/201833910} {\bibfield  {journal} {\bibinfo  {journal} {Astron. Astrophys.}\ }\textbf {\bibinfo {volume} {641}},\ \bibinfo {pages} {A6} (\bibinfo {year} {2020}{\natexlab{c}})},\ \bibinfo {note} {[Erratum: Astron.Astrophys. 652, C4 (2021)]},\ \Eprint {https://arxiv.org/abs/1807.06209} {arXiv:1807.06209 [astro-ph.CO]} \BibitemShut {NoStop}%
\bibitem [{\citenamefont {Riess}\ \emph {et~al.}(2022)\citenamefont {Riess} \emph {et~al.}}]{Riess:2021jrx}%
  \BibitemOpen
  \bibfield  {author} {\bibinfo {author} {\bibfnamefont {A.~G.}\ \bibnamefont {Riess}} \emph {et~al.},\ }\bibfield  {title} {\bibinfo {title} {{A Comprehensive Measurement of the Local Value of the Hubble Constant with 1 km s$^{-1}$ Mpc$^{-1}$ Uncertainty from the Hubble Space Telescope and the SH0ES Team}},\ }\href {https://doi.org/10.3847/2041-8213/ac5c5b} {\bibfield  {journal} {\bibinfo  {journal} {Astrophys. J. Lett.}\ }\textbf {\bibinfo {volume} {934}},\ \bibinfo {pages} {L7} (\bibinfo {year} {2022})},\ \Eprint {https://arxiv.org/abs/2112.04510} {arXiv:2112.04510 [astro-ph.CO]} \BibitemShut {NoStop}%
\bibitem [{\citenamefont {Aiola}\ \emph {et~al.}(2020)\citenamefont {Aiola} \emph {et~al.}}]{ACT:2020gnv}%
  \BibitemOpen
  \bibfield  {author} {\bibinfo {author} {\bibfnamefont {S.}~\bibnamefont {Aiola}} \emph {et~al.} (\bibinfo {collaboration} {ACT}),\ }\bibfield  {title} {\bibinfo {title} {{The Atacama Cosmology Telescope: DR4 Maps and Cosmological Parameters}},\ }\href {https://doi.org/10.1088/1475-7516/2020/12/047} {\bibfield  {journal} {\bibinfo  {journal} {JCAP}\ }\textbf {\bibinfo {volume} {12}},\ \bibinfo {pages} {047}},\ \Eprint {https://arxiv.org/abs/2007.07288} {arXiv:2007.07288 [astro-ph.CO]} \BibitemShut {NoStop}%
\bibitem [{\citenamefont {Balkenhol}\ \emph {et~al.}(2021)\citenamefont {Balkenhol} \emph {et~al.}}]{SPT-3G:2021wgf}%
  \BibitemOpen
  \bibfield  {author} {\bibinfo {author} {\bibfnamefont {L.}~\bibnamefont {Balkenhol}} \emph {et~al.} (\bibinfo {collaboration} {SPT-3G}),\ }\bibfield  {title} {\bibinfo {title} {{Constraints on \ensuremath{\Lambda}CDM extensions from the SPT-3G 2018 EE and TE power spectra}},\ }\href {https://doi.org/10.1103/PhysRevD.104.083509} {\bibfield  {journal} {\bibinfo  {journal} {Phys. Rev. D}\ }\textbf {\bibinfo {volume} {104}},\ \bibinfo {pages} {083509} (\bibinfo {year} {2021})},\ \Eprint {https://arxiv.org/abs/2103.13618} {arXiv:2103.13618 [astro-ph.CO]} \BibitemShut {NoStop}%
\bibitem [{\citenamefont {Freedman}\ \emph {et~al.}(2019)\citenamefont {Freedman} \emph {et~al.}}]{Freedman:2019jwv}%
  \BibitemOpen
  \bibfield  {author} {\bibinfo {author} {\bibfnamefont {W.~L.}\ \bibnamefont {Freedman}} \emph {et~al.},\ }\bibfield  {title} {\bibinfo {title} {{The Carnegie-Chicago Hubble Program. VIII. An Independent Determination of the Hubble Constant Based on the Tip of the Red Giant Branch}},\ }\href {https://doi.org/10.3847/1538-4357/ab2f73} {\bibfield  {journal} {\bibinfo  {journal} {Astrophys. J.}\ }\textbf {\bibinfo {volume} {882}},\ \bibinfo {pages} {34} (\bibinfo {year} {2019})},\ \Eprint {https://arxiv.org/abs/1907.05922} {arXiv:1907.05922 [astro-ph.CO]} \BibitemShut {NoStop}%
\bibitem [{\citenamefont {Ade}\ \emph {et~al.}(2014)\citenamefont {Ade} \emph {et~al.}}]{Planck:2013pxb}%
  \BibitemOpen
  \bibfield  {author} {\bibinfo {author} {\bibfnamefont {P.~A.~R.}\ \bibnamefont {Ade}} \emph {et~al.} (\bibinfo {collaboration} {Planck}),\ }\bibfield  {title} {\bibinfo {title} {{Planck 2013 results. XVI. Cosmological parameters}},\ }\href {https://doi.org/10.1051/0004-6361/201321591} {\bibfield  {journal} {\bibinfo  {journal} {Astron. Astrophys.}\ }\textbf {\bibinfo {volume} {571}},\ \bibinfo {pages} {A16} (\bibinfo {year} {2014})},\ \Eprint {https://arxiv.org/abs/1303.5076} {arXiv:1303.5076 [astro-ph.CO]} \BibitemShut {NoStop}%
\bibitem [{\citenamefont {Knox}\ and\ \citenamefont {Millea}(2020)}]{Knox:2019rjx}%
  \BibitemOpen
  \bibfield  {author} {\bibinfo {author} {\bibfnamefont {L.}~\bibnamefont {Knox}}\ and\ \bibinfo {author} {\bibfnamefont {M.}~\bibnamefont {Millea}},\ }\bibfield  {title} {\bibinfo {title} {{Hubble constant hunter\textquoteright{}s guide}},\ }\href {https://doi.org/10.1103/PhysRevD.101.043533} {\bibfield  {journal} {\bibinfo  {journal} {Phys. Rev. D}\ }\textbf {\bibinfo {volume} {101}},\ \bibinfo {pages} {043533} (\bibinfo {year} {2020})},\ \Eprint {https://arxiv.org/abs/1908.03663} {arXiv:1908.03663 [astro-ph.CO]} \BibitemShut {NoStop}%
\bibitem [{\citenamefont {Di~Valentino}\ \emph {et~al.}(2021{\natexlab{a}})\citenamefont {Di~Valentino} \emph {et~al.}}]{DiValentino:2020zio}%
  \BibitemOpen
  \bibfield  {author} {\bibinfo {author} {\bibfnamefont {E.}~\bibnamefont {Di~Valentino}} \emph {et~al.},\ }\bibfield  {title} {\bibinfo {title} {{Snowmass2021 - Letter of interest cosmology intertwined II: The hubble constant tension}},\ }\href {https://doi.org/10.1016/j.astropartphys.2021.102605} {\bibfield  {journal} {\bibinfo  {journal} {Astropart. Phys.}\ }\textbf {\bibinfo {volume} {131}},\ \bibinfo {pages} {102605} (\bibinfo {year} {2021}{\natexlab{a}})},\ \Eprint {https://arxiv.org/abs/2008.11284} {arXiv:2008.11284 [astro-ph.CO]} \BibitemShut {NoStop}%
\bibitem [{\citenamefont {Di~Valentino}\ \emph {et~al.}(2021{\natexlab{b}})\citenamefont {Di~Valentino}, \citenamefont {Mena}, \citenamefont {Pan}, \citenamefont {Visinelli}, \citenamefont {Yang}, \citenamefont {Melchiorri}, \citenamefont {Mota}, \citenamefont {Riess},\ and\ \citenamefont {Silk}}]{DiValentino:2021izs}%
  \BibitemOpen
  \bibfield  {author} {\bibinfo {author} {\bibfnamefont {E.}~\bibnamefont {Di~Valentino}}, \bibinfo {author} {\bibfnamefont {O.}~\bibnamefont {Mena}}, \bibinfo {author} {\bibfnamefont {S.}~\bibnamefont {Pan}}, \bibinfo {author} {\bibfnamefont {L.}~\bibnamefont {Visinelli}}, \bibinfo {author} {\bibfnamefont {W.}~\bibnamefont {Yang}}, \bibinfo {author} {\bibfnamefont {A.}~\bibnamefont {Melchiorri}}, \bibinfo {author} {\bibfnamefont {D.~F.}\ \bibnamefont {Mota}}, \bibinfo {author} {\bibfnamefont {A.~G.}\ \bibnamefont {Riess}},\ and\ \bibinfo {author} {\bibfnamefont {J.}~\bibnamefont {Silk}},\ }\bibfield  {title} {\bibinfo {title} {{In the realm of the Hubble tension\textemdash{}a review of solutions}},\ }\href {https://doi.org/10.1088/1361-6382/ac086d} {\bibfield  {journal} {\bibinfo  {journal} {Class. Quant. Grav.}\ }\textbf {\bibinfo {volume} {38}},\ \bibinfo {pages} {153001} (\bibinfo {year} {2021}{\natexlab{b}})},\ \Eprint {https://arxiv.org/abs/2103.01183} {arXiv:2103.01183 [astro-ph.CO]} \BibitemShut
  {NoStop}%
\bibitem [{\citenamefont {Perivolaropoulos}\ and\ \citenamefont {Skara}(2022)}]{Perivolaropoulos:2021jda}%
  \BibitemOpen
  \bibfield  {author} {\bibinfo {author} {\bibfnamefont {L.}~\bibnamefont {Perivolaropoulos}}\ and\ \bibinfo {author} {\bibfnamefont {F.}~\bibnamefont {Skara}},\ }\bibfield  {title} {\bibinfo {title} {{Challenges for $\Lambda$CDM: An update}},\ }\bibfield  {journal} {\bibinfo  {journal} {New Astron. Rev.}\ }\textbf {\bibinfo {volume} {95}},\ \href {https://doi.org/10.1016/j.newar.2022.101659} {10.1016/j.newar.2022.101659} (\bibinfo {year} {2022}),\ \Eprint {https://arxiv.org/abs/2105.05208} {arXiv:2105.05208 [astro-ph.CO]} \BibitemShut {NoStop}%
\bibitem [{\citenamefont {Sch\"oneberg}\ \emph {et~al.}(2022)\citenamefont {Sch\"oneberg}, \citenamefont {Franco~Abell\'an}, \citenamefont {P\'erez~S\'anchez}, \citenamefont {Witte}, \citenamefont {Poulin},\ and\ \citenamefont {Lesgourgues}}]{Schoneberg:2021qvd}%
  \BibitemOpen
  \bibfield  {author} {\bibinfo {author} {\bibfnamefont {N.}~\bibnamefont {Sch\"oneberg}}, \bibinfo {author} {\bibfnamefont {G.}~\bibnamefont {Franco~Abell\'an}}, \bibinfo {author} {\bibfnamefont {A.}~\bibnamefont {P\'erez~S\'anchez}}, \bibinfo {author} {\bibfnamefont {S.~J.}\ \bibnamefont {Witte}}, \bibinfo {author} {\bibfnamefont {V.}~\bibnamefont {Poulin}},\ and\ \bibinfo {author} {\bibfnamefont {J.}~\bibnamefont {Lesgourgues}},\ }\bibfield  {title} {\bibinfo {title} {{The H0 Olympics: A fair ranking of proposed models}},\ }\href {https://doi.org/10.1016/j.physrep.2022.07.001} {\bibfield  {journal} {\bibinfo  {journal} {Phys. Rept.}\ }\textbf {\bibinfo {volume} {984}},\ \bibinfo {pages} {1} (\bibinfo {year} {2022})},\ \Eprint {https://arxiv.org/abs/2107.10291} {arXiv:2107.10291 [astro-ph.CO]} \BibitemShut {NoStop}%
\bibitem [{\citenamefont {Shah}\ \emph {et~al.}(2021)\citenamefont {Shah}, \citenamefont {Lemos},\ and\ \citenamefont {Lahav}}]{Shah:2021onj}%
  \BibitemOpen
  \bibfield  {author} {\bibinfo {author} {\bibfnamefont {P.}~\bibnamefont {Shah}}, \bibinfo {author} {\bibfnamefont {P.}~\bibnamefont {Lemos}},\ and\ \bibinfo {author} {\bibfnamefont {O.}~\bibnamefont {Lahav}},\ }\bibfield  {title} {\bibinfo {title} {{A buyer\textquoteright{}s guide to the Hubble constant}},\ }\href {https://doi.org/10.1007/s00159-021-00137-4} {\bibfield  {journal} {\bibinfo  {journal} {Astron. Astrophys. Rev.}\ }\textbf {\bibinfo {volume} {29}},\ \bibinfo {pages} {9} (\bibinfo {year} {2021})},\ \Eprint {https://arxiv.org/abs/2109.01161} {arXiv:2109.01161 [astro-ph.CO]} \BibitemShut {NoStop}%
\bibitem [{\citenamefont {Abdalla}\ \emph {et~al.}(2022)\citenamefont {Abdalla} \emph {et~al.}}]{Abdalla:2022yfr}%
  \BibitemOpen
  \bibfield  {author} {\bibinfo {author} {\bibfnamefont {E.}~\bibnamefont {Abdalla}} \emph {et~al.},\ }\bibfield  {title} {\bibinfo {title} {{Cosmology intertwined: A review of the particle physics, astrophysics, and cosmology associated with the cosmological tensions and anomalies}},\ }\href {https://doi.org/10.1016/j.jheap.2022.04.002} {\bibfield  {journal} {\bibinfo  {journal} {JHEAp}\ }\textbf {\bibinfo {volume} {34}},\ \bibinfo {pages} {49} (\bibinfo {year} {2022})},\ \Eprint {https://arxiv.org/abs/2203.06142} {arXiv:2203.06142 [astro-ph.CO]} \BibitemShut {NoStop}%
\bibitem [{\citenamefont {Umilt\`a}\ \emph {et~al.}(2015)\citenamefont {Umilt\`a}, \citenamefont {Ballardini}, \citenamefont {Finelli},\ and\ \citenamefont {Paoletti}}]{Umilta:2015cta}%
  \BibitemOpen
  \bibfield  {author} {\bibinfo {author} {\bibfnamefont {C.}~\bibnamefont {Umilt\`a}}, \bibinfo {author} {\bibfnamefont {M.}~\bibnamefont {Ballardini}}, \bibinfo {author} {\bibfnamefont {F.}~\bibnamefont {Finelli}},\ and\ \bibinfo {author} {\bibfnamefont {D.}~\bibnamefont {Paoletti}},\ }\bibfield  {title} {\bibinfo {title} {{CMB and BAO constraints for an induced gravity dark energy model with a quartic potential}},\ }\href {https://doi.org/10.1088/1475-7516/2015/08/017} {\bibfield  {journal} {\bibinfo  {journal} {JCAP}\ }\textbf {\bibinfo {volume} {08}},\ \bibinfo {pages} {017}},\ \Eprint {https://arxiv.org/abs/1507.00718} {arXiv:1507.00718 [astro-ph.CO]} \BibitemShut {NoStop}%
\bibitem [{\citenamefont {Ballardini}\ \emph {et~al.}(2016)\citenamefont {Ballardini}, \citenamefont {Finelli}, \citenamefont {Umilt\`a},\ and\ \citenamefont {Paoletti}}]{Ballardini:2016cvy}%
  \BibitemOpen
  \bibfield  {author} {\bibinfo {author} {\bibfnamefont {M.}~\bibnamefont {Ballardini}}, \bibinfo {author} {\bibfnamefont {F.}~\bibnamefont {Finelli}}, \bibinfo {author} {\bibfnamefont {C.}~\bibnamefont {Umilt\`a}},\ and\ \bibinfo {author} {\bibfnamefont {D.}~\bibnamefont {Paoletti}},\ }\bibfield  {title} {\bibinfo {title} {{Cosmological constraints on induced gravity dark energy models}},\ }\href {https://doi.org/10.1088/1475-7516/2016/05/067} {\bibfield  {journal} {\bibinfo  {journal} {JCAP}\ }\textbf {\bibinfo {volume} {05}},\ \bibinfo {pages} {067}},\ \Eprint {https://arxiv.org/abs/1601.03387} {arXiv:1601.03387 [astro-ph.CO]} \BibitemShut {NoStop}%
\bibitem [{\citenamefont {Rossi}\ \emph {et~al.}(2019)\citenamefont {Rossi}, \citenamefont {Ballardini}, \citenamefont {Braglia}, \citenamefont {Finelli}, \citenamefont {Paoletti}, \citenamefont {Starobinsky},\ and\ \citenamefont {Umilt\`a}}]{Rossi:2019lgt}%
  \BibitemOpen
  \bibfield  {author} {\bibinfo {author} {\bibfnamefont {M.}~\bibnamefont {Rossi}}, \bibinfo {author} {\bibfnamefont {M.}~\bibnamefont {Ballardini}}, \bibinfo {author} {\bibfnamefont {M.}~\bibnamefont {Braglia}}, \bibinfo {author} {\bibfnamefont {F.}~\bibnamefont {Finelli}}, \bibinfo {author} {\bibfnamefont {D.}~\bibnamefont {Paoletti}}, \bibinfo {author} {\bibfnamefont {A.~A.}\ \bibnamefont {Starobinsky}},\ and\ \bibinfo {author} {\bibfnamefont {C.}~\bibnamefont {Umilt\`a}},\ }\bibfield  {title} {\bibinfo {title} {{Cosmological constraints on post-Newtonian parameters in effectively massless scalar-tensor theories of gravity}},\ }\href {https://doi.org/10.1103/PhysRevD.100.103524} {\bibfield  {journal} {\bibinfo  {journal} {Phys. Rev. D}\ }\textbf {\bibinfo {volume} {100}},\ \bibinfo {pages} {103524} (\bibinfo {year} {2019})},\ \Eprint {https://arxiv.org/abs/1906.10218} {arXiv:1906.10218 [astro-ph.CO]} \BibitemShut {NoStop}%
\bibitem [{\citenamefont {Ballesteros}\ \emph {et~al.}(2020)\citenamefont {Ballesteros}, \citenamefont {Notari},\ and\ \citenamefont {Rompineve}}]{Ballesteros:2020sik}%
  \BibitemOpen
  \bibfield  {author} {\bibinfo {author} {\bibfnamefont {G.}~\bibnamefont {Ballesteros}}, \bibinfo {author} {\bibfnamefont {A.}~\bibnamefont {Notari}},\ and\ \bibinfo {author} {\bibfnamefont {F.}~\bibnamefont {Rompineve}},\ }\bibfield  {title} {\bibinfo {title} {{The $H_0$ tension: $\Delta G_N$ vs. $\Delta N_{\rm eff}$}},\ }\href {https://doi.org/10.1088/1475-7516/2020/11/024} {\bibfield  {journal} {\bibinfo  {journal} {JCAP}\ }\textbf {\bibinfo {volume} {11}},\ \bibinfo {pages} {024}},\ \Eprint {https://arxiv.org/abs/2004.05049} {arXiv:2004.05049 [astro-ph.CO]} \BibitemShut {NoStop}%
\bibitem [{\citenamefont {Zumalacarregui}(2020)}]{Zumalacarregui:2020cjh}%
  \BibitemOpen
  \bibfield  {author} {\bibinfo {author} {\bibfnamefont {M.}~\bibnamefont {Zumalacarregui}},\ }\bibfield  {title} {\bibinfo {title} {{Gravity in the Era of Equality: Towards solutions to the Hubble problem without fine-tuned initial conditions}},\ }\href {https://doi.org/10.1103/PhysRevD.102.023523} {\bibfield  {journal} {\bibinfo  {journal} {Phys. Rev. D}\ }\textbf {\bibinfo {volume} {102}},\ \bibinfo {pages} {023523} (\bibinfo {year} {2020})},\ \Eprint {https://arxiv.org/abs/2003.06396} {arXiv:2003.06396 [astro-ph.CO]} \BibitemShut {NoStop}%
\bibitem [{\citenamefont {Braglia}\ \emph {et~al.}(2020)\citenamefont {Braglia}, \citenamefont {Ballardini}, \citenamefont {Emond}, \citenamefont {Finelli}, \citenamefont {Gumrukcuoglu}, \citenamefont {Koyama},\ and\ \citenamefont {Paoletti}}]{Braglia:2020iik}%
  \BibitemOpen
  \bibfield  {author} {\bibinfo {author} {\bibfnamefont {M.}~\bibnamefont {Braglia}}, \bibinfo {author} {\bibfnamefont {M.}~\bibnamefont {Ballardini}}, \bibinfo {author} {\bibfnamefont {W.~T.}\ \bibnamefont {Emond}}, \bibinfo {author} {\bibfnamefont {F.}~\bibnamefont {Finelli}}, \bibinfo {author} {\bibfnamefont {A.~E.}\ \bibnamefont {Gumrukcuoglu}}, \bibinfo {author} {\bibfnamefont {K.}~\bibnamefont {Koyama}},\ and\ \bibinfo {author} {\bibfnamefont {D.}~\bibnamefont {Paoletti}},\ }\bibfield  {title} {\bibinfo {title} {{Larger value for $H_0$ by an evolving gravitational constant}},\ }\href {https://doi.org/10.1103/PhysRevD.102.023529} {\bibfield  {journal} {\bibinfo  {journal} {Phys. Rev. D}\ }\textbf {\bibinfo {volume} {102}},\ \bibinfo {pages} {023529} (\bibinfo {year} {2020})},\ \Eprint {https://arxiv.org/abs/2004.11161} {arXiv:2004.11161 [astro-ph.CO]} \BibitemShut {NoStop}%
\bibitem [{\citenamefont {Ballardini}\ \emph {et~al.}(2020)\citenamefont {Ballardini}, \citenamefont {Braglia}, \citenamefont {Finelli}, \citenamefont {Paoletti}, \citenamefont {Starobinsky},\ and\ \citenamefont {Umilt\`a}}]{Ballardini:2020iws}%
  \BibitemOpen
  \bibfield  {author} {\bibinfo {author} {\bibfnamefont {M.}~\bibnamefont {Ballardini}}, \bibinfo {author} {\bibfnamefont {M.}~\bibnamefont {Braglia}}, \bibinfo {author} {\bibfnamefont {F.}~\bibnamefont {Finelli}}, \bibinfo {author} {\bibfnamefont {D.}~\bibnamefont {Paoletti}}, \bibinfo {author} {\bibfnamefont {A.~A.}\ \bibnamefont {Starobinsky}},\ and\ \bibinfo {author} {\bibfnamefont {C.}~\bibnamefont {Umilt\`a}},\ }\bibfield  {title} {\bibinfo {title} {{Scalar-tensor theories of gravity, neutrino physics, and the $H_0$ tension}},\ }\href {https://doi.org/10.1088/1475-7516/2020/10/044} {\bibfield  {journal} {\bibinfo  {journal} {JCAP}\ }\textbf {\bibinfo {volume} {10}},\ \bibinfo {pages} {044}},\ \Eprint {https://arxiv.org/abs/2004.14349} {arXiv:2004.14349 [astro-ph.CO]} \BibitemShut {NoStop}%
\bibitem [{\citenamefont {Braglia}\ \emph {et~al.}(2021)\citenamefont {Braglia}, \citenamefont {Ballardini}, \citenamefont {Finelli},\ and\ \citenamefont {Koyama}}]{Braglia:2020auw}%
  \BibitemOpen
  \bibfield  {author} {\bibinfo {author} {\bibfnamefont {M.}~\bibnamefont {Braglia}}, \bibinfo {author} {\bibfnamefont {M.}~\bibnamefont {Ballardini}}, \bibinfo {author} {\bibfnamefont {F.}~\bibnamefont {Finelli}},\ and\ \bibinfo {author} {\bibfnamefont {K.}~\bibnamefont {Koyama}},\ }\bibfield  {title} {\bibinfo {title} {{Early modified gravity in light of the $H_0$ tension and LSS data}},\ }\href {https://doi.org/10.1103/PhysRevD.103.043528} {\bibfield  {journal} {\bibinfo  {journal} {Phys. Rev. D}\ }\textbf {\bibinfo {volume} {103}},\ \bibinfo {pages} {043528} (\bibinfo {year} {2021})},\ \Eprint {https://arxiv.org/abs/2011.12934} {arXiv:2011.12934 [astro-ph.CO]} \BibitemShut {NoStop}%
\bibitem [{\citenamefont {Lee}\ \emph {et~al.}(2022)\citenamefont {Lee}, \citenamefont {Lee}, \citenamefont {Colg\'ain}, \citenamefont {Sheikh-Jabbari},\ and\ \citenamefont {Thakur}}]{Lee:2022cyh}%
  \BibitemOpen
  \bibfield  {author} {\bibinfo {author} {\bibfnamefont {B.-H.}\ \bibnamefont {Lee}}, \bibinfo {author} {\bibfnamefont {W.}~\bibnamefont {Lee}}, \bibinfo {author} {\bibfnamefont {E.~O.}\ \bibnamefont {Colg\'ain}}, \bibinfo {author} {\bibfnamefont {M.~M.}\ \bibnamefont {Sheikh-Jabbari}},\ and\ \bibinfo {author} {\bibfnamefont {S.}~\bibnamefont {Thakur}},\ }\bibfield  {title} {\bibinfo {title} {{Is local H $_{0}$ at odds with dark energy EFT?}},\ }\href {https://doi.org/10.1088/1475-7516/2022/04/004} {\bibfield  {journal} {\bibinfo  {journal} {JCAP}\ }\textbf {\bibinfo {volume} {04}}\bibfield  {number} {\bibinfo  {number} { (04)},\ \bibinfo {pages} {004}},\ }\Eprint {https://arxiv.org/abs/2202.03906} {arXiv:2202.03906 [astro-ph.CO]} \BibitemShut {NoStop}%
\bibitem [{\citenamefont {Bertotti}\ \emph {et~al.}(2003)\citenamefont {Bertotti}, \citenamefont {Iess},\ and\ \citenamefont {Tortora}}]{Bertotti:2003rm}%
  \BibitemOpen
  \bibfield  {author} {\bibinfo {author} {\bibfnamefont {B.}~\bibnamefont {Bertotti}}, \bibinfo {author} {\bibfnamefont {L.}~\bibnamefont {Iess}},\ and\ \bibinfo {author} {\bibfnamefont {P.}~\bibnamefont {Tortora}},\ }\bibfield  {title} {\bibinfo {title} {{A test of general relativity using radio links with the Cassini spacecraft}},\ }\href {https://doi.org/10.1038/nature01997} {\bibfield  {journal} {\bibinfo  {journal} {Nature}\ }\textbf {\bibinfo {volume} {425}},\ \bibinfo {pages} {374} (\bibinfo {year} {2003})}\BibitemShut {NoStop}%
\bibitem [{\citenamefont {Nicolis}\ \emph {et~al.}(2009)\citenamefont {Nicolis}, \citenamefont {Rattazzi},\ and\ \citenamefont {Trincherini}}]{Nicolis:2008in}%
  \BibitemOpen
  \bibfield  {author} {\bibinfo {author} {\bibfnamefont {A.}~\bibnamefont {Nicolis}}, \bibinfo {author} {\bibfnamefont {R.}~\bibnamefont {Rattazzi}},\ and\ \bibinfo {author} {\bibfnamefont {E.}~\bibnamefont {Trincherini}},\ }\bibfield  {title} {\bibinfo {title} {{The Galileon as a local modification of gravity}},\ }\href {https://doi.org/10.1103/PhysRevD.79.064036} {\bibfield  {journal} {\bibinfo  {journal} {Phys. Rev. D}\ }\textbf {\bibinfo {volume} {79}},\ \bibinfo {pages} {064036} (\bibinfo {year} {2009})},\ \Eprint {https://arxiv.org/abs/0811.2197} {arXiv:0811.2197 [hep-th]} \BibitemShut {NoStop}%
\bibitem [{\citenamefont {Deffayet}\ \emph {et~al.}(2009)\citenamefont {Deffayet}, \citenamefont {Esposito-Farese},\ and\ \citenamefont {Vikman}}]{Deffayet:2009wt}%
  \BibitemOpen
  \bibfield  {author} {\bibinfo {author} {\bibfnamefont {C.}~\bibnamefont {Deffayet}}, \bibinfo {author} {\bibfnamefont {G.}~\bibnamefont {Esposito-Farese}},\ and\ \bibinfo {author} {\bibfnamefont {A.}~\bibnamefont {Vikman}},\ }\bibfield  {title} {\bibinfo {title} {{Covariant Galileon}},\ }\href {https://doi.org/10.1103/PhysRevD.79.084003} {\bibfield  {journal} {\bibinfo  {journal} {Phys. Rev. D}\ }\textbf {\bibinfo {volume} {79}},\ \bibinfo {pages} {084003} (\bibinfo {year} {2009})},\ \Eprint {https://arxiv.org/abs/0901.1314} {arXiv:0901.1314 [hep-th]} \BibitemShut {NoStop}%
\bibitem [{\citenamefont {Kobayashi}\ \emph {et~al.}(2010)\citenamefont {Kobayashi}, \citenamefont {Tashiro},\ and\ \citenamefont {Suzuki}}]{Kobayashi:2009wr}%
  \BibitemOpen
  \bibfield  {author} {\bibinfo {author} {\bibfnamefont {T.}~\bibnamefont {Kobayashi}}, \bibinfo {author} {\bibfnamefont {H.}~\bibnamefont {Tashiro}},\ and\ \bibinfo {author} {\bibfnamefont {D.}~\bibnamefont {Suzuki}},\ }\bibfield  {title} {\bibinfo {title} {{Evolution of linear cosmological perturbations and its observational implications in Galileon-type modified gravity}},\ }\href {https://doi.org/10.1103/PhysRevD.81.063513} {\bibfield  {journal} {\bibinfo  {journal} {Phys. Rev. D}\ }\textbf {\bibinfo {volume} {81}},\ \bibinfo {pages} {063513} (\bibinfo {year} {2010})},\ \Eprint {https://arxiv.org/abs/0912.4641} {arXiv:0912.4641 [astro-ph.CO]} \BibitemShut {NoStop}%
\bibitem [{\citenamefont {Barreira}\ \emph {et~al.}(2014)\citenamefont {Barreira}, \citenamefont {Li}, \citenamefont {Baugh},\ and\ \citenamefont {Pascoli}}]{Barreira:2014jha}%
  \BibitemOpen
  \bibfield  {author} {\bibinfo {author} {\bibfnamefont {A.}~\bibnamefont {Barreira}}, \bibinfo {author} {\bibfnamefont {B.}~\bibnamefont {Li}}, \bibinfo {author} {\bibfnamefont {C.}~\bibnamefont {Baugh}},\ and\ \bibinfo {author} {\bibfnamefont {S.}~\bibnamefont {Pascoli}},\ }\bibfield  {title} {\bibinfo {title} {{The observational status of Galileon gravity after Planck}},\ }\href {https://doi.org/10.1088/1475-7516/2014/08/059} {\bibfield  {journal} {\bibinfo  {journal} {JCAP}\ }\textbf {\bibinfo {volume} {08}},\ \bibinfo {pages} {059}},\ \Eprint {https://arxiv.org/abs/1406.0485} {arXiv:1406.0485 [astro-ph.CO]} \BibitemShut {NoStop}%
\bibitem [{\citenamefont {Renk}\ \emph {et~al.}(2016)\citenamefont {Renk}, \citenamefont {Zumalacarregui},\ and\ \citenamefont {Montanari}}]{Renk:2016olm}%
  \BibitemOpen
  \bibfield  {author} {\bibinfo {author} {\bibfnamefont {J.}~\bibnamefont {Renk}}, \bibinfo {author} {\bibfnamefont {M.}~\bibnamefont {Zumalacarregui}},\ and\ \bibinfo {author} {\bibfnamefont {F.}~\bibnamefont {Montanari}},\ }\bibfield  {title} {\bibinfo {title} {{Gravity at the horizon: on relativistic effects, CMB-LSS correlations and ultra-large scales in Horndeski's theory}},\ }\href {https://doi.org/10.1088/1475-7516/2016/07/040} {\bibfield  {journal} {\bibinfo  {journal} {JCAP}\ }\textbf {\bibinfo {volume} {07}},\ \bibinfo {pages} {040}},\ \Eprint {https://arxiv.org/abs/1604.03487} {arXiv:1604.03487 [astro-ph.CO]} \BibitemShut {NoStop}%
\bibitem [{\citenamefont {Frusciante}\ \emph {et~al.}(2020)\citenamefont {Frusciante}, \citenamefont {Peirone}, \citenamefont {Atayde},\ and\ \citenamefont {De~Felice}}]{Frusciante:2019puu}%
  \BibitemOpen
  \bibfield  {author} {\bibinfo {author} {\bibfnamefont {N.}~\bibnamefont {Frusciante}}, \bibinfo {author} {\bibfnamefont {S.}~\bibnamefont {Peirone}}, \bibinfo {author} {\bibfnamefont {L.}~\bibnamefont {Atayde}},\ and\ \bibinfo {author} {\bibfnamefont {A.}~\bibnamefont {De~Felice}},\ }\bibfield  {title} {\bibinfo {title} {{Phenomenology of the generalized cubic covariant Galileon model and cosmological bounds}},\ }\href {https://doi.org/10.1103/PhysRevD.101.064001} {\bibfield  {journal} {\bibinfo  {journal} {Phys. Rev. D}\ }\textbf {\bibinfo {volume} {101}},\ \bibinfo {pages} {064001} (\bibinfo {year} {2020})},\ \Eprint {https://arxiv.org/abs/1912.07586} {arXiv:1912.07586 [astro-ph.CO]} \BibitemShut {NoStop}%
\bibitem [{\citenamefont {Horndeski}(1974)}]{Horndeski:1974wa}%
  \BibitemOpen
  \bibfield  {author} {\bibinfo {author} {\bibfnamefont {G.~W.}\ \bibnamefont {Horndeski}},\ }\bibfield  {title} {\bibinfo {title} {{Second-order scalar-tensor field equations in a four-dimensional space}},\ }\href {https://doi.org/10.1007/BF01807638} {\bibfield  {journal} {\bibinfo  {journal} {Int. J. Theor. Phys.}\ }\textbf {\bibinfo {volume} {10}},\ \bibinfo {pages} {363} (\bibinfo {year} {1974})}\BibitemShut {NoStop}%
\bibitem [{\citenamefont {Deffayet}\ \emph {et~al.}(2011)\citenamefont {Deffayet}, \citenamefont {Gao}, \citenamefont {Steer},\ and\ \citenamefont {Zahariade}}]{Deffayet:2011gz}%
  \BibitemOpen
  \bibfield  {author} {\bibinfo {author} {\bibfnamefont {C.}~\bibnamefont {Deffayet}}, \bibinfo {author} {\bibfnamefont {X.}~\bibnamefont {Gao}}, \bibinfo {author} {\bibfnamefont {D.~A.}\ \bibnamefont {Steer}},\ and\ \bibinfo {author} {\bibfnamefont {G.}~\bibnamefont {Zahariade}},\ }\bibfield  {title} {\bibinfo {title} {{From k-essence to generalised Galileons}},\ }\href {https://doi.org/10.1103/PhysRevD.84.064039} {\bibfield  {journal} {\bibinfo  {journal} {Phys. Rev. D}\ }\textbf {\bibinfo {volume} {84}},\ \bibinfo {pages} {064039} (\bibinfo {year} {2011})},\ \Eprint {https://arxiv.org/abs/1103.3260} {arXiv:1103.3260 [hep-th]} \BibitemShut {NoStop}%
\bibitem [{\citenamefont {Kobayashi}\ \emph {et~al.}(2011)\citenamefont {Kobayashi}, \citenamefont {Yamaguchi},\ and\ \citenamefont {Yokoyama}}]{Kobayashi:2011nu}%
  \BibitemOpen
  \bibfield  {author} {\bibinfo {author} {\bibfnamefont {T.}~\bibnamefont {Kobayashi}}, \bibinfo {author} {\bibfnamefont {M.}~\bibnamefont {Yamaguchi}},\ and\ \bibinfo {author} {\bibfnamefont {J.}~\bibnamefont {Yokoyama}},\ }\bibfield  {title} {\bibinfo {title} {{Generalized G-inflation: Inflation with the most general second-order field equations}},\ }\href {https://doi.org/10.1143/PTP.126.511} {\bibfield  {journal} {\bibinfo  {journal} {Prog. Theor. Phys.}\ }\textbf {\bibinfo {volume} {126}},\ \bibinfo {pages} {511} (\bibinfo {year} {2011})},\ \Eprint {https://arxiv.org/abs/1105.5723} {arXiv:1105.5723 [hep-th]} \BibitemShut {NoStop}%
\bibitem [{\citenamefont {Kase}\ and\ \citenamefont {Tsujikawa}(2019)}]{Kase:2018aps}%
  \BibitemOpen
  \bibfield  {author} {\bibinfo {author} {\bibfnamefont {R.}~\bibnamefont {Kase}}\ and\ \bibinfo {author} {\bibfnamefont {S.}~\bibnamefont {Tsujikawa}},\ }\bibfield  {title} {\bibinfo {title} {{Dark energy in Horndeski theories after GW170817: A review}},\ }\href {https://doi.org/10.1142/S0218271819420057} {\bibfield  {journal} {\bibinfo  {journal} {Int. J. Mod. Phys. D}\ }\textbf {\bibinfo {volume} {28}},\ \bibinfo {pages} {1942005} (\bibinfo {year} {2019})},\ \Eprint {https://arxiv.org/abs/1809.08735} {arXiv:1809.08735 [gr-qc]} \BibitemShut {NoStop}%
\bibitem [{\citenamefont {Abbott}\ \emph {et~al.}(2017)\citenamefont {Abbott} \emph {et~al.}}]{LIGOScientific:2017zic}%
  \BibitemOpen
  \bibfield  {author} {\bibinfo {author} {\bibfnamefont {B.~P.}\ \bibnamefont {Abbott}} \emph {et~al.} (\bibinfo {collaboration} {LIGO Scientific, Virgo, Fermi-GBM, INTEGRAL}),\ }\bibfield  {title} {\bibinfo {title} {{Gravitational Waves and Gamma-rays from a Binary Neutron Star Merger: GW170817 and GRB 170817A}},\ }\href {https://doi.org/10.3847/2041-8213/aa920c} {\bibfield  {journal} {\bibinfo  {journal} {Astrophys. J. Lett.}\ }\textbf {\bibinfo {volume} {848}},\ \bibinfo {pages} {L13} (\bibinfo {year} {2017})},\ \Eprint {https://arxiv.org/abs/1710.05834} {arXiv:1710.05834 [astro-ph.HE]} \BibitemShut {NoStop}%
\bibitem [{\citenamefont {Jordan}(1955)}]{jordan55}%
  \BibitemOpen
  \bibfield  {author} {\bibinfo {author} {\bibfnamefont {P.}~\bibnamefont {Jordan}},\ }\href {https://books.google.it/books?id=snJTcgAACAAJ} {\emph {\bibinfo {title} {Schwerkraft und Weltall: Grundlagen der theoretischen Kosmologie}}},\ Die Wissenschaft\ (\bibinfo  {publisher} {F. Vieweg},\ \bibinfo {year} {1955})\BibitemShut {NoStop}%
\bibitem [{\citenamefont {Brans}\ and\ \citenamefont {Dicke}(1961)}]{PhysRev.124.925}%
  \BibitemOpen
  \bibfield  {author} {\bibinfo {author} {\bibfnamefont {C.}~\bibnamefont {Brans}}\ and\ \bibinfo {author} {\bibfnamefont {R.~H.}\ \bibnamefont {Dicke}},\ }\bibfield  {title} {\bibinfo {title} {Mach's principle and a relativistic theory of gravitation},\ }\href {https://doi.org/10.1103/PhysRev.124.925} {\bibfield  {journal} {\bibinfo  {journal} {Phys. Rev.}\ }\textbf {\bibinfo {volume} {124}},\ \bibinfo {pages} {925} (\bibinfo {year} {1961})}\BibitemShut {NoStop}%
\bibitem [{\citenamefont {Zee}(1979)}]{PhysRevLett.42.417}%
  \BibitemOpen
  \bibfield  {author} {\bibinfo {author} {\bibfnamefont {A.}~\bibnamefont {Zee}},\ }\bibfield  {title} {\bibinfo {title} {Broken-symmetric theory of gravity},\ }\href {https://doi.org/10.1103/PhysRevLett.42.417} {\bibfield  {journal} {\bibinfo  {journal} {Phys. Rev. Lett.}\ }\textbf {\bibinfo {volume} {42}},\ \bibinfo {pages} {417} (\bibinfo {year} {1979})}\BibitemShut {NoStop}%
\bibitem [{\citenamefont {Silva}\ and\ \citenamefont {Koyama}(2009)}]{Silva:2009km}%
  \BibitemOpen
  \bibfield  {author} {\bibinfo {author} {\bibfnamefont {F.~P.}\ \bibnamefont {Silva}}\ and\ \bibinfo {author} {\bibfnamefont {K.}~\bibnamefont {Koyama}},\ }\bibfield  {title} {\bibinfo {title} {Self-accelerating universe in galileon cosmology},\ }\href {https://doi.org/10.1103/PhysRevD.80.121301} {\bibfield  {journal} {\bibinfo  {journal} {Phys. Rev. D}\ }\textbf {\bibinfo {volume} {80}},\ \bibinfo {pages} {121301} (\bibinfo {year} {2009})},\ \Eprint {https://arxiv.org/abs/0909.4538} {arXiv:0909.4538 [astro-ph.CO]} \BibitemShut {NoStop}%
\bibitem [{\citenamefont {Finelli}\ \emph {et~al.}(2008)\citenamefont {Finelli}, \citenamefont {Tronconi},\ and\ \citenamefont {Venturi}}]{Finelli:2007wb}%
  \BibitemOpen
  \bibfield  {author} {\bibinfo {author} {\bibfnamefont {F.}~\bibnamefont {Finelli}}, \bibinfo {author} {\bibfnamefont {A.}~\bibnamefont {Tronconi}},\ and\ \bibinfo {author} {\bibfnamefont {G.}~\bibnamefont {Venturi}},\ }\bibfield  {title} {\bibinfo {title} {{Dark Energy, Induced Gravity and Broken Scale Invariance}},\ }\href {https://doi.org/10.1016/j.physletb.2007.11.053} {\bibfield  {journal} {\bibinfo  {journal} {Phys. Lett. B}\ }\textbf {\bibinfo {volume} {659}},\ \bibinfo {pages} {466} (\bibinfo {year} {2008})},\ \Eprint {https://arxiv.org/abs/0710.2741} {arXiv:0710.2741 [astro-ph]} \BibitemShut {NoStop}%
\bibitem [{\citenamefont {Torres}(2002)}]{Torres:2002pe}%
  \BibitemOpen
  \bibfield  {author} {\bibinfo {author} {\bibfnamefont {D.~F.}\ \bibnamefont {Torres}},\ }\bibfield  {title} {\bibinfo {title} {{Quintessence, superquintessence and observable quantities in Brans-Dicke and nonminimally coupled theories}},\ }\href {https://doi.org/10.1103/PhysRevD.66.043522} {\bibfield  {journal} {\bibinfo  {journal} {Phys. Rev. D}\ }\textbf {\bibinfo {volume} {66}},\ \bibinfo {pages} {043522} (\bibinfo {year} {2002})},\ \Eprint {https://arxiv.org/abs/astro-ph/0204504} {arXiv:astro-ph/0204504} \BibitemShut {NoStop}%
\bibitem [{\citenamefont {Gannouji}\ \emph {et~al.}(2006)\citenamefont {Gannouji}, \citenamefont {Polarski}, \citenamefont {Ranquet},\ and\ \citenamefont {Starobinsky}}]{Gannouji:2006jm}%
  \BibitemOpen
  \bibfield  {author} {\bibinfo {author} {\bibfnamefont {R.}~\bibnamefont {Gannouji}}, \bibinfo {author} {\bibfnamefont {D.}~\bibnamefont {Polarski}}, \bibinfo {author} {\bibfnamefont {A.}~\bibnamefont {Ranquet}},\ and\ \bibinfo {author} {\bibfnamefont {A.~A.}\ \bibnamefont {Starobinsky}},\ }\bibfield  {title} {\bibinfo {title} {{Scalar-Tensor Models of Normal and Phantom Dark Energy}},\ }\href {https://doi.org/10.1088/1475-7516/2006/09/016} {\bibfield  {journal} {\bibinfo  {journal} {JCAP}\ }\textbf {\bibinfo {volume} {09}},\ \bibinfo {pages} {016}},\ \Eprint {https://arxiv.org/abs/astro-ph/0606287} {arXiv:astro-ph/0606287} \BibitemShut {NoStop}%
\bibitem [{\citenamefont {Cerioni}(2011)}]{CerioniPhd}%
  \BibitemOpen
  \bibfield  {author} {\bibinfo {author} {\bibfnamefont {A.}~\bibnamefont {Cerioni}},\ }\emph {\bibinfo {title} {Cosmological perturbations in generalized theories of gravity}},\ \href {http://amsdottorato.unibo.it/3562/} {Ph.D. thesis},\ \bibinfo  {school} {Alma Mater Studiorum - Università di Bologna} (\bibinfo {year} {2011})\BibitemShut {NoStop}%
\bibitem [{\citenamefont {Hohmann}(2015)}]{Hohmann:2015kra}%
  \BibitemOpen
  \bibfield  {author} {\bibinfo {author} {\bibfnamefont {M.}~\bibnamefont {Hohmann}},\ }\bibfield  {title} {\bibinfo {title} {{Parametrized post-Newtonian limit of Horndeski\textquoteright{}s gravity theory}},\ }\href {https://doi.org/10.1103/PhysRevD.92.064019} {\bibfield  {journal} {\bibinfo  {journal} {Phys. Rev. D}\ }\textbf {\bibinfo {volume} {92}},\ \bibinfo {pages} {064019} (\bibinfo {year} {2015})},\ \Eprint {https://arxiv.org/abs/1506.04253} {arXiv:1506.04253 [gr-qc]} \BibitemShut {NoStop}%
\bibitem [{\citenamefont {Quiros}\ \emph {et~al.}(2020)\citenamefont {Quiros}, \citenamefont {De~Arcia}, \citenamefont {Garc\'\i{}a-Salcedo}, \citenamefont {Gonzalez},\ and\ \citenamefont {Horta-Rangel}}]{Quiros:2019gbt}%
  \BibitemOpen
  \bibfield  {author} {\bibinfo {author} {\bibfnamefont {I.}~\bibnamefont {Quiros}}, \bibinfo {author} {\bibfnamefont {R.}~\bibnamefont {De~Arcia}}, \bibinfo {author} {\bibfnamefont {R.}~\bibnamefont {Garc\'\i{}a-Salcedo}}, \bibinfo {author} {\bibfnamefont {T.}~\bibnamefont {Gonzalez}},\ and\ \bibinfo {author} {\bibfnamefont {F.~A.}\ \bibnamefont {Horta-Rangel}},\ }\bibfield  {title} {\bibinfo {title} {{An issue with the classification of the scalar\textendash{}tensor theories of gravity}},\ }\href {https://doi.org/10.1142/S0218271820500479} {\bibfield  {journal} {\bibinfo  {journal} {Int. J. Mod. Phys. D}\ }\textbf {\bibinfo {volume} {29}},\ \bibinfo {pages} {2050047} (\bibinfo {year} {2020})},\ \Eprint {https://arxiv.org/abs/1905.08177} {arXiv:1905.08177 [gr-qc]} \BibitemShut {NoStop}%
\bibitem [{\citenamefont {Boisseau}\ \emph {et~al.}(2000)\citenamefont {Boisseau}, \citenamefont {Esposito-Farese}, \citenamefont {Polarski},\ and\ \citenamefont {Starobinsky}}]{Boisseau:2000pr}%
  \BibitemOpen
  \bibfield  {author} {\bibinfo {author} {\bibfnamefont {B.}~\bibnamefont {Boisseau}}, \bibinfo {author} {\bibfnamefont {G.}~\bibnamefont {Esposito-Farese}}, \bibinfo {author} {\bibfnamefont {D.}~\bibnamefont {Polarski}},\ and\ \bibinfo {author} {\bibfnamefont {A.~A.}\ \bibnamefont {Starobinsky}},\ }\bibfield  {title} {\bibinfo {title} {{Reconstruction of a scalar tensor theory of gravity in an accelerating universe}},\ }\href {https://doi.org/10.1103/PhysRevLett.85.2236} {\bibfield  {journal} {\bibinfo  {journal} {Phys. Rev. Lett.}\ }\textbf {\bibinfo {volume} {85}},\ \bibinfo {pages} {2236} (\bibinfo {year} {2000})},\ \Eprint {https://arxiv.org/abs/gr-qc/0001066} {arXiv:gr-qc/0001066} \BibitemShut {NoStop}%
\bibitem [{\citenamefont {Chow}\ and\ \citenamefont {Khoury}(2009)}]{Chow:2009fm}%
  \BibitemOpen
  \bibfield  {author} {\bibinfo {author} {\bibfnamefont {N.}~\bibnamefont {Chow}}\ and\ \bibinfo {author} {\bibfnamefont {J.}~\bibnamefont {Khoury}},\ }\bibfield  {title} {\bibinfo {title} {{Galileon Cosmology}},\ }\href {https://doi.org/10.1103/PhysRevD.80.024037} {\bibfield  {journal} {\bibinfo  {journal} {Phys. Rev. D}\ }\textbf {\bibinfo {volume} {80}},\ \bibinfo {pages} {024037} (\bibinfo {year} {2009})},\ \Eprint {https://arxiv.org/abs/0905.1325} {arXiv:0905.1325 [hep-th]} \BibitemShut {NoStop}%
\bibitem [{\citenamefont {Kimura}\ \emph {et~al.}(2012)\citenamefont {Kimura}, \citenamefont {Kobayashi},\ and\ \citenamefont {Yamamoto}}]{Kimura:2011dc}%
  \BibitemOpen
  \bibfield  {author} {\bibinfo {author} {\bibfnamefont {R.}~\bibnamefont {Kimura}}, \bibinfo {author} {\bibfnamefont {T.}~\bibnamefont {Kobayashi}},\ and\ \bibinfo {author} {\bibfnamefont {K.}~\bibnamefont {Yamamoto}},\ }\bibfield  {title} {\bibinfo {title} {{Vainshtein screening in a cosmological background in the most general second-order scalar-tensor theory}},\ }\href {https://doi.org/10.1103/PhysRevD.85.024023} {\bibfield  {journal} {\bibinfo  {journal} {Phys. Rev. D}\ }\textbf {\bibinfo {volume} {85}},\ \bibinfo {pages} {024023} (\bibinfo {year} {2012})},\ \Eprint {https://arxiv.org/abs/1111.6749} {arXiv:1111.6749 [astro-ph.CO]} \BibitemShut {NoStop}%
\bibitem [{\citenamefont {Bellini}\ and\ \citenamefont {Sawicki}(2014)}]{Bellini:2014fua}%
  \BibitemOpen
  \bibfield  {author} {\bibinfo {author} {\bibfnamefont {E.}~\bibnamefont {Bellini}}\ and\ \bibinfo {author} {\bibfnamefont {I.}~\bibnamefont {Sawicki}},\ }\bibfield  {title} {\bibinfo {title} {{Maximal freedom at minimum cost: linear large-scale structure in general modifications of gravity}},\ }\href {https://doi.org/10.1088/1475-7516/2014/07/050} {\bibfield  {journal} {\bibinfo  {journal} {JCAP}\ }\textbf {\bibinfo {volume} {07}},\ \bibinfo {pages} {050}},\ \Eprint {https://arxiv.org/abs/1404.3713} {arXiv:1404.3713 [astro-ph.CO]} \BibitemShut {NoStop}%
\bibitem [{\citenamefont {De~Felice}\ and\ \citenamefont {Tsujikawa}(2012)}]{DeFelice:2011bh}%
  \BibitemOpen
  \bibfield  {author} {\bibinfo {author} {\bibfnamefont {A.}~\bibnamefont {De~Felice}}\ and\ \bibinfo {author} {\bibfnamefont {S.}~\bibnamefont {Tsujikawa}},\ }\bibfield  {title} {\bibinfo {title} {{Conditions for the cosmological viability of the most general scalar-tensor theories and their applications to extended Galileon dark energy models}},\ }\href {https://doi.org/10.1088/1475-7516/2012/02/007} {\bibfield  {journal} {\bibinfo  {journal} {JCAP}\ }\textbf {\bibinfo {volume} {02}},\ \bibinfo {pages} {007}},\ \Eprint {https://arxiv.org/abs/1110.3878} {arXiv:1110.3878 [gr-qc]} \BibitemShut {NoStop}%
\bibitem [{\citenamefont {Babichev}\ \emph {et~al.}(2008)\citenamefont {Babichev}, \citenamefont {Mukhanov},\ and\ \citenamefont {Vikman}}]{Babichev:2007dw}%
  \BibitemOpen
  \bibfield  {author} {\bibinfo {author} {\bibfnamefont {E.}~\bibnamefont {Babichev}}, \bibinfo {author} {\bibfnamefont {V.}~\bibnamefont {Mukhanov}},\ and\ \bibinfo {author} {\bibfnamefont {A.}~\bibnamefont {Vikman}},\ }\bibfield  {title} {\bibinfo {title} {{k-Essence, superluminal propagation, causality and emergent geometry}},\ }\href {https://doi.org/10.1088/1126-6708/2008/02/101} {\bibfield  {journal} {\bibinfo  {journal} {JHEP}\ }\textbf {\bibinfo {volume} {02}},\ \bibinfo {pages} {101}},\ \Eprint {https://arxiv.org/abs/0708.0561} {arXiv:0708.0561 [hep-th]} \BibitemShut {NoStop}%
\bibitem [{\citenamefont {Gorini}\ \emph {et~al.}(2008)\citenamefont {Gorini}, \citenamefont {Kamenshchik}, \citenamefont {Moschella}, \citenamefont {Piattella},\ and\ \citenamefont {Starobinsky}}]{Gorini:2007ta}%
  \BibitemOpen
  \bibfield  {author} {\bibinfo {author} {\bibfnamefont {V.}~\bibnamefont {Gorini}}, \bibinfo {author} {\bibfnamefont {A.~Y.}\ \bibnamefont {Kamenshchik}}, \bibinfo {author} {\bibfnamefont {U.}~\bibnamefont {Moschella}}, \bibinfo {author} {\bibfnamefont {O.~F.}\ \bibnamefont {Piattella}},\ and\ \bibinfo {author} {\bibfnamefont {A.~A.}\ \bibnamefont {Starobinsky}},\ }\bibfield  {title} {\bibinfo {title} {{Gauge-invariant analysis of perturbations in Chaplygin gas unified models of dark matter and dark energy}},\ }\href {https://doi.org/10.1088/1475-7516/2008/02/016} {\bibfield  {journal} {\bibinfo  {journal} {JCAP}\ }\textbf {\bibinfo {volume} {02}},\ \bibinfo {pages} {016}},\ \Eprint {https://arxiv.org/abs/0711.4242} {arXiv:0711.4242 [astro-ph]} \BibitemShut {NoStop}%
\bibitem [{\citenamefont {Adams}\ \emph {et~al.}(2006)\citenamefont {Adams}, \citenamefont {Arkani-Hamed}, \citenamefont {Dubovsky}, \citenamefont {Nicolis},\ and\ \citenamefont {Rattazzi}}]{Adams:2006sv}%
  \BibitemOpen
  \bibfield  {author} {\bibinfo {author} {\bibfnamefont {A.}~\bibnamefont {Adams}}, \bibinfo {author} {\bibfnamefont {N.}~\bibnamefont {Arkani-Hamed}}, \bibinfo {author} {\bibfnamefont {S.}~\bibnamefont {Dubovsky}}, \bibinfo {author} {\bibfnamefont {A.}~\bibnamefont {Nicolis}},\ and\ \bibinfo {author} {\bibfnamefont {R.}~\bibnamefont {Rattazzi}},\ }\bibfield  {title} {\bibinfo {title} {{Causality, analyticity and an IR obstruction to UV completion}},\ }\href {https://doi.org/10.1088/1126-6708/2006/10/014} {\bibfield  {journal} {\bibinfo  {journal} {JHEP}\ }\textbf {\bibinfo {volume} {10}},\ \bibinfo {pages} {014}},\ \Eprint {https://arxiv.org/abs/hep-th/0602178} {arXiv:hep-th/0602178} \BibitemShut {NoStop}%
\bibitem [{\citenamefont {Bonvin}\ \emph {et~al.}(2006)\citenamefont {Bonvin}, \citenamefont {Caprini},\ and\ \citenamefont {Durrer}}]{Bonvin:2006vc}%
  \BibitemOpen
  \bibfield  {author} {\bibinfo {author} {\bibfnamefont {C.}~\bibnamefont {Bonvin}}, \bibinfo {author} {\bibfnamefont {C.}~\bibnamefont {Caprini}},\ and\ \bibinfo {author} {\bibfnamefont {R.}~\bibnamefont {Durrer}},\ }\bibfield  {title} {\bibinfo {title} {{A no-go theorem for k-essence dark energy}},\ }\href {https://doi.org/10.1103/PhysRevLett.97.081303} {\bibfield  {journal} {\bibinfo  {journal} {Phys. Rev. Lett.}\ }\textbf {\bibinfo {volume} {97}},\ \bibinfo {pages} {081303} (\bibinfo {year} {2006})},\ \Eprint {https://arxiv.org/abs/astro-ph/0606584} {arXiv:astro-ph/0606584} \BibitemShut {NoStop}%
\bibitem [{\citenamefont {Ellis}\ \emph {et~al.}(2007)\citenamefont {Ellis}, \citenamefont {Maartens},\ and\ \citenamefont {MacCallum}}]{Ellis:2007ic}%
  \BibitemOpen
  \bibfield  {author} {\bibinfo {author} {\bibfnamefont {G.}~\bibnamefont {Ellis}}, \bibinfo {author} {\bibfnamefont {R.}~\bibnamefont {Maartens}},\ and\ \bibinfo {author} {\bibfnamefont {M.~A.~H.}\ \bibnamefont {MacCallum}},\ }\bibfield  {title} {\bibinfo {title} {{Causality and the speed of sound}},\ }\href {https://doi.org/10.1007/s10714-007-0479-2} {\bibfield  {journal} {\bibinfo  {journal} {Gen. Rel. Grav.}\ }\textbf {\bibinfo {volume} {39}},\ \bibinfo {pages} {1651} (\bibinfo {year} {2007})},\ \Eprint {https://arxiv.org/abs/gr-qc/0703121} {arXiv:gr-qc/0703121} \BibitemShut {NoStop}%
\bibitem [{\citenamefont {Ma}\ and\ \citenamefont {Bertschinger}(1995)}]{Ma_1995}%
  \BibitemOpen
  \bibfield  {author} {\bibinfo {author} {\bibfnamefont {C.-P.}\ \bibnamefont {Ma}}\ and\ \bibinfo {author} {\bibfnamefont {E.}~\bibnamefont {Bertschinger}},\ }\bibfield  {title} {\bibinfo {title} {Cosmological perturbation theory in the synchronous and conformal newtonian gauges},\ }\href {https://doi.org/10.1086/176550} {\bibfield  {journal} {\bibinfo  {journal} {The Astrophysical Journal}\ }\textbf {\bibinfo {volume} {455}},\ \bibinfo {pages} {7} (\bibinfo {year} {1995})}\BibitemShut {NoStop}%
\bibitem [{\citenamefont {Paoletti}\ \emph {et~al.}(2019)\citenamefont {Paoletti}, \citenamefont {Braglia}, \citenamefont {Finelli}, \citenamefont {Ballardini},\ and\ \citenamefont {Umilt\`a}}]{Paoletti:2018xet}%
  \BibitemOpen
  \bibfield  {author} {\bibinfo {author} {\bibfnamefont {D.}~\bibnamefont {Paoletti}}, \bibinfo {author} {\bibfnamefont {M.}~\bibnamefont {Braglia}}, \bibinfo {author} {\bibfnamefont {F.}~\bibnamefont {Finelli}}, \bibinfo {author} {\bibfnamefont {M.}~\bibnamefont {Ballardini}},\ and\ \bibinfo {author} {\bibfnamefont {C.}~\bibnamefont {Umilt\`a}},\ }\bibfield  {title} {\bibinfo {title} {{Isocurvature fluctuations in the effective Newton\textquoteright{}s constant}},\ }\href {https://doi.org/10.1016/j.dark.2019.100307} {\bibfield  {journal} {\bibinfo  {journal} {Phys. Dark Univ.}\ }\textbf {\bibinfo {volume} {25}},\ \bibinfo {pages} {100307} (\bibinfo {year} {2019})},\ \Eprint {https://arxiv.org/abs/1809.03201} {arXiv:1809.03201 [astro-ph.CO]} \BibitemShut {NoStop}%
\bibitem [{\citenamefont {Ballardini}\ \emph {et~al.}(2023)\citenamefont {Ballardini}, \citenamefont {Ferrari},\ and\ \citenamefont {Finelli}}]{Ballardini:2023mzm}%
  \BibitemOpen
  \bibfield  {author} {\bibinfo {author} {\bibfnamefont {M.}~\bibnamefont {Ballardini}}, \bibinfo {author} {\bibfnamefont {A.~G.}\ \bibnamefont {Ferrari}},\ and\ \bibinfo {author} {\bibfnamefont {F.}~\bibnamefont {Finelli}},\ }\bibfield  {title} {\bibinfo {title} {{Phantom scalar-tensor models and cosmological tensions}},\ }\href {https://doi.org/10.1088/1475-7516/2023/04/029} {\bibfield  {journal} {\bibinfo  {journal} {JCAP}\ }\textbf {\bibinfo {volume} {04}},\ \bibinfo {pages} {029}},\ \Eprint {https://arxiv.org/abs/2302.05291} {arXiv:2302.05291 [astro-ph.CO]} \BibitemShut {NoStop}%
\bibitem [{\citenamefont {{Lesgourgues}}(2011)}]{class1}%
  \BibitemOpen
  \bibfield  {author} {\bibinfo {author} {\bibfnamefont {J.}~\bibnamefont {{Lesgourgues}}},\ }\bibfield  {title} {\bibinfo {title} {{The Cosmic Linear Anisotropy Solving System (CLASS) I: Overview}},\ }\href@noop {} {\bibfield  {journal} {\bibinfo  {journal} {arXiv e-prints}\ ,\ \bibinfo {eid} {arXiv:1104.2932}} (\bibinfo {year} {2011})},\ \Eprint {https://arxiv.org/abs/1104.2932} {arXiv:1104.2932 [astro-ph.IM]} \BibitemShut {NoStop}%
\bibitem [{\citenamefont {Audren}\ \emph {et~al.}(2013)\citenamefont {Audren}, \citenamefont {Lesgourgues}, \citenamefont {Benabed},\ and\ \citenamefont {Prunet}}]{Audren_2013}%
  \BibitemOpen
  \bibfield  {author} {\bibinfo {author} {\bibfnamefont {B.}~\bibnamefont {Audren}}, \bibinfo {author} {\bibfnamefont {J.}~\bibnamefont {Lesgourgues}}, \bibinfo {author} {\bibfnamefont {K.}~\bibnamefont {Benabed}},\ and\ \bibinfo {author} {\bibfnamefont {S.}~\bibnamefont {Prunet}},\ }\bibfield  {title} {\bibinfo {title} {Conservative constraints on early cosmology: an illustration of the monte python cosmological parameter inference code},\ }\href {https://doi.org/10.1088/1475-7516/2013/02/001} {\bibfield  {journal} {\bibinfo  {journal} {Journal of Cosmology and Astroparticle Physics}\ }\textbf {\bibinfo {volume} {2013}}\bibinfo  {number} { (02)},\ \bibinfo {pages} {001}}\BibitemShut {NoStop}%
\bibitem [{\citenamefont {Brinckmann}\ and\ \citenamefont {Lesgourgues}(2019)}]{Brinckmann:2018cvx}%
  \BibitemOpen
\bibfield  {number} {  }\bibfield  {author} {\bibinfo {author} {\bibfnamefont {T.}~\bibnamefont {Brinckmann}}\ and\ \bibinfo {author} {\bibfnamefont {J.}~\bibnamefont {Lesgourgues}},\ }\bibfield  {title} {\bibinfo {title} {{MontePython 3: boosted MCMC sampler and other features}},\ }\href {https://doi.org/10.1016/j.dark.2018.100260} {\bibfield  {journal} {\bibinfo  {journal} {Phys. Dark Univ.}\ }\textbf {\bibinfo {volume} {24}},\ \bibinfo {pages} {100260} (\bibinfo {year} {2019})},\ \Eprint {https://arxiv.org/abs/1804.07261} {arXiv:1804.07261 [astro-ph.CO]} \BibitemShut {NoStop}%
\bibitem [{\citenamefont {Gelman}\ and\ \citenamefont {Rubin}(1992)}]{Gelman:1992zz}%
  \BibitemOpen
  \bibfield  {author} {\bibinfo {author} {\bibfnamefont {A.}~\bibnamefont {Gelman}}\ and\ \bibinfo {author} {\bibfnamefont {D.~B.}\ \bibnamefont {Rubin}},\ }\bibfield  {title} {\bibinfo {title} {{Inference from Iterative Simulation Using Multiple Sequences}},\ }\href {https://doi.org/10.1214/ss/1177011136} {\bibfield  {journal} {\bibinfo  {journal} {Statist. Sci.}\ }\textbf {\bibinfo {volume} {7}},\ \bibinfo {pages} {457} (\bibinfo {year} {1992})}\BibitemShut {NoStop}%
\bibitem [{\citenamefont {{Lewis}}(2019)}]{Lewis:2019xzd}%
  \BibitemOpen
  \bibfield  {author} {\bibinfo {author} {\bibfnamefont {A.}~\bibnamefont {{Lewis}}},\ }\bibfield  {title} {\bibinfo {title} {{GetDist: a Python package for analysing Monte Carlo samples}},\ }\href@noop {} {\bibfield  {journal} {\bibinfo  {journal} {arXiv e-prints}\ ,\ \bibinfo {eid} {arXiv:1910.13970}} (\bibinfo {year} {2019})},\ \Eprint {https://arxiv.org/abs/1910.13970} {arXiv:1910.13970 [astro-ph.IM]} \BibitemShut {NoStop}%
\bibitem [{\citenamefont {Aghanim}\ \emph {et~al.}(2020{\natexlab{d}})\citenamefont {Aghanim} \emph {et~al.}}]{Planck:2019nip}%
  \BibitemOpen
  \bibfield  {author} {\bibinfo {author} {\bibfnamefont {N.}~\bibnamefont {Aghanim}} \emph {et~al.} (\bibinfo {collaboration} {Planck}),\ }\bibfield  {title} {\bibinfo {title} {{Planck 2018 results. V. CMB power spectra and likelihoods}},\ }\href {https://doi.org/10.1051/0004-6361/201936386} {\bibfield  {journal} {\bibinfo  {journal} {Astron. Astrophys.}\ }\textbf {\bibinfo {volume} {641}},\ \bibinfo {pages} {A5} (\bibinfo {year} {2020}{\natexlab{d}})},\ \Eprint {https://arxiv.org/abs/1907.12875} {arXiv:1907.12875 [astro-ph.CO]} \BibitemShut {NoStop}%
\bibitem [{\citenamefont {Alam}\ \emph {et~al.}(2017)\citenamefont {Alam} \emph {et~al.}}]{BOSS:2016wmc}%
  \BibitemOpen
  \bibfield  {author} {\bibinfo {author} {\bibfnamefont {S.}~\bibnamefont {Alam}} \emph {et~al.} (\bibinfo {collaboration} {BOSS}),\ }\bibfield  {title} {\bibinfo {title} {{The clustering of galaxies in the completed SDSS-III Baryon Oscillation Spectroscopic Survey: cosmological analysis of the DR12 galaxy sample}},\ }\href {https://doi.org/10.1093/mnras/stx721} {\bibfield  {journal} {\bibinfo  {journal} {Mon. Not. Roy. Astron. Soc.}\ }\textbf {\bibinfo {volume} {470}},\ \bibinfo {pages} {2617} (\bibinfo {year} {2017})},\ \Eprint {https://arxiv.org/abs/1607.03155} {arXiv:1607.03155 [astro-ph.CO]} \BibitemShut {NoStop}%
\bibitem [{\citenamefont {Beutler}\ \emph {et~al.}(2011)\citenamefont {Beutler}, \citenamefont {Blake}, \citenamefont {Colless}, \citenamefont {Jones}, \citenamefont {Staveley-Smith}, \citenamefont {Campbell}, \citenamefont {Parker}, \citenamefont {Saunders},\ and\ \citenamefont {Watson}}]{Beutler_2011}%
  \BibitemOpen
  \bibfield  {author} {\bibinfo {author} {\bibfnamefont {F.}~\bibnamefont {Beutler}}, \bibinfo {author} {\bibfnamefont {C.}~\bibnamefont {Blake}}, \bibinfo {author} {\bibfnamefont {M.}~\bibnamefont {Colless}}, \bibinfo {author} {\bibfnamefont {D.~H.}\ \bibnamefont {Jones}}, \bibinfo {author} {\bibfnamefont {L.}~\bibnamefont {Staveley-Smith}}, \bibinfo {author} {\bibfnamefont {L.}~\bibnamefont {Campbell}}, \bibinfo {author} {\bibfnamefont {Q.}~\bibnamefont {Parker}}, \bibinfo {author} {\bibfnamefont {W.}~\bibnamefont {Saunders}},\ and\ \bibinfo {author} {\bibfnamefont {F.}~\bibnamefont {Watson}},\ }\bibfield  {title} {\bibinfo {title} {The 6df galaxy survey: baryon acoustic oscillations and the local hubble constant},\ }\href {https://doi.org/10.1111/j.1365-2966.2011.19250.x} {\bibfield  {journal} {\bibinfo  {journal} {Monthly Notices of the Royal Astronomical Society}\ }\textbf {\bibinfo {volume} {416}},\ \bibinfo {pages} {3017} (\bibinfo {year} {2011})}\BibitemShut {NoStop}%
\bibitem [{\citenamefont {Ross}\ \emph {et~al.}(2015)\citenamefont {Ross}, \citenamefont {Samushia}, \citenamefont {Howlett}, \citenamefont {Percival}, \citenamefont {Burden},\ and\ \citenamefont {Manera}}]{Ross:2014qpa}%
  \BibitemOpen
  \bibfield  {author} {\bibinfo {author} {\bibfnamefont {A.~J.}\ \bibnamefont {Ross}}, \bibinfo {author} {\bibfnamefont {L.}~\bibnamefont {Samushia}}, \bibinfo {author} {\bibfnamefont {C.}~\bibnamefont {Howlett}}, \bibinfo {author} {\bibfnamefont {W.~J.}\ \bibnamefont {Percival}}, \bibinfo {author} {\bibfnamefont {A.}~\bibnamefont {Burden}},\ and\ \bibinfo {author} {\bibfnamefont {M.}~\bibnamefont {Manera}},\ }\bibfield  {title} {\bibinfo {title} {{The clustering of the SDSS DR7 main Galaxy sample \textendash{} I. A 4 per cent distance measure at $z = 0.15$}},\ }\href {https://doi.org/10.1093/mnras/stv154} {\bibfield  {journal} {\bibinfo  {journal} {Mon. Not. Roy. Astron. Soc.}\ }\textbf {\bibinfo {volume} {449}},\ \bibinfo {pages} {835} (\bibinfo {year} {2015})},\ \Eprint {https://arxiv.org/abs/1409.3242} {arXiv:1409.3242 [astro-ph.CO]} \BibitemShut {NoStop}%
\bibitem [{\citenamefont {de~Sainte~Agathe}\ \emph {et~al.}(2019)\citenamefont {de~Sainte~Agathe} \emph {et~al.}}]{deSainteAgathe:2019voe}%
  \BibitemOpen
  \bibfield  {author} {\bibinfo {author} {\bibfnamefont {V.}~\bibnamefont {de~Sainte~Agathe}} \emph {et~al.},\ }\bibfield  {title} {\bibinfo {title} {{Baryon acoustic oscillations at z = 2.34 from the correlations of Ly$\alpha$ absorption in eBOSS DR14}},\ }\href {https://doi.org/10.1051/0004-6361/201935638} {\bibfield  {journal} {\bibinfo  {journal} {Astron. Astrophys.}\ }\textbf {\bibinfo {volume} {629}},\ \bibinfo {pages} {A85} (\bibinfo {year} {2019})},\ \Eprint {https://arxiv.org/abs/1904.03400} {arXiv:1904.03400 [astro-ph.CO]} \BibitemShut {NoStop}%
\bibitem [{\citenamefont {Blomqvist}\ \emph {et~al.}(2019)\citenamefont {Blomqvist} \emph {et~al.}}]{Blomqvist:2019rah}%
  \BibitemOpen
  \bibfield  {author} {\bibinfo {author} {\bibfnamefont {M.}~\bibnamefont {Blomqvist}} \emph {et~al.},\ }\bibfield  {title} {\bibinfo {title} {{Baryon acoustic oscillations from the cross-correlation of Ly$\alpha$ absorption and quasars in eBOSS DR14}},\ }\href {https://doi.org/10.1051/0004-6361/201935641} {\bibfield  {journal} {\bibinfo  {journal} {Astron. Astrophys.}\ }\textbf {\bibinfo {volume} {629}},\ \bibinfo {pages} {A86} (\bibinfo {year} {2019})},\ \Eprint {https://arxiv.org/abs/1904.03430} {arXiv:1904.03430 [astro-ph.CO]} \BibitemShut {NoStop}%
\bibitem [{\citenamefont {Cuceu}\ \emph {et~al.}(2019)\citenamefont {Cuceu}, \citenamefont {Farr}, \citenamefont {Lemos},\ and\ \citenamefont {Font-Ribera}}]{Cuceu:2019for}%
  \BibitemOpen
  \bibfield  {author} {\bibinfo {author} {\bibfnamefont {A.}~\bibnamefont {Cuceu}}, \bibinfo {author} {\bibfnamefont {J.}~\bibnamefont {Farr}}, \bibinfo {author} {\bibfnamefont {P.}~\bibnamefont {Lemos}},\ and\ \bibinfo {author} {\bibfnamefont {A.}~\bibnamefont {Font-Ribera}},\ }\bibfield  {title} {\bibinfo {title} {{Baryon Acoustic Oscillations and the Hubble Constant: Past, Present and Future}},\ }\href {https://doi.org/10.1088/1475-7516/2019/10/044} {\bibfield  {journal} {\bibinfo  {journal} {JCAP}\ }\textbf {\bibinfo {volume} {10}},\ \bibinfo {pages} {044}},\ \Eprint {https://arxiv.org/abs/1906.11628} {arXiv:1906.11628 [astro-ph.CO]} \BibitemShut {NoStop}%
\bibitem [{\citenamefont {Scolnic}\ \emph {et~al.}(2018)\citenamefont {Scolnic} \emph {et~al.}}]{Pan-STARRS1:2017jku}%
  \BibitemOpen
  \bibfield  {author} {\bibinfo {author} {\bibfnamefont {D.~M.}\ \bibnamefont {Scolnic}} \emph {et~al.} (\bibinfo {collaboration} {Pan-STARRS1}),\ }\bibfield  {title} {\bibinfo {title} {{The Complete Light-curve Sample of Spectroscopically Confirmed SNe Ia from Pan-STARRS1 and Cosmological Constraints from the Combined Pantheon Sample}},\ }\href {https://doi.org/10.3847/1538-4357/aab9bb} {\bibfield  {journal} {\bibinfo  {journal} {Astrophys. J.}\ }\textbf {\bibinfo {volume} {859}},\ \bibinfo {pages} {101} (\bibinfo {year} {2018})},\ \Eprint {https://arxiv.org/abs/1710.00845} {arXiv:1710.00845 [astro-ph.CO]} \BibitemShut {NoStop}%
\bibitem [{\citenamefont {Camarena}\ and\ \citenamefont {Marra}(2021)}]{Camarena:2021jlr}%
  \BibitemOpen
  \bibfield  {author} {\bibinfo {author} {\bibfnamefont {D.}~\bibnamefont {Camarena}}\ and\ \bibinfo {author} {\bibfnamefont {V.}~\bibnamefont {Marra}},\ }\bibfield  {title} {\bibinfo {title} {{On the use of the local prior on the absolute magnitude of Type Ia supernovae in cosmological inference}},\ }\href {https://doi.org/10.1093/mnras/stab1200} {\bibfield  {journal} {\bibinfo  {journal} {Mon. Not. Roy. Astron. Soc.}\ }\textbf {\bibinfo {volume} {504}},\ \bibinfo {pages} {5164} (\bibinfo {year} {2021})},\ \Eprint {https://arxiv.org/abs/2101.08641} {arXiv:2101.08641 [astro-ph.CO]} \BibitemShut {NoStop}%
\bibitem [{\citenamefont {Pisanti}\ \emph {et~al.}(2008)\citenamefont {Pisanti}, \citenamefont {Cirillo}, \citenamefont {Esposito}, \citenamefont {Iocco}, \citenamefont {Mangano}, \citenamefont {Miele},\ and\ \citenamefont {Serpico}}]{Pisanti:2007hk}%
  \BibitemOpen
  \bibfield  {author} {\bibinfo {author} {\bibfnamefont {O.}~\bibnamefont {Pisanti}}, \bibinfo {author} {\bibfnamefont {A.}~\bibnamefont {Cirillo}}, \bibinfo {author} {\bibfnamefont {S.}~\bibnamefont {Esposito}}, \bibinfo {author} {\bibfnamefont {F.}~\bibnamefont {Iocco}}, \bibinfo {author} {\bibfnamefont {G.}~\bibnamefont {Mangano}}, \bibinfo {author} {\bibfnamefont {G.}~\bibnamefont {Miele}},\ and\ \bibinfo {author} {\bibfnamefont {P.~D.}\ \bibnamefont {Serpico}},\ }\bibfield  {title} {\bibinfo {title} {{PArthENoPE: Public Algorithm Evaluating the Nucleosynthesis of Primordial Elements}},\ }\href {https://doi.org/10.1016/j.cpc.2008.02.015} {\bibfield  {journal} {\bibinfo  {journal} {Comput. Phys. Commun.}\ }\textbf {\bibinfo {volume} {178}},\ \bibinfo {pages} {956} (\bibinfo {year} {2008})},\ \Eprint {https://arxiv.org/abs/0705.0290} {arXiv:0705.0290 [astro-ph]} \BibitemShut {NoStop}%
\bibitem [{\citenamefont {Consiglio}\ \emph {et~al.}(2018)\citenamefont {Consiglio}, \citenamefont {de~Salas}, \citenamefont {Mangano}, \citenamefont {Miele}, \citenamefont {Pastor},\ and\ \citenamefont {Pisanti}}]{Consiglio:2017pot}%
  \BibitemOpen
  \bibfield  {author} {\bibinfo {author} {\bibfnamefont {R.}~\bibnamefont {Consiglio}}, \bibinfo {author} {\bibfnamefont {P.~F.}\ \bibnamefont {de~Salas}}, \bibinfo {author} {\bibfnamefont {G.}~\bibnamefont {Mangano}}, \bibinfo {author} {\bibfnamefont {G.}~\bibnamefont {Miele}}, \bibinfo {author} {\bibfnamefont {S.}~\bibnamefont {Pastor}},\ and\ \bibinfo {author} {\bibfnamefont {O.}~\bibnamefont {Pisanti}},\ }\bibfield  {title} {\bibinfo {title} {{PArthENoPE reloaded}},\ }\href {https://doi.org/10.1016/j.cpc.2018.06.022} {\bibfield  {journal} {\bibinfo  {journal} {Comput. Phys. Commun.}\ }\textbf {\bibinfo {volume} {233}},\ \bibinfo {pages} {237} (\bibinfo {year} {2018})},\ \Eprint {https://arxiv.org/abs/1712.04378} {arXiv:1712.04378 [astro-ph.CO]} \BibitemShut {NoStop}%
\bibitem [{\citenamefont {Akaike}(1974)}]{1100705}%
  \BibitemOpen
  \bibfield  {author} {\bibinfo {author} {\bibfnamefont {H.}~\bibnamefont {Akaike}},\ }\bibfield  {title} {\bibinfo {title} {A new look at the statistical model identification},\ }\href {https://doi.org/10.1109/TAC.1974.1100705} {\bibfield  {journal} {\bibinfo  {journal} {IEEE Transactions on Automatic Control}\ }\textbf {\bibinfo {volume} {19}},\ \bibinfo {pages} {716} (\bibinfo {year} {1974})}\BibitemShut {NoStop}%
\bibitem [{\citenamefont {Ballardini}\ \emph {et~al.}(2022)\citenamefont {Ballardini}, \citenamefont {Finelli},\ and\ \citenamefont {Sapone}}]{Ballardini:2021evv}%
  \BibitemOpen
  \bibfield  {author} {\bibinfo {author} {\bibfnamefont {M.}~\bibnamefont {Ballardini}}, \bibinfo {author} {\bibfnamefont {F.}~\bibnamefont {Finelli}},\ and\ \bibinfo {author} {\bibfnamefont {D.}~\bibnamefont {Sapone}},\ }\bibfield  {title} {\bibinfo {title} {{Cosmological constraints on the gravitational constant}},\ }\href {https://doi.org/10.1088/1475-7516/2022/06/004} {\bibfield  {journal} {\bibinfo  {journal} {JCAP}\ }\textbf {\bibinfo {volume} {06}}\bibfield  {number} {\bibinfo  {number} { (06)},\ \bibinfo {pages} {004}},\ }\Eprint {https://arxiv.org/abs/2111.09168} {arXiv:2111.09168 [astro-ph.CO]} \BibitemShut {NoStop}%
\bibitem [{\citenamefont {Ballardini}\ and\ \citenamefont {Finelli}(2022)}]{Ballardini:2021eox}%
  \BibitemOpen
  \bibfield  {author} {\bibinfo {author} {\bibfnamefont {M.}~\bibnamefont {Ballardini}}\ and\ \bibinfo {author} {\bibfnamefont {F.}~\bibnamefont {Finelli}},\ }\bibfield  {title} {\bibinfo {title} {{Type Ia supernovae data with scalar-tensor gravity}},\ }\href {https://doi.org/10.1103/PhysRevD.106.063531} {\bibfield  {journal} {\bibinfo  {journal} {Phys. Rev. D}\ }\textbf {\bibinfo {volume} {106}},\ \bibinfo {pages} {063531} (\bibinfo {year} {2022})},\ \Eprint {https://arxiv.org/abs/2112.15126} {arXiv:2112.15126 [astro-ph.CO]} \BibitemShut {NoStop}%
\bibitem [{\citenamefont {Di~Valentino}\ \emph {et~al.}(2021{\natexlab{c}})\citenamefont {Di~Valentino} \emph {et~al.}}]{DiValentino:2020vvd}%
  \BibitemOpen
  \bibfield  {author} {\bibinfo {author} {\bibfnamefont {E.}~\bibnamefont {Di~Valentino}} \emph {et~al.},\ }\bibfield  {title} {\bibinfo {title} {{Cosmology intertwined III: $f\sigma_8$ and $S_8$}},\ }\href {https://doi.org/10.1016/j.astropartphys.2021.102604} {\bibfield  {journal} {\bibinfo  {journal} {Astropart. Phys.}\ }\textbf {\bibinfo {volume} {131}},\ \bibinfo {pages} {102604} (\bibinfo {year} {2021}{\natexlab{c}})},\ \Eprint {https://arxiv.org/abs/2008.11285} {arXiv:2008.11285 [astro-ph.CO]} \BibitemShut {NoStop}%
\end{thebibliography}%
\end{document}